\documentclass[11pt]{article}

\usepackage{algorithm}  
\usepackage{algorithmic}
\usepackage{epsfig,subfigure}
\usepackage{amssymb, amsmath}
\usepackage{color}
\usepackage{graphicx}

\usepackage{amssymb}
\usepackage{amsmath,amsfonts,amssymb}
\usepackage{verbatim}
\usepackage{stfloats}
\usepackage[bookmarks=false]{}

\newtheorem{theorem}{Theorem}
\newtheorem{corollary}{Corollary}

\newtheorem{remark}{Remark}

\newtheorem{example}{Example}

\newcommand\blfootnote[1]{%
  \begingroup
  \renewcommand\thefootnote{}\footnote{#1}%
  \addtocounter{footnote}{-1}%
  \endgroup
}

\begin{document}
\date{}
\title{On the Optimality of Treating Interference as Noise: \\Compound Interference Networks
\author{ \normalsize Chunhua Geng and Syed A. Jafar  \\
}}

\maketitle
\blfootnote{Chunhua Geng (email: chunhug@uci.edu) and Syed A. Jafar (email: syed@uci.edu) are with the Center of Pervasive Communications and Computing (CPCC) in the Department of Electrical Engineering and Computer Science (EECS) at the University of California Irvine. }

\begin{abstract}
In a $K$-user Gaussian interference channel, it has been shown by Geng et al. that if for each user the desired signal strength is no less than the sum of the strengths of the strongest interference from this user and the strongest interference to this user (all values in dB scale), then power control and treating interference as noise (TIN) is optimal from the perspective of generalized degrees of freedom (GDoF) and achieves the entire channel capacity region to within a constant gap. In this work, we  generalize the optimality of TIN to compound networks. We show that for a  $K$-user compound Gaussian interference channel, if in every possible state for each receiver, the channel always satisfies the TIN-optimality condition identified by Geng et al., then the GDoF region of the compound channel is the intersection of the GDoF regions of all possible network realizations, which is achievable by power control and TIN. Furthermore, we demonstrate that for a general $K$-user compound interference channel, regardless of the number of states of each receiver, we can always construct a counterpart $K$-user regular interference channel that has the same TIN region as the original compound channel. The regular interference channel has only one state for each receiver, which may be different from all of the original states. Solving the GDoF-based power control problem for the compound channel is equivalent to solving the same problem in its regular counterpart. Exploring the power control problem further we develop a centralized power control scheme  for  $K$-user compound interference channels, to achieve all the Pareto optimal GDoF tuples. Finally, based on this scheme, we devise an iterative power control algorithm which requires at most $K$ updates  to obtain the globally optimal power allocation for  any feasible GDoF tuple.
\end{abstract}

\newpage

\section{Introduction}

It is well known (see, e.g., \cite{Motahari_Khandani_TIN, Annapureddy_Veeravalli_TIN, Kramer_TIN, Annapureddy_Veeravalli_MIMO, Shang_Kramer_vector, Shang_Kramer_parallel, Tse_GDoF, Jafar_GDoF,Huang_GDoF_X})  that the simple scheme of treating interference as noise (TIN) in Gaussian interference networks is information theoretically optimal when the interference is sufficiently weak. Such results are quite remarkable since they provide exact capacity characterizations, which are  rare in network information theory. However, the difficulty of pursuing the exact capacity metric manifests itself through various limitations --- the results are typically limited to sum-capacity (as opposed to capacity region),  power control is not considered (all transmitting nodes use all available power), and the regime where the exact optimality of TIN is established tends to be rather small. 

In contrast, by pursuing approximate capacity characterizations, recent work by Geng et al. in  \cite{Geng_TIN} identifies a much broader regime where TIN is optimal in the generalized degrees of freedom (GDoF) sense, and  within a constant gap to exact capacity. Remarkably, the approximate optimality of TIN established in \cite{Geng_TIN} is not only for sum-capacity but for the entire capacity region, and it fully incorporates power control. In fact power control is a crucial aspect of the TIN scheme, since it is generally not optimal for all nodes to transmit at full power if interference is treated as noise. Specifically, it is shown in \cite{Geng_TIN} that in a $K$-user Gaussian interference network, if for each user the desired signal strength is larger than or equal to the sum of the strengths of the strongest interference from this user and the strongest interference to this user (all values in dB scale), then TIN (with power control) is optimal from the perspective of GDoF and achieves the entire  capacity region to within a constant gap. Furthermore, by essentially a Fourier-Motzkin elimination of the power control variables (accomplished in \cite{Geng_TIN} by applying the potential theorem), a direct representation of the GDoF region achievable through TIN  is obtained. This is particularly useful because while rate regions achievable through TIN have  been investigated for decades because of their obvious practical significance, the main complication has been the coupling of achievable rates and the  transmit powers, which requires joint optimization over both. De-coupling the rate (GDoF) region from power control variables allows direct rate optimizations, seemingly ignoring power control variables while in fact automatically optimizing over those as well.

\begin{figure}[h]
\begin{center}
 \includegraphics[width= 5 cm]{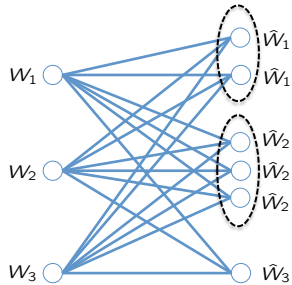}
 \caption{A $3$-user compound interference channel, where Receiver 1, 2, and 3 have 2, 3 and 1 possible  states, respectively. For each receiver, the multiple potential states may also be interpreted as different users which intend to decode the common message from the corresponding transmitter. For this compound interference channel,  there are totally $2\times3\times1 =6$ possible network realizations (or states),  each of which corresponds to a $3$-user regular interference channel. }
\label{compound1}
\end{center}
\end{figure}

In this work, we generalize the question of optimality of TIN to compound networks \cite{Lapidoth_Compound, Compound_BC, Gou_Compound, Ali_Compound, Raja_Compound_IC}, with focus on interference channels. In a compound network, each receiver is associated with a set of states (a receiver's state is identified by the channel realizations associated with that receiver). The sets of possible states for each receiver are globally known a-priori, however the transmitters are unaware of the particular realization chosen from within these sets. Thus, a reliable coding scheme must guarantee vanishing probability of error for every possible realization of each receiver from its given set of possible states. Equivalently, the compound interference channel can be described as having potentially multiple receivers for each message. This is known as the multiple multicast, or multiple groupcast setting \cite{Maleki_Cadambe_Jafar_Index} and is of interest in its own right.  An example of a $3$-user compound interference channel is illustrated in Fig.~\ref{compound1}. 

Of particular interest for this work are compound networks where each possible state of the network individually satisfies the TIN-optimality condition of Geng et al.  \cite{Geng_TIN}. Is TIN still optimal for such a compound interference network as a whole, when all possible states are simultaneously considered? More generally,  the implications of the compound setting on the GDoF region achievable through TIN (even if the TIN optimality conditions are not satisfied) and  GDoF-optimal power control are also of interest and will be explored in this work. 

We begin with an overview of what makes the optimality of TIN for compound networks non-trivial, followed by a summary of the main contributions of this work.

\subsection{Challenges Posed by the Compound Setting}
\subsubsection{Optimality of TIN}
Consider a $K$-user compound interference channel which satisfies the TIN-optimality condition of \cite{Geng_TIN} in each possible state, individually.  Recall that with the simple TIN scheme,  each transmitter uses a point-to-point Gaussian codebook with an appropriate power level and each receiver only decodes the signal from its corresponding transmitter and treats the incoming interference as Gaussian noise. Denote by $\mathcal{P}$ the set of all the valid power allocations, $\mathcal{H}$ the set of all the possible network realizations, and $\mathcal{D}(\mathbf{P},\mathbf{H})$ the achievable GDoF region through the TIN scheme for the network realization $\mathbf{H}\in\mathcal{H}$ with power allocation $\mathbf{P}\in\mathcal{P}$. 

First, consider the achievability argument. Since the rate for each message  in the TIN scheme is only limited by the minimum SINR (signal to interference and noise power ratio) across all states of the intended receiver, it is evident that for any valid power allocation $\mathbf{P}\in\mathcal{P}$, $\cap_{\mathbf{H}\in\mathcal{H}}\mathcal{D}(\mathbf{P},\mathbf{H})$ is achievable. For instance, consider a $2$-user compound interference channel, where each receiver has $2$ generic states, so that  there are $4$ possible network realizations. For the transmit power allocation $\mathbf{P}$, assume that the obtained GDoF values are $d_1'$ and $d_1''$ at the two states of Receiver $1$, and $d_2'$ and $d_2''$ at the two states of Receiver $2$. Without loss of generality, we also assume that $d_1'\leq d_1''$ and $d_2'\leq d_2''$. With this power allocation all the GDoF tuples in the region $\{(d_1,d_2): 0\leq d_1\leq d_1', 0\leq d_2\leq d_2'\}$ are achievable for the compound channel, which is evidently the intersection of the achievable GDoF regions of all the four possible network realizations with power allocation $\mathbf{P}$. Taking the union of achievable rates over  all the valid power allocations, we obtain the following inner bound on the GDoF region.
\begin{align}
\mathrm{Inner~Bound}= \cup_{\mathbf{P}\in\mathcal{P}}\cap_{\mathbf{H}\in\mathcal{H}}\mathcal{D}(\mathbf{P},\mathbf{H})\label{INBound}
\end{align}

Next, consider the converse argument. Recall the assumption that  the compound network satisfies the TIN-optimality condition of  \cite{Geng_TIN} in each possible state. In other words, each possible network state, by itself is TIN-optimal. For a given network state $\mathbf{H}\in\mathcal{H}$, since  TIN is optimal, the entire GDoF region is achievable through power control and TIN, and it can be expressed as $\cup_{\mathbf{P}\in\mathcal{P}}\mathcal{D}(\mathbf{P},\mathbf{H})$. Furthermore, since the GDoF region for each state is an outer bound for the GDoF region of the compound setting, by taking the intersection of the GDoF regions of all network states, we get the following outer bound on the GDoF region.
\begin{align}
\mathrm{Outer~Bound}= \cap_{\mathbf{H}\in\mathcal{H}}\cup_{\mathbf{P}\in\mathcal{P}}\mathcal{D}(\mathbf{P},\mathbf{H})\label{OTBound}
\end{align}

Note that while both the inner bound and the outer bound for the GDoF region involve union over power allocations and intersection over network states, the inner bound is the union of intersections whereas the outer bound is the intersection of unions. In general a union of intersections is (possibly strictly) smaller than or equal to an intersection of unions, consistent with their roles as inner and outer bounds, respectively.\footnote{Consider a simple example where $\mathcal{P}=\{\mathbf{P_1},\mathbf{P_2}\}$, $\mathcal{H}=\{\mathbf{H_1},\mathbf{H_2}\}$, $\mathcal{D}(\mathbf{P_1},\mathbf{H_1})=\{1\}$, $\mathcal{D}(\mathbf{P_1},\mathbf{H_2})=\{2\}$, $\mathcal{D}(\mathbf{P_2},\mathbf{H_1})=\{2\}$, and $\mathcal{D}(\mathbf{P_2},\mathbf{H_2})=\{1\}$.  It is easy to check that the right hand sides of (\ref{INBound}) and (\ref{OTBound}) are $\phi$ and $\{1,2\}$, respectively, which are not equal to each other.} So the main challenge in settling the optimality of TIN for compound interference networks is to prove that, in our context, the two are indeed, identical. 

\subsubsection{Optimal Power Control}

Transmit power control is of paramount importance for interference management in wireless networks using the simple scheme of TIN.  There is an abundance of literature on the optimization of power allocations. References \cite{Zander_PC, Zander_DPC} consider the carrier-to-interference (C/I) balancing problem, which intends to maximize the minimal achievable C/I ratios of all the users with the minimized overall power consumption. For a given feasible SINR target, distributed power control algorithms are devised in \cite{Foschini_DPC,Yates_SIF,Bambos_DPC} to obtain the optimal transmit power allocation.  In particular, there is much interest in joint SINR (or rate) assignment (e.g., for sum-rate maximization) and power control problem, studied recently in \cite{Boyd_Genie, Chiang_PC, Luo_Spectrum, Chiang_Sumrate}. The joint optimization problem is difficult to solve because: (i) although subclasses solvable in polynomial time are identified, it is non-convex and NP-hard in general \cite{Luo_Spectrum}, and one standard approach to deal with this issue is using high SINR approximations to formulate a convex optimization problem \cite{Chiang_PC}; (ii) the feasible SINR region is highly coupled with power allocations across all users. There are several studies attempting to derive the entire feasible SINR (or rate) region \cite{Chiang_SIR_assignment,Khandani_SINR_region, Chara_TIN}, and a well-known result is the log-convexity of the feasible SINR region \cite{Sung_Log_convex}.\footnote{It is noteworthy that the feasible SINR region considered in \cite{Sung_Log_convex} corresponds to the case where all users in the network are active and thus attain strictly positive SINRs. This is because the concerned matrix $\mathbf{B}$ in \cite{Sung_Log_convex}  is irreducible. In fact, under the GDoF framework, this feasible SINR region in the log scale essentially corresponds to the counterpart polyhedral TIN region $\mathcal{P}$ defined in \cite{Geng_TIN}, which is also always convex. However, as shown in \cite{Geng_TIN}, in general the feasible GDoF region is not convex without time-sharing, if we allow certain users in the network to be silent.}  The sum-rate maximization problem, is so far only solved in the $K$-user fully symmetric network \cite{Hanly_Symmetric} and $2$-user asymmetric network \cite{Khandani_2user,Gesbert_2user}.  For the general $K$-user asymmetric networks, various algorithms based on geometric programming, game theory, etc. (see \cite{Chiang_GP,Poor_GT} and references therein), are developed. 

Most relevant to this work are the  results of \cite{Geng_TIN} for  fully asymmetric $K$-user interference channels, where the entire TIN GDoF region is fully characterized.  Within this GDoF  framework, a remarkable advantage as shown in \cite{Geng_TIN} is that using (essentially) Fourier-Motzkin Elimination, the power allocation variables can be entirely eliminated and thus the feasible GDoF  region characterization and power control problem are decoupled. Therefore, a closed-form feasible GDoF region depending only upon the effective channel gains is derived, with which the optimal GDoF assignment under a given objective function (e.g., to maximize sum-GDoF) is relatively easy to solve. The only problem left is power control, i.e., how to allocate transmit powers (efficiently, since multiple solutions may be possible in general) to achieve the target GDoF tuples. %A fixed point iterative algorithm for optimal power allocation is presented in \cite{Geng_TIN} and is guaranteed to converge.

The compound setting adds another layer of complexity on the optimal power allocation problem, since it is not clear a-priori where the bottlenecks lie. To identify the bottlenecks, simplify the compound network setting as much as possible, and to allocate power optimally to achieve any desired GDoF objective function are the goals that we pursue in this context.

% {\color{blue} In general the compound setting does not reduce to just the worst case among all the possible network states. Indeed, as we will show in this work, it is possible that the bottlenecks may not lie in any one of the possible network states, but rather in a non-trivial mixture of states.}

\subsection{Main Contributions}

\subsubsection{The optimality of TIN for compound interference channels} We show that for  $K$-user compound Gaussian interference channels, if  each possible network realization  satisfies the TIN-optimality condition of \cite{Geng_TIN}, then  TIN  achieves the entire GDoF region of the compound setting, which is the intersection of the GDoF regions of all possible network realizations (see Theorem \ref{T1_IC}). The result is derived from a non-trivial argument based on the potential theorem in \cite{Potential}, which builds upon the application of the potential theorem in \cite{Geng_TIN}. 
 For a general compound interference channel which may be not TIN-optimal, the entire TIN region (i.e., the achievable GDoF region through the TIN scheme) is also fully characterized (see Remark \ref{remark1}). 

\subsubsection{Simplification of the compound interference channel into a regular counterpart} 
\begin{figure}[h]
\centering
\subfigure[]{
\includegraphics[width= 6 cm]{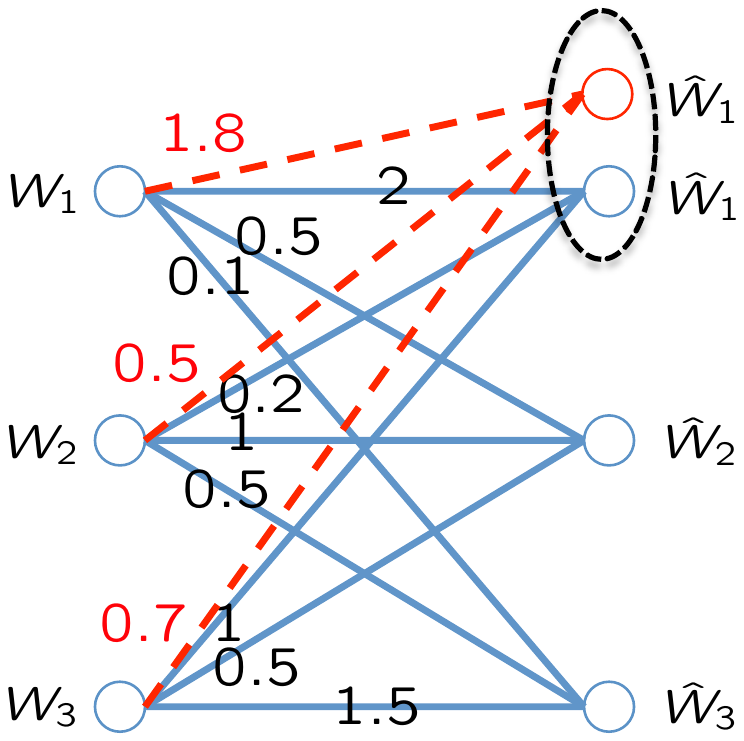}
\label{TIN_ex1_2_2}}
\subfigure[]{
\includegraphics[width= 4.8 cm]{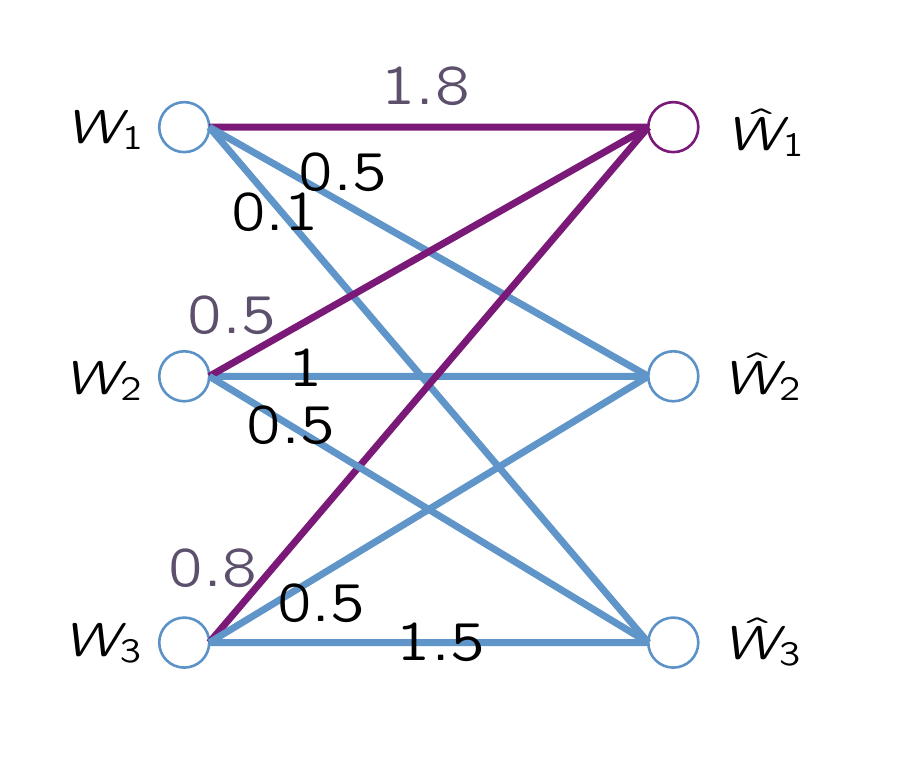}
\label{reg_mtv}}
\caption[]{
\subref{TIN_ex1_2_2} A $3$-user compound interference channel. The value on each link denotes its channel strength level, which is defined in Section \ref{S_2} formally; \subref{reg_mtv} The counterpart $3$-user regular interference channel for the compound channel in Fig.~\ref{TIN_ex1_2_2}. } 
\end{figure}

We  show that for any $K$-user compound interference channel, regardless of the number of states for each receiver, we can always construct a counterpart $K$-user regular (where the network has only one state) interference channel, which has the same TIN region as that  compound channel. Remarkably, the counterpart regular interference channel is in general none of the possible network realizations of the compound channel. We show that to solve the GDoF-optimal power control problem for a $K$-user compound interference channel, we only need to solve the problem in its regular counterpart (see Theorem \ref{th-equ}). In other words, from the GDoF perspective, the power control and TIN problems for compound and regular interference channels are equivalent, which  significantly reduces the computational complexity of the GDoF-based power control in the compound setting.  For instance, for the $3$-user compound interference channel in Fig.~\ref{TIN_ex1_2_2},  we can construct its counterpart $3$-user regular interference channel, which is given in Fig.~\ref{reg_mtv}.   Not only the two channels in Fig.~\ref{TIN_ex1_2_2} and \ref{reg_mtv} have the same TIN region, but also solving the GDoF-based power control problem for one is equivalent to solving the problem for the other. However, it should be noted that the compound channel and its regular counterpart are only equivalent in terms of the achievability of the TIN scheme. In Section \ref{sec_counterexp}, we give an example in which  the compound interference channel is \emph{not} TIN-optimal (i.e., there exist other achievable schemes outperforming the TIN scheme), while its regular counterpart satisfies the TIN-optimality condition of  \cite{Geng_TIN}.   

\subsubsection{GDoF-based power control algorithms} Taking advantage of the simplification of the compound setting to the regular case, and using additional insights from the potential theorem, we design GDoF-based power control algorithms for compound networks. Specifically, for general $K$-user compound interference channels, we propose a centralized power control scheme to achieve any \emph{Pareto optimal} GDoF tuple (see Corollary \ref{PGPC}).  This scheme is equivalent to solving a single-source shortest path problem in a carefully designed complete digraph with $K+1$ vertices. Based on this scheme, we further develop an iterative power control algorithm with at most $K$ updates for all the achievable GDoF tuples to obtain their \emph{globally optimal} power allocations (see Theorem \ref{GPU_converge} and Corollary  \ref{coro_GPU_converge}). 

\bigskip

\subsection{Notation} Throughout this paper, for any positive integer $Z$, $\langle Z  \rangle$ denotes the set $\{1,2,...,Z\}$, and for any real number $a$, $(a)^+$ and $\max\{0,a\}$ are used interchangeably. For two vectors $\mathbf{u}$ and $\mathbf{v}$, we say that $\mathbf{u}$ dominates $\mathbf{v}$ if $\mathbf{u}\geq \mathbf{v}$, where $\geq$ denotes componentwise inequality. In addition, unless otherwise stated, all logarithms are to the base $2$. 

\section{Channel Model and Preliminaries}\label{S_2}
Consider a $K$-user Gaussian interference channel consisting of $K$ transmitter-receiver pairs.  The channel coefficients associated with Receiver $k$, $\forall k\in \langle K \rangle$, are denoted as a vector $(\tilde{h}_{k1}, \tilde{h}_{k2},...,\tilde{h}_{kK})$, which is drawn from a finite set $\mathcal{L}_k$ with cardinality $L_k$. We assume that the channel coefficients  remain fixed during the transmission. In addition, while the transmitters are unaware of the specific channel realizations, knowledge of $\mathcal{L}_k$ is  assumed to be globally available. The receivers are assumed to have perfect channel state information. In this compound setting, reliable communication needs to be guaranteed simultaneously for all possible channel realizations. 

For Receiver $k$, the received signal in state $l_k$ is given by
\begin{equation}
\label{ori_channel}
Y_k^{[l_k]}(t)=\sum_{i=1}^K\tilde{h}_{ki}^{[l_k]}\tilde{X}_i(t)+Z_k^{[l_k]}(t),~~~\forall k\in \langle K\rangle, \forall l_k\in \langle L_k\rangle
\end{equation}
where $\tilde{h}_{ki}^{[l_k]}$ is the channel gain value from Transmitter $i$ to Receiver $k$, $\tilde{X}_i(t)$ and $Z_k^{[l_k]}(t)$ are the transmitted symbol of Transmitter $i$ and the additive white circularly symmetric Gaussian noise with zero mean and unit variance at Receiver $k$, respectively, in the $t$-th channel use.  All symbols are complex.  Each Transmitter $i$ is subject to the average power constraint $\mathbb{E}[|\tilde{X}_i(t)|^2]\leq P_i$.

To facilitate the GDoF studies, the standard channel model (\ref{ori_channel}) is translated into an equivalent normalized form following similar approaches in \cite{Tse_GDoF,Geng_TIN}. Define\footnote{Similar to \cite{Geng_TIN}, it is not hard to verify that avoiding negative $\alpha$'s has no impact on the GDoF and constant gap results.}
\begin{align}
\label{ratio}
\alpha_{ki}^{[l_k]}\triangleq\frac{\log(\max\{1,|\tilde{h}_{ki}^{[l_k]}|^2P_i\})}{\log P},~~\forall i,k\in \langle K\rangle, \forall l_k\in\langle L_k\rangle
\end{align}
where $P>1$ is a nominal power value. Then the original channel model (\ref{ori_channel}) can be presented in the following form,
\begin{equation}
\label{equ_channel}
\begin{aligned}
Y_k^{[l_k]}(t)=&\sum_{i=1}^Kh_{ki}^{[l_k]}X_i(t)+Z_k^{[l_k]}(t)\\
=&\sum_{i=1}^K\sqrt{P^{\alpha_{ki}^{[l_k]}}}e^{j\theta_{ki}^{[l_k]}}X_i(t)+Z_k^{[l_k]}(t),~~~\forall k\in \langle K\rangle, \forall l_k\in \langle L_k\rangle
\end{aligned}
\end{equation}
where $X_i(t)$ is the normalized transmit symbol of Transmitter $i$, subject to the unit power constraint, i.e., $\mathbb{E}[|X_i(t)|^2]\leq 1$. $\sqrt{P^{\alpha_{ki}^{[l_k]}}}$ and $\theta_{ki}^{[l_k]}$ are the channel magnitude and phase between Transmitter $i$ and Receiver $k$  under state $l_k$, respectively. As in \cite{Geng_TIN}, the exponent $\alpha_{ki}^{[l_k]}$ is called the channel strength level. In the rest of this paper we will consider the equivalent channel model in (\ref{equ_channel}).

In this $K$-user compound interference channel, Transmitter $i$ intends to send one message $W_i$ to Receiver $i$, $\forall i\in \langle K \rangle$.  All the messages are independent. The size of the message set $\{W_{i}\}$ is denoted by $|W_{i}|$. For codewords spanning $n$ channel uses, the rates $R_{i}=\frac{\log|W_{i}|}{n}$ are achievable if all messages can be decoded simultaneously with arbitrarily small error probability as $n$ grows to infinity regardless of the channel realizations. The channel capacity region $\mathcal{C}$ is the closure of the set of all achievable rate tuples. The sum channel capacity is defined as $R_{\Sigma}\triangleq\max_{\mathcal{C}}\sum_{i=1}^KR_{i}$.  The GDoF region of the $K$-user compound  interference channel is given by
\begin{equation}
\begin{aligned}
\mathcal{D}\triangleq \Big\{(d_{1},d_{2},...,d_{K}):~d_{k}=\lim_{P\rightarrow\infty}\frac{R_k}{\log P},~\forall k\in \langle K\rangle,~(R_{1},R_{2},...,R_{K})\in \mathcal{C}\Big\},
\end{aligned}
\end{equation}
and its sum-GDoF is defined as
\begin{align}
d_{\Sigma}\triangleq\max_{\mathcal{D}}\sum_{i=1}^Kd_{k}
\end{align}

\begin{remark}
The compound interference channel can be also modeled in more general or less general ways. The model adopted in this work assumes that each receiver has multiple possible states, i.e., 
\begin{align}
(\tilde{h}_{k1},...,\tilde{h}_{kK})\in \mathcal{L}_k, ~~\forall k\in\langle K\rangle
\end{align}
A more general way is to allow for all the channel parameters $\tilde{h}_{ki}$'s to be jointly taken from a finite set $\mathcal{L}$, i.e.,
\begin{align}
(\tilde{h}_{11},...,\tilde{h}_{1K},....,\tilde{h}_{K1},...,\tilde{h}_{KK})\in  \mathcal{L}
\end{align}
As explained in \cite{Raja_Compound_IC} (see Proposition 2 in \cite{Raja_Compound_IC}), since the receivers cannot cooperate, it turns out that the latter is no more general than the model assumed in this work. Consider a $2$-user compound interference channel with $|\mathcal{L}|=2$ as an example, where
\begin{align}\label{com_def}
\mathcal{L}=\{(\tilde{h}_{11}^{[1]},\tilde{h}_{12}^{[1]},\tilde{h}_{21}^{[1]},\tilde{h}_{22}^{[1]}),(\tilde{h}_{11}^{[2]},\tilde{h}_{12}^{[2]},\tilde{h}_{21}^{[2]},\tilde{h}_{22}^{[2]})\}
\end{align}
The capacity region of the compound interference channel defined by the set $\mathcal{L}$ in (\ref{com_def}) is the same as that of the $2$-user compound interference channel defined in (\ref{ori_channel}) with $L_1=L_2=2$ and thus $L_1\times L_2 =4$ potential network states, where 
\begin{align}
\mathcal{L}_1 &= \{(\tilde{h}_{11}^{[1]},\tilde{h}_{12}^{[1]}), (\tilde{h}_{11}^{[2]},\tilde{h}_{12}^{[2]}) \}\\
\mathcal{L}_2 &= \{(\tilde{h}_{21}^{[1]},\tilde{h}_{22}^{[1]}), (\tilde{h}_{21}^{[2]},\tilde{h}_{22}^{[2]}) \}
\end{align}
On the other hand, a less general model for the compound setting than the one adopted in this work, is to assume that each channel parameter $\tilde{h}_{ij}$ belongs to a set $\mathcal{S}_{ij}$. This model turns out to be rather trivial from a TIN perspective, since we only need to consider the ``worst case", which is readily attainable:  for $i=j$, $\tilde{h}_{ij}= \min\{|s_{ij}|: s_{ij}\in\mathcal{S}_{ij}\}$, and for $i\neq j$, $\tilde{h}_{ij}= \max\{|s_{ij}|: s_{ij}\in\mathcal{S}_{ij}\}$. 
\end{remark}

\begin{figure}[h]
\begin{center}
\includegraphics[width= 12 cm]{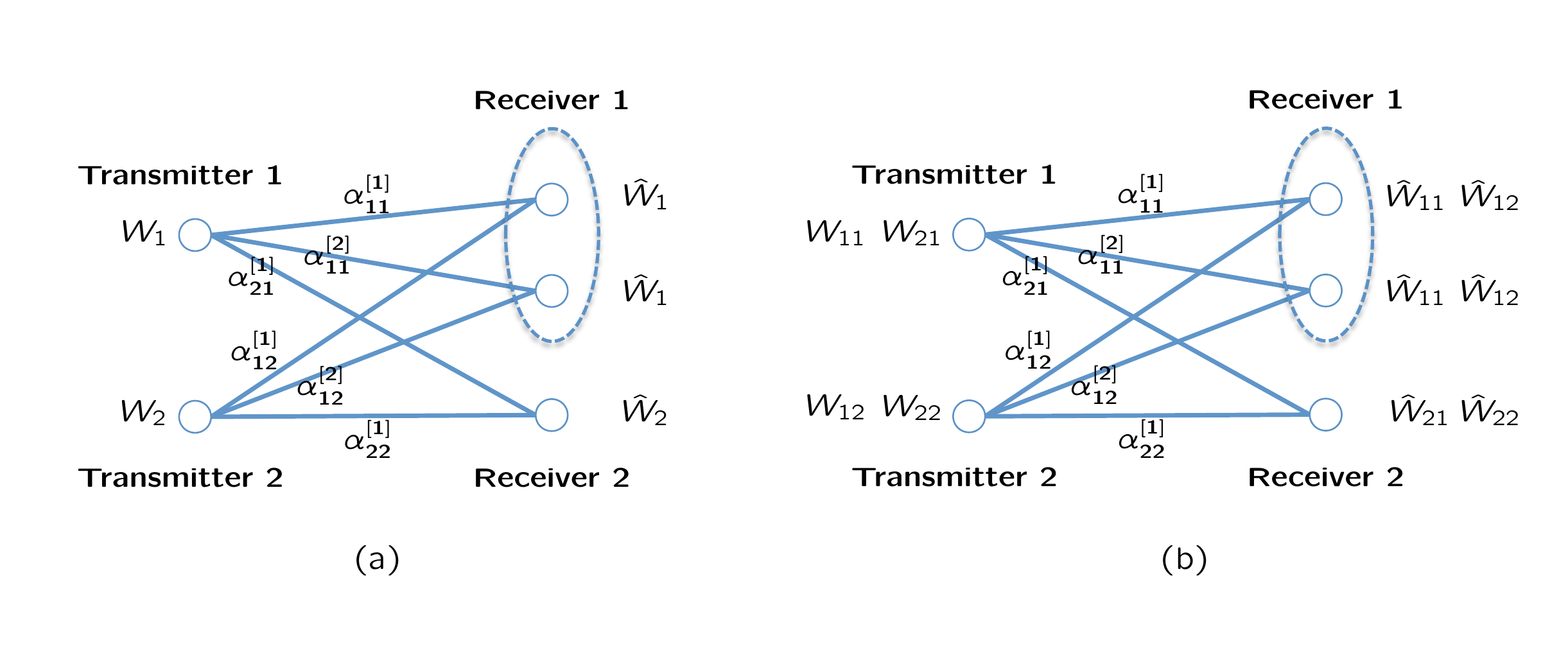}
 \caption{(a) A $2$-user compound interference channel with $L_1=2$ and $L_2=1$; (b) A $2\times 2$ compound $X$ channel with $L_1=2$ and $L_2=1$.}
\label{ex1}
\end{center}
\end{figure}

\begin{example}\label{ex1b}
\emph{A $2$-user compound interference channel with $L_1=2$ and $L_2=1$ is illustrated in Fig.~\ref{ex1}(a). The compound channel can be extended to other message settings, e.g., the $2\times 2$ compound  $X$ channel with $4$ messages as depicted in Fig.~\ref{ex1}(b). }
\end{example}

\subsection{The Achievable GDoF Region through TIN}\label{S_PTIN}
In the $K$-user compound interference channel, we assume the power allocated to Transmitter $k$ is $P^{r_k}$.  Due to the unit power constraint, $r_k\leq 0$, $\forall k\in \langle K\rangle$. The rate achievable by User $k$ through TIN is limited by the smallest SINR across all states possible for this user. So User $k$ can achieve any rate $R_k$ such that, 
\begin{align}
R_k\leq \min_{l_k\in\langle L_k\rangle}\bigg\{\log\bigg(1+\frac{P^{r_k}\times P^{\alpha_{kk}^{[l_k]}}}{1+\sum_{j=1,j\neq k}^KP^{r_j}\times P^{\alpha_{kj}^{[l_k]}}}\bigg)\bigg\},~~\forall k\in \langle K\rangle
\end{align}  
In the GDoF sense, we have 
%\begin{align}
%d_k&=\min_{l_k\in\langle L_k\rangle}\big\{ \max\{0,\alpha_{kk}^{[l_k]}+r_k-(\max_{j:j\neq k}\{\alpha_{kj}^{[l_k]}+r_j\})^+\}\big\},~~\forall k\in \langle K\rangle
%\end{align}
%Since the achievable GDoF value in (\ref{Rop1_g}) can always be reduced, user $k$ can achieve any GDoF value satisfying
\begin{align}\label{Rop1_g}
0\leq d_k&\leq\min_{l_k\in\langle L_k\rangle}\big\{ \max\{0,\alpha_{kk}^{[l_k]}+r_k-(\max_{j:j\neq k}\{\alpha_{kj}^{[l_k]}+r_j\})^+\}\big\},~~ \forall k\in \langle K\rangle
\end{align}
The TIN region, i.e., the achievable GDoF region through the TIN scheme, which is denoted by $\mathcal{P}^*$, is the set of all GDoF tuples $(d_1,d_2,..., d_K)$ for which there exist $r_k$'s,  such that%\footnote{To make the TIN scheme meaningful in the compound setting, we relax the definition of TIN scheme here compared with \cite{Geng_TIN}.}
\begin{alignat*}{2}
r_k&\leq 0,&&\forall k\in \langle K\rangle\\
d_k&\geq 0,&&\forall k\in \langle K\rangle\\
d_k&\leq\min_{l_k\in\langle L_k\rangle}\big\{ \max\{0,\alpha_{kk}^{[l_k]}+r_k-(\max_{j:j\neq k}\{\alpha_{kj}^{[l_k]}+r_j\})^+\}\big\},\:\:&& \forall k\in \langle K\rangle
\end{alignat*}
We will refer to $(P^{r_1}, P^{r_2},...,P^{r_K})$ and $(r_1,r_2,...,r_K)$  as the transmit power vector and transmit power exponent vector, respectively.

Similar to \cite{Geng_TIN},  we also introduce a polyhedral version of the TIN scheme, which is called the \emph{polyhedral TIN scheme}. By requiring $\min_{l_k\in\langle L_k\rangle}\big\{ \alpha_{kk}^{[l_k]}+r_k-(\max_{j:j\neq k}\{\alpha_{kj}^{[l_k]}+r_j\})^+\big\}$ to be no less than $0$ for all users, we can ignore the first $\max\{0,.\}$ term in (\ref{Rop1_g}) and obtain
%\begin{align}\label{P_TIN}
%d_k&=\min_{l_k\in\langle L_k\rangle}\big\{ \alpha_{kk}^{[l_k]}+r_k-(\max_{j:j\neq k}\{\alpha_{kj}^{[l_k]}+r_j\})^+\big\}, ~~\forall k\in \langle K\rangle
%\end{align}
%Again, since the achievable GDoF value in (\ref{P_TIN}) can always be reduced, in the polyhedral TIN scheme user $k$ can achieve any GDoF value satisfying
\begin{align}\label{P_TIN}
0\leq d_k\leq\min_{l_k\in\langle L_k\rangle}\big\{ \alpha_{kk}^{[l_k]}+r_k-(\max_{j:j\neq k}\{\alpha_{kj}^{[l_k]}+r_j\})^+\big\}, ~~\forall k\in \langle K\rangle
\end{align}
We call the new achievable GDoF region after the above modification the \emph{polyhedral TIN region} $\mathcal{P}$.  So $\mathcal{P}$ is the set of all GDoF tuples $(d_1,d_2,...,d_K)$ for which there exist $r_k$'s, $k\in \langle K \rangle$, such that
\begin{alignat*}{2}
r_k&\leq 0,&&\forall k\in \langle K\rangle\\
d_k&\geq 0,&&\forall k\in \langle K\rangle\\
d_k\leq \alpha_{kk}^{[l_k]}+r_k \Leftrightarrow r_k&\geq d_k-\alpha_{kk}^{[l_k]},&&\forall k\in \langle K\rangle,\forall l_k\in\langle L_k\rangle\\
d_k\leq \alpha_{kk}^{[l_k]}+r_k-(\alpha_{kj}^{[l_k]}+r_j) \Leftrightarrow r_k-r_j&\geq (\alpha_{kj}^{[l_k]}-\alpha_{kk}^{[l_k]})+d_k,\:\:&&\forall k,j\in \langle K\rangle, k\neq j, \forall l_k\in\langle L_k\rangle
\end{alignat*}
In general, with this modification we require that the right hand side of (\ref{P_TIN}) is non-negative for all users $k\in \langle K\rangle$, hence we put more constraints on the power exponents $r_i$'s besides the constraints of $r_i\leq 0$, which can only shrink the achievable GDoF region via power control and TIN. In other words, $\mathcal{P}\subseteq \mathcal{P}^*$. Later our results will show that when certain conditions hold, $\mathcal{P}=\mathcal{P}^*$, i.e., the above modification incurs no loss.

\subsection{Potential Graph}\label{S_PG}
For any $K$-user compound interference channel, we can construct a complete digraph $D_p=(V,E)$ with $\sum_kL_k+1$ vertices, where $V$ and $E$ are the sets  of vertices and edges, respectively, and 
\begin{alignat*}{2}
V&=\{v_1^{[1]},...,v_k^{[l_k]},...,v_K^{[L_K]},u\},~~~~~&&\forall k\in\langle K\rangle, \forall l_k\in\langle L_k\rangle\\
E&=E_1\cup E_2 \cup E_3\cup E_4\\
E_1&=\{(v_k^{[l_k]},v_k^{[l_k']})\}, &&\forall k\in\langle K\rangle, \forall l_k,l_k'\in\langle L_k\rangle, l_k\neq l_k'\\%\{(v_1^{[1]},v_1^{[2]}),(v_1^{[2]},v_1^{[1]})\}\\
E_2&=\{(v_k^{[l_k]},v_j^{[l_j]})\},  &&\forall k,j\in\langle K\rangle, k\neq j, \forall l_k\in\langle L_k\rangle,\forall l_j\in\langle L_j\rangle\\ 
E_3&=\{(v_k^{[l_k]},u)\}, &&\forall k\in\langle K\rangle, \forall l_k\in\langle L_k\rangle\\
E_4&=\{(u,v_k^{[l_k]})\}, &&\forall k\in\langle K\rangle, \forall l_k\in\langle L_k\rangle
\end{alignat*}
We assign a length $l(e)$ to each edge $e\in E$ as follows,  
\begin{alignat}{2}
l(v_k^{[l_k]},v_k^{[l_k']})&=0,&&\forall k\in\langle K\rangle, \forall l_k,l_k'\in\langle L_k\rangle, l_k\neq l_k'\label{eq:sameuser}\\
l(v_k^{[l_k]},v_j^{[l_j]})&=(\alpha_{kk}^{[l_k]}-\alpha_{kj}^{[l_k]})-d_k,~~~~~&&\forall k,j\in\langle K\rangle, k\neq j, \forall l_k\in\langle L_k\rangle,\forall l_j\in\langle L_j\rangle\label{eq:samelength}\\
l(v_k^{[l_k]},u)&=\alpha_{kk}^{[l_k]}-d_k,&&\forall k\in\langle K\rangle, \forall l_k\in\langle L_k\rangle\label{eq_lst1}\\
l(u,v_k^{[l_k]})&=0,&&\forall k\in\langle K\rangle, \forall l_k\in\langle L_k\rangle\label{eq:neg}
\end{alignat}
Note that we  denote by $(a,b)$ the edge \emph{from} vertex $a$ \emph{to} vertex $b$.

Henceforth, we call such a complete digraph $D_p$ the \emph{potential graph}.
 In the potential graph $D_p$, denote the lengths of the shortest paths from the vertex $u$ to each vertex ${v}_k^{[1]}$ ($\forall k\in\langle K\rangle$) as  $l_{k,\mathrm{dst}}$. 
 It is notable that due to (\ref{eq:sameuser}) the lengths of the shortest paths from the vertex $u$ to the vertices $v_k^{[l_k]}$ and $v_k^{[l_k']}$ are the same. Also, due to (\ref{eq:neg}), $l_{k,\mathrm{dst}}\leq 0$ for all $k\in\langle K\rangle$.

%In fact, the nomenclature is borrowed from \emph{potential function} in combinatorial optimization \cite{Potential}.

\begin{figure}[h]
\begin{center}
 \includegraphics[width= 6 cm]{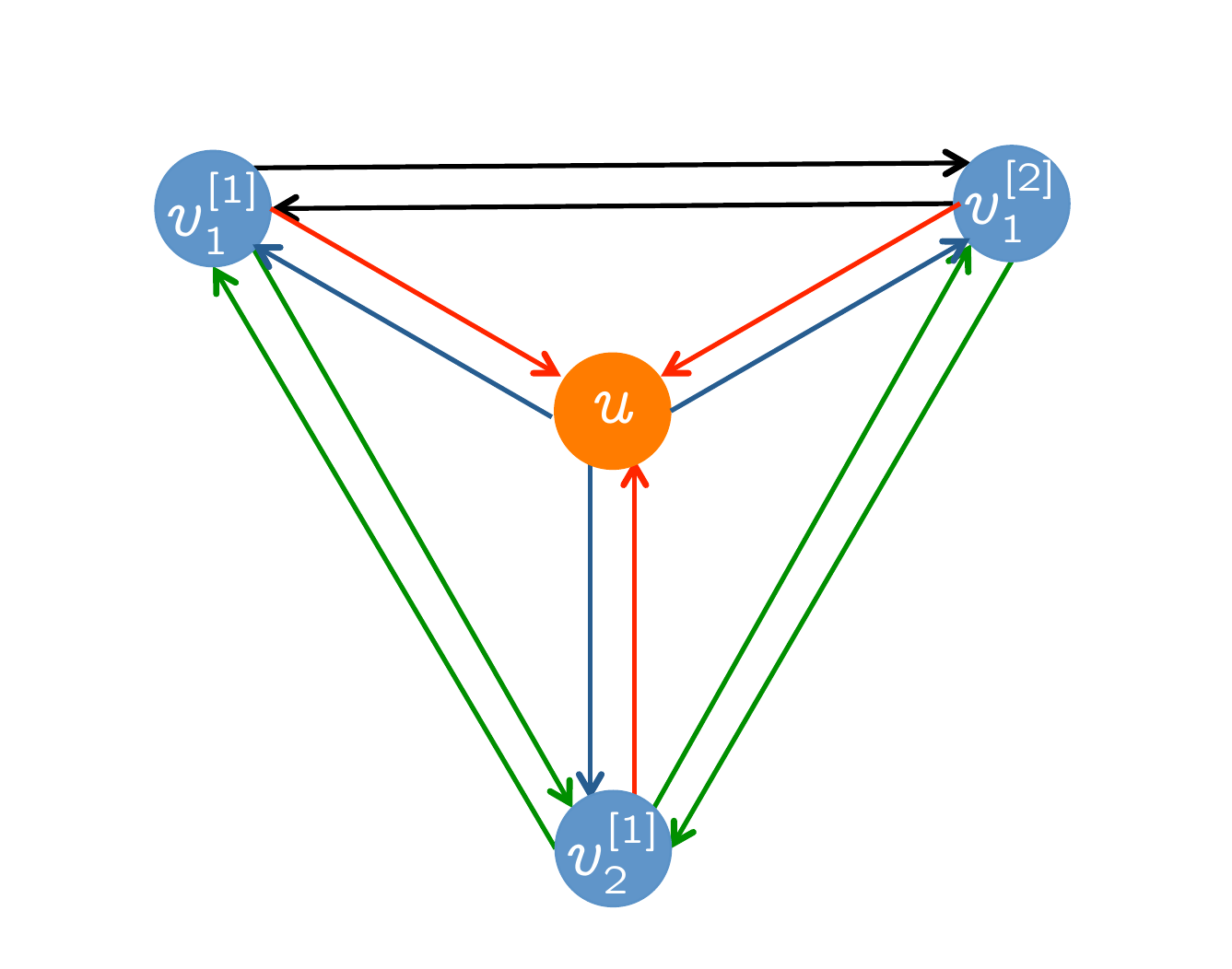}
 \caption{The potential graph $D_p$ for the $2$-user compound interference channel in Fig.~\ref{ex1}(a).  It is comprised of $\sum_{k=1}^2L_k+1=4$ vertices, black edges $E_1$, green edges $E_2$, red edges $E_3$ and blue edges $E_4$.}
\label{potential_graph1}
\end{center}
\end{figure}

\begin{example}\label{ex2b}
\emph{Consider the $2$-user compound interference channel as shown in Fig.~\ref{ex1}(a). Its potential graph $D_p=(V,E)$ is depicted in Fig.~\ref{potential_graph1},  where
\begin{align*}
V&=\{v_1^{[1]},v_1^{[2]},v_2^{[1]},u\}\\
E&=E_1\cup E_2 \cup E_3\cup E_4\\
E_1&=\{(v_1^{[1]},v_1^{[2]}),(v_1^{[2]},v_1^{[1]})\}\\
E_2&=\{(v_1^{[1]},v_2^{[1]}),(v_1^{[2]},v_2^{[1]}),(v_2^{[1]},v_1^{[1]}),(v_2^{[1]},v_1^{[2]})\}\\
E_3&=\{(v_1^{[1]},u),(v_1^{[2]},u),(v_2^{[1]},u)\}\\
E_4&=\{(u,v_1^{[1]}),(u,v_1^{[2]}),(u,v_2^{[1]})\}
\end{align*}
The length $l(e)$ for each edge $e\in E$ is
\begin{align*}
l(v_1^{[1]},v_1^{[2]})&=l(v_1^{[2]},v_1^{[1]})=0\\ 
l(v_1^{[1]},v_2^{[1]})&=(\alpha_{11}^{[1]}-\alpha_{12}^{[1]})-d_1\\
l(v_1^{[2]},v_2^{[1]})&=(\alpha_{11}^{[2]}-\alpha_{12}^{[2]})-d_1\\
l(v_2^{[1]},v_1^{[1]})&=l(v_2^{[1]},v_1^{[2]})=(\alpha_{22}^{[1]}-\alpha_{21}^{[1]})-d_2\\
l(v_i^{[l_i]},u)&=\alpha_{ii}^{[l_i]}-d_i,~~~\forall i\in \langle2\rangle, \forall l_i\in\langle L_i \rangle\\
l(u,v_i^{[l_i]})&=0,~~~~~~~~~~~~~\forall i\in\langle2\rangle, \forall l_i\in\langle L_i\rangle
\end{align*}}
\end{example}

\section{Results} \label{s_result}
In this section, we present the main results of this work with remarks and illustrative examples. The proofs are relegated to the Appendix.

The following theorem provides a broadly applicable condition under which power control and TIN achieves the entire GDoF region of a $K$-user compound interference channel. 
\begin{theorem}\label{T1_IC}
In a $K$-user compound interference channel, if the following condition is satisfied,
\begin{align}\label{TIN_cond}
\alpha_{ii}^{[l_i]}\geq \max_{j:j\neq i}\{\alpha_{ji}^{[l_j]}\}+\max_{k:k\neq i}\{\alpha_{ik}^{[l_i]}\},~~\forall i,j,k\in\langle K\rangle,\forall l_i\in\langle L_i\rangle, \forall l_j\in\langle L_j\rangle  
\end{align}
 then power control and TIN achieves the entire GDoF region,  which is the intersection of the GDoF regions of all the possible network realizations and  includes all the GDoF tuples $(d_1,d_2,...,d_K)$ satisfying
 \begin{alignat}{2}
0\leq d_i&\leq \alpha_{ii}^{[l_i]},\:\:&&\forall i\in\langle K\rangle, \forall l_i\in\langle L_i\rangle \label{eqq111}\\
\sum_{j=1}^{m}d_{i_j}&\leq \sum_{j=1}^{m}(\alpha_{i_ji_j}^{[l_{i_j}]}-\alpha_{i_{j}i_{j+1}}^{[l_{i_{j}}]}),\:\:&&\forall (i_1,i_2,...,i_m)\in\Pi_K,\forall m\in\{2,3,...,K\}, \forall l_{i_j}\in\langle L_{i_j}\rangle\label{eqq112}
\end{alignat}
where $\Pi_K$ is the set of all possible cyclic sequences\footnote{Each cyclic sequence in $\Pi_K$ is a cyclically ordered subset of user indices, without repetitions. For instance, $\Pi_3=\big\{(1,2),(1,3),(2,3),(1,2,3),(1,3,2)\big\}$.} of all subsets of $\langle K\rangle$ with cardinality no less than $2$, and the modulo-$m$ arithmetic is implicitly used on the user indices, e.g., $i_m=i_0$.
\end{theorem}

In the above theorem, by setting $L_k=1$, $\forall k\in\langle K\rangle$, Theorem 1 in \cite{Geng_TIN} is readily recovered for regular interference channels. From the proof of Theorem \ref{T1_IC} in Appendix \ref{S1_IC}, we also know that  the polyhedral TIN region $\mathcal{P}$ we defined in Section \ref{S_PTIN} is fully characterized by (\ref{eqq111})-(\ref{eqq112}), which is in fact the intersection of the polyhedral TIN regions for all possible network states. In other words, \emph{for a general compound interference channel, the intersection of the polyhedral TIN regions for all its potential network states is  always achievable via power control and TIN}.

\begin{figure}[h]
\begin{center}
 \includegraphics[width= 12 cm]{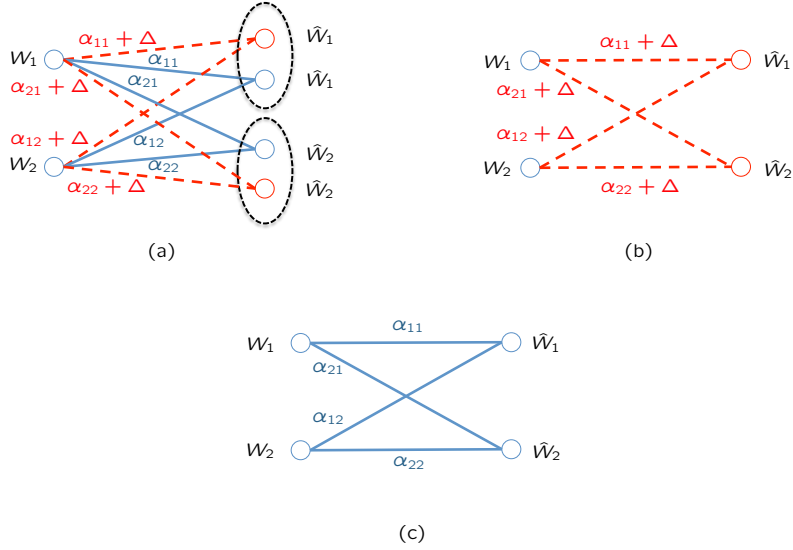}
 \caption{(a) A $2$-user compound interference channel with $L_1=2$ and $L_2=2$, where the value on each link denotes the channel strength level; (b) The possible network realization violating the TIN-optimality condition; (c) The possible network realization providing the tight GDoF outer bounds.}
\label{necessity}
\end{center}
\end{figure}

%\begin{remark}\label{remark2}
In Theorem \ref{T1_IC}, the condition (\ref{TIN_cond}) means that in each possible network realization, the channel satisfies the TIN-optimality condition identified in \cite{Geng_TIN}. It should be noted that (\ref{TIN_cond}) is only a sufficient condition for the TIN scheme to be optimal for the entire GDoF region of a $K$-user compound interference channel, but not a necessary condition. Consider the $2$-user compound interference channel in Fig~\ref{necessity}(a), where $\alpha_{22}\geq \alpha_{11}=\alpha_{12}+\alpha_{21}$ and $\Delta > 0$. Clearly, it does not satisfy condition (\ref{TIN_cond}), as one possible network realization shown in Fig.~\ref{necessity}(b) is not TIN-optimal.   Following the proof of Theorem \ref{T1_IC}, it is not hard to verify that the GDoF region $\{(d_1,d_2):0\leq d_1\leq \alpha_{11}, d_2\geq 0, d_1+d_2\leq \alpha_{22} \}$ is achievable via power control and TIN, which matches the outer bounds provided by the TIN-optimal network realization depicted in Fig.~\ref{necessity}(c). Therefore, power control and TIN still achieves its entire GDoF region. As mentioned before, a key observation we make from the proof of Theorem \ref{T1_IC} is that for the $K$-user compound interference channel with arbitrary channel strength levels, the intersection of the polyhedral TIN regions of all possible network realizations is always achievable through power control and TIN. When this achievable region is the intersection of the GDoF regions of several possible TIN-optimal network realizations, then even if condition (\ref{TIN_cond}) is not satisfied (in other words, some other network realizations are not TIN-optimal), power control and TIN still achieves the whole GDoF region of this compound channel. 
%\end{remark}

\begin{remark} \label{remark1}
Based on the result of Theorem \ref{T1_IC},  we further obtain the TIN region $\mathcal{P}^*$ for general $K$-user compound interference channels.  Following along the lines of  Theorem 5  in \cite{Geng_TIN}, we get that in general $\mathcal{P}^*$ is a union of $2^K$ polyhedral TIN regions $\mathcal{P_S}$, each of which corresponds to the case where the users in $\mathcal{S}$ $($any subset of $\langle K\rangle$$)$ are deactivated and thus removed from the network. Note that for general compound interference channels, $\mathcal{P}_{\phi}$ (i.e., the polyhedral TIN region for the case in which all the users are active) is the same as the polyhedral TIN region $\mathcal{P}$ defined in Section \ref{S_PTIN}, i.e., $\mathcal{P}=\mathcal{P}_{\phi}$. When (\ref{TIN_cond}) is satisfied, the polyhedral TIN region $\mathcal{P}_{\phi}$ subsumes all the others and $\mathcal{P}^*=\mathcal{P}$.
\end{remark}

\begin{remark}\label{remark4}
It is also not hard to extend the result of Theorem \ref{T1_IC} to the sum-GDoF optimality of TIN for $M\times N$ compound $X$ channels following \cite{Geng_TIN_X}. In addition, based on the bounding techniques provided in \cite{Geng_TIN, Geng_TIN_X}, it is easy to prove that under the same condition (\ref{TIN_cond}), power control and TIN is sufficient to achieve a constant gap to the capacity region of the $K$-user compound interference channel, and the constant gap result can also be generalized to compound $X$ channels in terms of the sum capacity. 
\end{remark}

\begin{figure}[h]
\begin{center}
 \includegraphics[width= 12.6 cm]{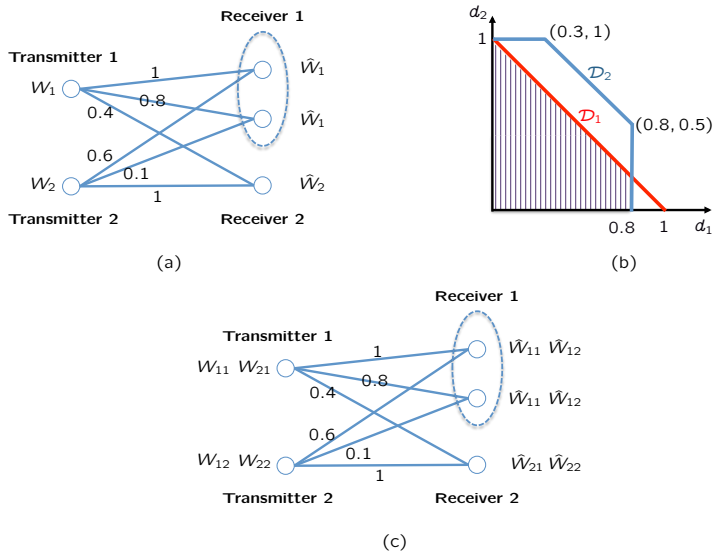}
 \caption{(a) A $2$-user compound interference channel with $L_1=2$ and $L_2=1$, where the value on each link denotes the channel strength level;  (b) The shadowed region is the GDoF region of the $2$-user compound interference channel in Fig.~\ref{ex2}(a); (c) A $2\times 2$ compound $X$ channel with $L_1=2$ and $L_2=1$, which has the same channel strength levels as the compound interference channel in Fig.~\ref{ex2}(a).}
\label{ex2}
\end{center}
\end{figure}

\begin{example}\label{ex3b}
\emph{Consider the $2$-user compound interference channel with $L_1=2$ and $L_2=1$ as illustrated in Fig \ref{ex2}(a).  The value on each link denotes the channel strength level. It is easy to check that in each of the two possible network realizations the channel is TIN-optimal, and $\mathcal{D}_1$ and $\mathcal{D}_2$ in Fig.~\ref{ex2}(b) are their GDoF regions, respectively.  According to Theorem \ref{T1_IC}, the GDoF region of this compound interference channel is $\mathcal{D}_1\cap \mathcal{D}_2$ (the shadowed region in Fig. \ref{ex2}(b)), which is achievable via power control and TIN.  And it is easy to get its sum-GDoF value is $1$. Next, we expand the message set to the $X$ setting as shown in Fig.~\ref{ex2}(c), where there are $4$ messages totally. According to Remark \ref{remark4}, following \cite{Geng_TIN_X}, we know that even if the message set is increased, the sum-GDoF of this compound $X$ channel remains as $1$, which is apparently achievable by setting $W_{12}=W_{21}=\phi$, sending only $\{W_{11},W_{22}\}$ through the channel with appropriate power levels and treating interference as noise at each receiver. }
\end{example}

The above results show that under certain conditions, power control and TIN is GDoF-optimal for compound channels.  However, the optimal power control solution, i.e., the optimal power exponent vector for the target GDoF tuple is not given explicitly. In the sequel, we will devise power control algorithms from the GDoF perspective. Without loss of generality, for the power control problem we will only consider the GDoF tuples $(d_1,d_2,...,d_K)$ where $d_i>0$,  $\forall i\in\langle K\rangle$. Obviously, if certain $d_i=0$, we just need to set $r_i=-\infty$ and then can remove User $i$ from the network without affecting others. Therefore, in the following, we only deal with the polyhedral TIN scheme,  as the polyhedral TIN region $\mathcal{P}$ includes all the GDoF tuples we are interested in. And implicitly we only need to consider a subset of the valid power allocations for the polyhedral TIN scheme, i.e., the power exponent vectors which guarantee the right hand side of (\ref{P_TIN}) to be positive for all users.

For the GDoF-based power control problem, we are mainly interested in the \emph{globally optimal} power allocation, in which to achieve the target GDoF tuple none of the users can lower their transmit powers. Note that the desired solution should be \emph{locally optimal} at first, in the sense that no single user can \emph{unilaterally} lower its own transmit power and still gain its target GDoF value.\footnote{For instance, in a $3$-user interference channel, with the power exponent vector $(r_1^*, r_2^*, r_3^*)$, through the TIN scheme we obtain a GDoF tuple $(d_1^*,d_2^*,d_3^*)$ in which $d_i^*>0$ for $i\in\{1,2,3\}$. Then the power allocation $(r_1^*, r_2^*, r_3^*)$ is locally optimal for the tuple $(d_1^*,d_2^*,d_3^*)$.  Using the terminology of game theory, each locally optimal power control solution refers to a \emph{Nash equilibrium}.} For one GDoF tuple, while there may exist multiple locally optimal power allocations in general, there is only one globally optimal power control solution, which is also the unique Pareto optimal solution as shown later.

Before proceeding to the details of power control algorithms, we first show how to simplify the GDoF-based power control problem for general compound interference channels.  Denote the $K$-user compound interference channel defined in (\ref{equ_channel}) as $\mathcal{IC}_C$. Based on $\mathcal{IC}_C$, we construct a counterpart $K$-user regular interference channel $\mathcal{IC}_R$ with the input-output relationship 
\begin{align}
Y_k(t)=\sum_{i=1}^K\sqrt{P^{\bar{\alpha}_{ki}}}e^{j\theta_{ki}}X_i(t)+\bar{Z}_k(t),~~~ \forall k\in\langle K\rangle, 
\end{align}
where 
\begin{alignat}{2}
\bar{\alpha}_{kk}&=\min_{l_k\in\langle L_k\rangle}\{\alpha_{kk}^{[l_k]}\},&& \forall k\in\langle K \rangle\\
\bar{\alpha}_{kj}&=\min_{l_k\in\langle L_k\rangle}\{\alpha_{kk}^{[l_k]}\}-\min_{l_k\in\langle L_k\rangle}\{\alpha_{kk}^{[l_k]}-\alpha_{kj}^{[l_k]}\}, ~~&&\forall k,j\in\langle K\rangle, j\neq k, \label{reg_c2}
\end{alignat}
and $\bar{Z}_k(t)\sim\mathcal{CN}(0,1)$. We  rewrite (\ref{reg_c2}) as 
\begin{align}
\bar{\alpha}_{kk}-\bar{\alpha}_{kj}=\min_{l_k\in\langle L_k\rangle}\{\alpha_{kk}^{[l_k]}-\alpha_{kj}^{[l_k]}\}, ~~~~~\forall k,j\in\langle K\rangle, j\neq k \label{reg_c3}
\end{align}
Denote $\alpha_{kk}^{[l_k]}-\alpha_{kj}^{[l_k]}$ as the power level ``gain" of User $k$ against User $j$ under state $l_k$, $k,j\in\langle K \rangle$. In words, for a general $K$-user compound interference channel, in its regular counterpart, the channel strength level of the direct link for User $k$ is equal to that of the weakest direct link among all states of User $k$ in the compound setting, and the power level gain of User $k$ against User $j$ (i.e., $\bar{\alpha}_{kk}-\bar{\alpha}_{kj}$) is equal to the minimal power level gain of User $k$ against User $j$ among all states of User $k$ in the compound channel. Notably, for User $k$ in the original compound channel, the state with the weakest direct link and the state with the minimal power level gain of User $k$ against User $j$ are not the same in general. Therefore, the regular interference channel $\mathcal{IC}_R$ is a non-trivial mixture of states of the compound channel $\mathcal{IC}_C$. 

The following theorem shows that to solve the GDoF-based power control problem for the $K$-user compound interference channel $\mathcal{IC}_C$, we only need to consider its counterpart $K$-user regular interference channel $\mathcal{IC}_R$. 

\begin{theorem}\label{th-equ}
The $K$-user compound interference channel $\mathcal{IC}_C$ and its counterpart $K$-user regular interference channel $\mathcal{IC}_R$ have the same TIN region $\mathcal{P}^*$ and the same polyhedral TIN region $\mathcal{P}$. Moreover, for any feasible GDoF tuple in $\mathcal{P}$, $\mathcal{IC}_C$ and $\mathcal{IC}_R$ have the same set of locally optimal power allocations.  
\end{theorem}

\begin{remark}\label{remark_equ}
There are several ways to prove that $\mathcal{IC}_C$ and $\mathcal{IC}_R$ have the same TIN region and polyhedral TIN region. To help understand the result, besides the proof presented in Appendix \ref{app_equ}, here we also give an explanation based on potential theorem.  From the potential graph $D_p$ of $\mathcal{IC}_C$ defined in Section \ref{S_PG}, we can construct another complete digraph $\bar{D}_p=\{\bar{V},\bar{E}\}$ with $K+1$ vertices, where 
\begin{alignat*}{2}
\bar{V}&=\{\bar{v}_1,\bar{v}_2,...,\bar{v}_K,u\},~~~~~&&\forall k\in\langle K\rangle\\
\bar{E}&=\bar{E}_1\cup \bar{E}_2 \cup \bar{E}_3\\
\bar{E}_1&=\{(\bar{v}_k,\bar{v}_j)\},  &&\forall k,j\in\langle K\rangle, k\neq j\\ 
\bar{E}_2&=\{(\bar{v}_k,u)\}, &&\forall k\in\langle K\rangle\\
\bar{E}_3&=\{(u,\bar{v}_k)\}, &&\forall k\in\langle K\rangle
\end{alignat*}
The length $l(\bar{e})$ of each edge $\bar{e}\in \bar{E}$ is assigned as follows,  
\begin{alignat*}{2}
l(\bar{v}_k,\bar{v}_j)&=\min_{l_k\in\langle L_k\rangle}\{\alpha_{kk}^{[l_k]}-\alpha_{kj}^{[l_k]}\}-d_k,~~~~~&&\forall k,j\in\langle K\rangle, k\neq j\\
l(\bar{v}_k,u)&=\min_{l_k\in\langle L_k\rangle}\{\alpha_{kk}^{[l_k]}\}-d_k,&&\forall k\in\langle K\rangle\\
l(u,\bar{v}_k)&=0,&&\forall k\in\langle K\rangle
\end{alignat*}
It is not hard to verify that $\bar{D}_p$ is the potential graph of $\mathcal{IC}_R$. Also note that
\begin{align*}
l(\bar{v}_k,\bar{v}_j)&=\min_{l_k\in\langle L_k\rangle}\{l(v_k^{[l_k]},v_j^{[l_j]})\},  \\
l(\bar{v}_k,u)&=\min_{l_k\in\langle L_k\rangle}\{l(v_k^{[l_k]},u)\}.
\end{align*}
It is easy to check that the length of the shortest path from the vertex $v_k^{[l_k]}$ to the vertex $v_j^{[l_j]}$ in $D_p$ is equal to that of the shortest path from $\bar{v}_k$ to $\bar{v}_j$ in $\bar{D}_p$, $\forall k,j\in\langle K\rangle$, $j\neq k$. Similarly, the length of the shortest path from $u$ (or $v_k^{[l_k]}$) to $v_k^{[l_k]}$ (or u) in $D_p$ is equal to that of the shortest path from $u$ (or $\bar{v}_k$) to $\bar{v}_k$ (or $u$) in $\bar{D}_p$, $\forall k\in\langle K\rangle$. Denote the polyhedral TIN regions of $\mathcal{IC}_C$ and $\mathcal{IC}_R$ as $\mathcal{P}_C$ and $\mathcal{P}_R$, respectively. In the $K$-user compound interference channel $\mathcal{IC}_C$, for any GDoF tuple $(d_1^*,d_2^*,...,d_K^*)\in\mathcal{P}_C$, from the proof of Theorem \ref{T1_IC}, we know that in its corresponding potential graph $D_p$ (recall that in this case $d_k=d_k^*$ in (\ref{eq:samelength})-(\ref{eq_lst1}), $ k\in\langle K\rangle$), there exists no circuit with a negative length.  According to the above observation, it is easy to check that for its regular counterpart $\mathcal{IC}_R$, when the target GDoF tuple is $(d_1^*,d_2^*,...,d_K^*)$, all the circuits in $\bar{D}_p$ have non-negative lengths as well. Hence we have $(d_1^*,d_2^*,...,d_K^*)\in\mathcal{P}_R$ and $\mathcal{P}_C \subseteq \mathcal{P}_R$. Similarly, we obtain that $\mathcal{P}_R \subseteq \mathcal{P}_C$. Therefore, we establish that $\mathcal{IC}_C$ and $\mathcal{IC}_R$ have the same polyhedral TIN region. Next, based on Remark \ref{remark1}, it is not hard to argue that they also have the same TIN region. 
\end{remark}
 
Next, we explore some properties of the potential graph, which will help us develop efficient GDoF-based power control algorithms. 
\begin{theorem}\label{PGPC_NP}
In a $K$-user compound interference channel, for any achievable GDoF tuple $(d_1,d_2,...,d_K)$ in the polyhedral TIN region $\mathcal{P}$, using $(l_{1,\mathrm{dst}},l_{2,\mathrm{dst}},...,l_{K,\mathrm{dst}})$ in the potential graph $D_p$ as the transmit power allocation and treating interference as noise at each receiver, we obtain a GDoF tuple $(\hat{d}_1,\hat{d}_2,...,\hat{d}_K)\in\mathcal{P}$, which dominates $(d_1,d_2,...,d_K)$, i.e., $\hat{d}_k\geq d_k$, $\forall k\in\langle K\rangle$. 
\end{theorem}

%\begin{remark} \label{Bellman-Ford remark} 
As mentioned in Remark \ref{remark_equ}, to obtain the shortest path results in the potential graph $D_p$ of the original compound channel $\mathcal{IC}_C$, we only need to deal with a corresponding single-source shortest path problem in $\bar{D}_p$.  This problem can be solved efficiently via some classical algorithms, e.g., a $K$-step Bellman-Ford algorithm with complexity $O(K^3)$, which is independent of the state number of each receiver in the original compound channel. It is also well known that Bellman-Ford algorithm can detect the negative-length circuits in a digraph with arbitrary-length edges  \cite{Potential}. From the proof of Theorem \ref{T1_IC} and Remark \ref{remark_equ}, we  know that for a $K$-user compound interference channel, when the target GDoF tuple is out of the polyhedral TIN region $\mathcal{P}$, i.e., it is not achievable via the polyhedral TIN scheme, $D_p$ and $\bar{D}_p$ are with negative-length circuits. Bellman-Ford algorithm thus can also serve as the feasibility test for the target GDoF tuples. Correspondingly, in the conventional SINR-based power control algorithms for regular interference channels, the feasibility of SINR targets can be checked through centralized computations based on non-negative matrix theory \cite{Bambos_DPC}. How to test its feasibility in a distributed way is still a difficult task.
%\end{remark}

%\begin{figure}[h]
%\begin{center}
 %\includegraphics[width= 11 cm]{ex5}
 %\caption{(a) A $3$-user regular interference channel, and (b) its potential graph $D_p$. }
%\label{ex5}
%\end{center}
%\end{figure}

%\begin{example}
%Consider a $3$-user interference channel depicted in Fig.~\ref{ex5}(a). Its potential graph is shown in Fig.~\ref{ex5}(b). Consider a feasible GDoF tuple $(0.4, 0.1, 1)$. In its potential graph $D_p$, we obtain $(l_{1,\mathrm{dst}},l_{2,\mathrm{dst}},l_{3,\mathrm{dst}})=(-0.3,-0.4,0)$. When we use $(-0.3,-0.4,0)$ as the transmit power and treat interference as noise at each receiver, we obtain a GDoF tuple $(0.5,0.1,1)$. Similarly, start with another achievable GDoF tuple $(1,0.1,0.3)$. Repeating the above procedure, we end up with a GDoF tuple $(1,0.3,0.5)$. 
%\end{example}

To optimize the system performance, we are interested in the \emph{Pareto optimal} GDoF tuples in the feasible GDoF region.  If a GDoF tuple $(d_1,d_2,...,d_K)$ is Pareto optimal for the polyhedral TIN region $\mathcal{P}$, it indicates that in $\mathcal{P}$ there does not exist another different tuple $(\hat{d}_1,\hat{d}_2,...,\hat{d}_K)$ such that $\hat{d}_k\geq d_k$, $\forall k\in\langle K\rangle$. Therefore, based on Theorem \ref{PGPC_NP} and Remark \ref{remark_equ}, we  obtain the following corollary. %which illustrates how to obtain a locally optimal power control solution for Pareto optimal GDoF tuples based on the potential graph. 

\begin{corollary}\label{PGPC}
In a $K$-user compound interference channel $\mathcal{IC}_C$, for any Pareto optimal GDoF tuple in its polyhedral TIN region $\mathcal{P}$, $(l_{1,\mathrm{dst}},l_{2,\mathrm{dst}},...,l_{K,\mathrm{dst}})$ in the potential graph $\bar{D}_p$ of its regular counterpart $\mathcal{IC}_R$ is a locally optimal transmit power allocation. 
\end{corollary}

Since for a $K$-user compound interference channel in which (\ref{TIN_cond}) is satisfied, its GDoF region is exactly the polyhedral TIN region $\mathcal{P}$, we get the following result straightforwardly. 

\begin{corollary}\label{PGPC-optimal}
In a $K$-user compound interference channel $\mathcal{IC}_C$, if the condition (\ref{TIN_cond}) is satisfied, for any Pareto optimal GDoF tuple in its GDoF region, $(l_{1,\mathrm{dst}},l_{2,\mathrm{dst}},...,l_{K,\mathrm{dst}})$ in the potential graph $\bar{D}_p$ of its regular counterpart $\mathcal{IC}_R$ is  a locally optimal transmit power allocation. 
\end{corollary}

\begin{figure}[h]
\centering
\subfigure[]{
\includegraphics[width= 5.5 cm]{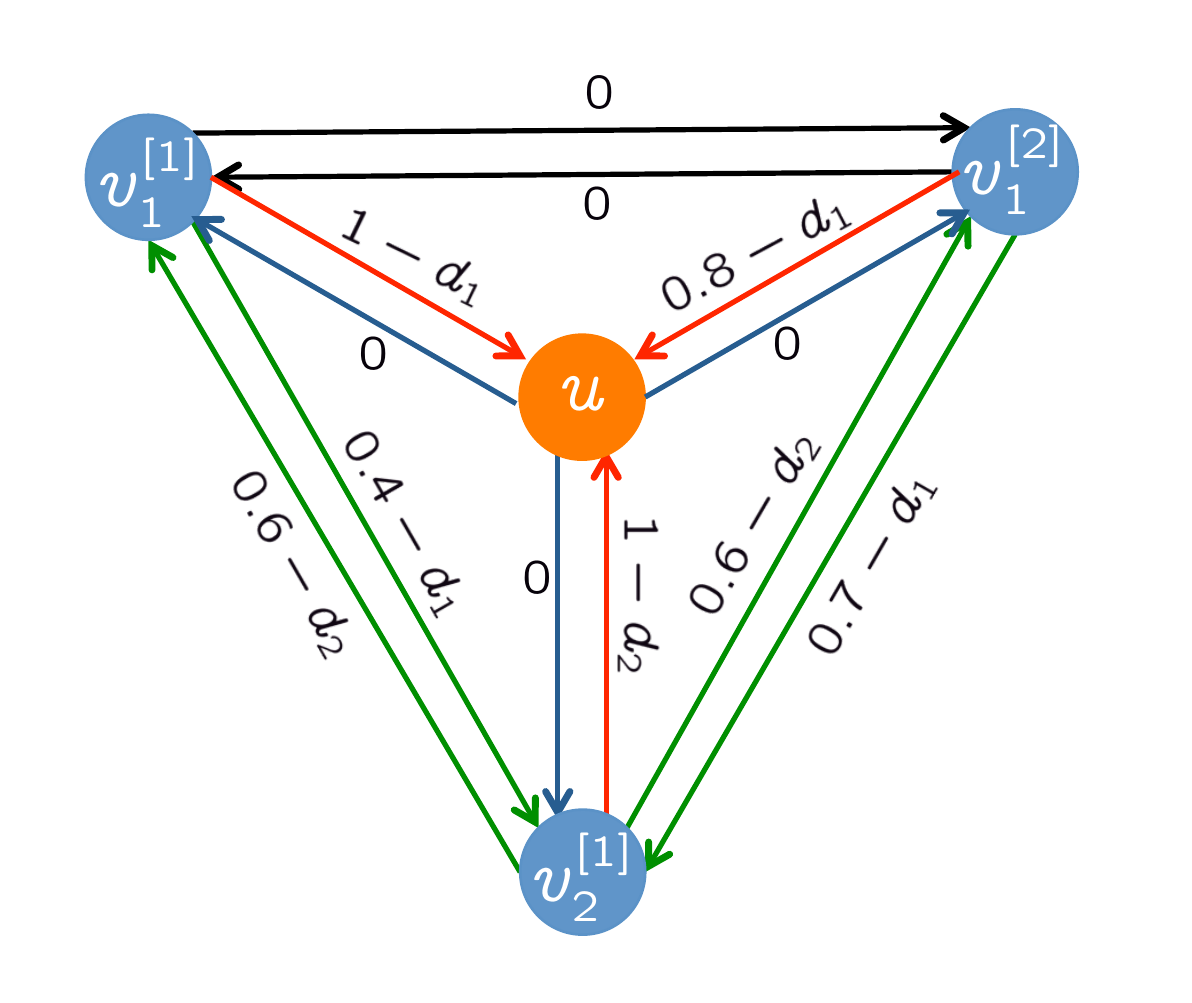}
\label{ex3}}
\hspace{1in}
\subfigure[]{
\includegraphics[width = 5.5 cm]{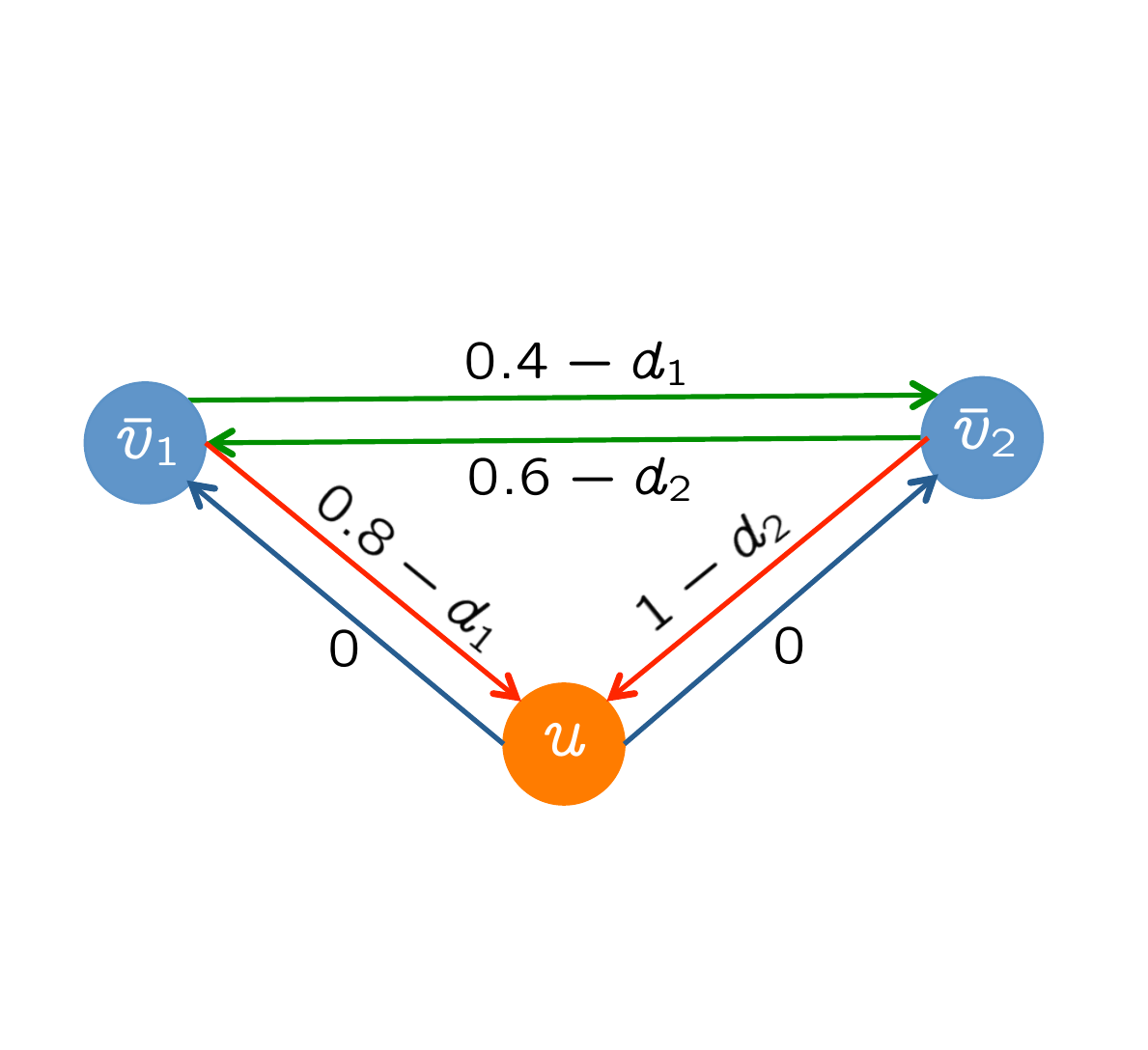}
\label{reduced_PG}}
\caption[]{
\subref{ex3} The potential graph $D_p$ for the $2$-user compound interference channel in Example \ref{ex3b};  \subref{reduced_PG} The potential graph $\bar{D}_p$ for its counterpart $2$-user regular interference channel}
\end{figure}

\begin{example}\label{ex4b}
\emph{Consider the $2$-user compound interference channel with $L_1=2$ and $L_2=1$ in Example \ref{ex3b}. Its potential graph $D_p$ is depicted in Fig.~\ref{ex3}. The potential graph $\bar{D}_p$ for its counterpart $2$-user regular interference channel is shown in Fig.~\ref{reduced_PG}. From Fig.~\ref{ex2}(b), one can find that the GDoF tuple $(0.7, 0.3)$ is Pareto optimal. For this tuple, it is easy to check that in $\bar{D}_p$, the length of the shortest paths from $u$ to $\bar{v}_1$ and $\bar{v}_2$ are $0$ and $-0.3$, respectively.  From Corollary \ref{PGPC-optimal}, we obtain that $(0, -0.3)$ is a locally optimal power allocation for the target GDoF tuple $(0.7,0.3)$ in the original $2$-user compound interference channel. It is not hard to verify that $(0,-0.3)$ is not the globally optimal solution, since the power allocation $(-0.1,-0.4)$ also achieves the target tuple. }
\end{example}

Finally, we show how to obtain the globally optimal power allocations for all the achievable GDoF tuples. Notably, there exist some remarkable differences between the power control problems for target \emph{rate} tuples and \emph{GDoF} tuples. For any achievable \emph{rate} tuple, there only exists a unique locally optimal transmit power vector, which is thus the globally optimal power allocation naturally. However, as said before, for one achievable \emph{GDoF} tuple, in general there are multiple locally optimal power allocations. Also, it is well known that for any feasible target \emph{rate} tuple, a celebrated synchronous fixed-point power control algorithm developed by Foschini and Miljanic \cite{Foschini_DPC} provides the globally optimal power allocation. In Appendix \ref{app_FP}, we show that  for an achievable \emph{GDoF} tuple, through a similar GDoF-based synchronous fixed-point power control (GSFPC) algorithm, we obtain a locally optimal solution, which is not globally optimal in general. Consider a motivating example first.

\begin{example}
\emph{For the 3-user interference channel in Fig.~\ref{GGPC_exp}, we intend to achieve the GDoF tuple $(d_1,d_2,d_3)=(0.5,0.6,0.7)$.  We start with the initial power allocation given by the shortest path result in the potential graph $\big(r_1(0),r_2(0),r_3(0)\big)=(-0.1,0,-0.1)$. In this network, we find that if all transmitters reduce their transmit powers by the same amount  below a threshold $\Delta(0)$, we still achieve an acceptable GDoF tuple, where
\begin{align*}
\Delta(0)=\min_i\big\{r_i(0)+\alpha_{ii}-d_i\big\}=0.4,~~i\in\{1,2,3\}
\end{align*}}

\begin{figure}[h]
\begin{center}
 \includegraphics[width= 12 cm]{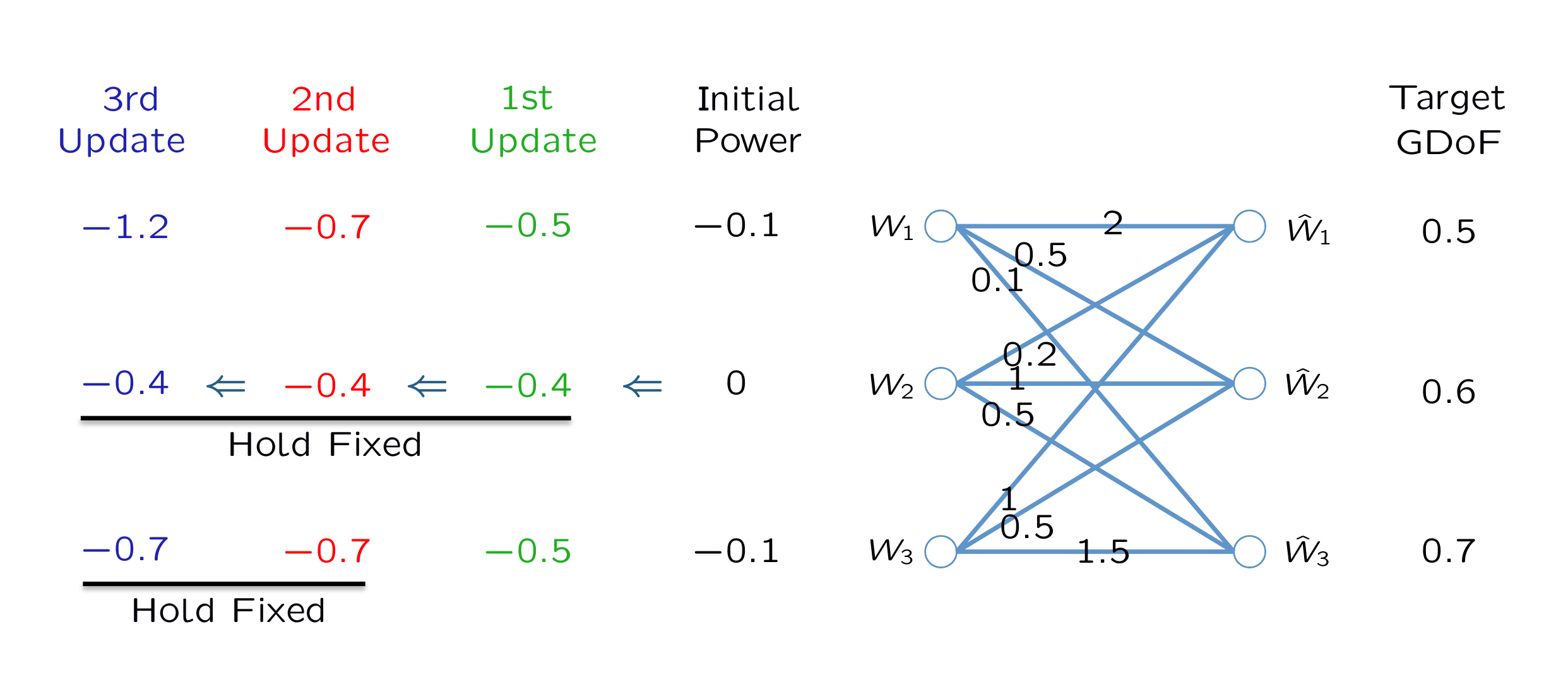}
 \caption{The transmit power updates for a $3$-user interference channel }
\label{GGPC_exp}
\end{center}
\end{figure}

\emph{Thus in the first update, each user lowers its transmit power by $0.4$. The resulting transmit power allocation is now $\big(r_1(1),r_2(1),r_3(1)\big)=(-0.5,-0.4,-0.5)$, and the achieved GDoF tuple is $(1,0.6,0.9)$. At this point, User 2 cannot reduce its transmit power any more in order to maintain the desired GDoF value $0.6$. A consequence is that in the following update, Transmitter $2$ exerts fixed interference levels to Receiver $1$ and $3$, whose strengths are $\max\{0,\alpha_{12}+r_2(1))\}=0$ and $\max\{0,\alpha_{32}+r_2(1)\}=0.1$, respectively.  In the second update, the other two users (User 1 and 3) can further reduce their transmit power simultaneously by the same amount below a threshold $\Delta(1)$ to obtain an acceptable GDoF tuple, where
\begin{align*}
\Delta(1)=\min_i\big\{r_i(1)+\alpha_{ii}-d_i-\max\{0,\alpha_{i2}+r_2(1))\}\big\}= 0.2, ~~i\in\{1,3\} 
\end{align*}
Hence after the second update, keeping the transmit power of User $2$ and reducing the transmit powers of User $1$ and $3$ by $0.2$, we obtain the power control solution $\big(r_1(2),r_2(2),r_3(2)\big)=(-0.7,-0.4,-0.7)$ and its corresponding achieved GDoF tuple $(1,0.6,0.7)$. Now User $3$ cannot further reduce its transmit power to maintain the desired GDoF value $0.7$. Thus in the following update, Transmitter $3$ exerts a fixed interference level to Receiver $1$ with strength $\max\{0,\alpha_{13}+r_3(2)\}=0.3$. To obtain an acceptable GDoF tuple, at this point only User $1$ can further lower its transmit power by an amount below the threshold 
\begin{align*}
\Delta(2)=r_1(2)+\alpha_{11}-d_1-\max\{0,\alpha_{1j}+r_j(2))\}= 0.5,~~j\in\{2,3\}
\end{align*}
Finally, we obtain the power allocation $\big(r_1(3),r_2(3),r_3(3)\big)=(-1.2,-0.4,-0.7)$ and its corresponding achieved GDoF tuple $(0.5,0.6,0.7)$. In what follows, we will show that the obtained power allocation $(-1.2,-0.4,-0.7)$ is globally optimal for the target GDoF tuple $(0.5,0.6,0.7)$}.
\end{example}

For a $K$-user interference channel, to obtain the globally optimal power control solution for any achievable GDoF tuple $(d_1,d_2,...,d_K)$ in the polyhedral TIN region, we develop an iterative algorithm with at most $K$ updates, which is called GDoF-based globally-optimal power control (GGPC) algorithm and  specified at the top of next page. 
% \\Non-cmpound algorithm
\begin{algorithm}
\caption{GDoF-based globally-optimal power control (GGPC)}
\begin{algorithmic}
\STATE 1) Initialize: set $\mathcal{I}=\langle K\rangle$, $\mathcal{M}=\phi$, and $r_i(0)=l_{i,dst}$ for $i\in\langle K\rangle$; 

\STATE 2) Update:
\begin{alignat}{2}
&\Delta(n)=\min_i\Big\{r_i(n)+\alpha_{ii}-d_i-\max_{m\neq i}\{0,\alpha_{im}+r_m(n)\}\Big\},\:\:&&i\in \mathcal{I}, m\in \mathcal{M}\\
%&\Delta(n)=\min_i\{\max_{j\neq i}\{0,\alpha_{ij}+r_j(n)\}-\max_{m\neq i}\{0,\alpha_{im}+r_m(n)\}\},\:\:&&i\in I, j\in [K], m\in M\\
&\mathcal{N}=\arg\min_i\Big\{r_i(n)+\alpha_{ii}-d_i-\max_{m\neq i}\{0,\alpha_{im}+r_m(n)\}\Big\},\:\:&&i\in \mathcal{I}, m\in \mathcal{M}\\
%&k=\arg\min_i\{\max_{j\neq i}\{0,\alpha_{ij}+r_j(n)\}-\max_{m\neq i}\{0,\alpha_{im}+r_m(n)\}\},\:\:&&i\in I, j\in [K], m\in M\\
&r_i(n+1)=r_i(n)-\Delta(n),&&i\in \mathcal{I}\\
&r_m(n+1)=r_m(n),&& m\in \mathcal{M}\\
&\mathcal{I}=\mathcal{I}\backslash \mathcal{N}, \mathcal{M}=\mathcal{M}\cup \mathcal{N}
\end{alignat}

\STATE where $n$ indexes discrete time slots. The update phase terminates when $\mathcal{I}=\phi$.
\end{algorithmic}
\end{algorithm}

As mentioned before, we can use Bellman-Ford algorithm to calculate the lengths of the shortest paths in the initialize phase, which also serves as the feasibility test for the target GDoF tuple. For the feasible GDoF tuple, we repeat the update phase until $\mathcal{I}=\phi$ to obtain its power control solution. Starting from the initial power allocation, in each update the GGPC algorithm reduces the transmit powers of certain users to their limits (these users also achieve the target GDoF value exactly after the update) and still guarantees to achieve an acceptable GDoF tuple dominating the target one. Then these users hold their powers fixed, which in turn exert fixed interference levels for the remaining users participating in the following updates. These fixed interference levels will limit how much power can be further reduced by the remaining users. The optimality of the GGPC algorithm is illustrated by the following theorem. 

\begin{theorem}\label{GPU_converge}
In a $K$-user interference channel, for any achievable GDoF tuple in its polyhedral TIN region $\mathcal{P}$, the GGPC algorithm obtains the globally optimal transmit power allocation.  
\end{theorem}

Combining Theorem \ref{th-equ} and \ref{GPU_converge}, we get the following corollary. 
\begin{corollary}\label{coro_GPU_converge}
In a $K$-user compound interference channel $\mathcal{IC}_C$, for any achievable GDoF tuple in its polyhedral TIN region $\mathcal{P}$, applying the GGPC algorithm to its counterpart $K$-user regular interference channel $\mathcal{IC}_R$ obtains the globally optimal transmit power allocation.  
\end{corollary}

Correspondingly, for the $K$-user TIN-optimal compound interference channel satisfying the condition (\ref{TIN_cond}), we have the following result.
\begin{corollary}
In a $K$-user compound interference channel $\mathcal{IC}_C$, when the condition (\ref{TIN_cond}) is satisfied, for any achievable GDoF tuple in its GDoF region, applying the GGPC algorithm to its counterpart $K$-user regular interference channel $\mathcal{IC}_R$ obtains the globally optimal transmit power allocation. 
\end{corollary}

%\begin{remark}
In fact, the obtained globally optimal power allocation is the unique Pareto optimal solution for the GDoF-based power control problem. We prove it by contradiction. Denote by $\mathcal{R}$ the set of all the power allocations that can achieve the target GDoF tuple and $\mathbf{r}^*\in\mathcal{R}$ the globally optimal power allocation. Then we know that we cannot find a $\mathbf{r}'\in\mathcal{R}$ in which there exists an $i\in\langle K\rangle$ such that $r'_i<r^*_i$ ($r'_i$ and $r^*_i$ denote the $i$-th entry of $\mathbf{r}'$ and $\mathbf{r}^*$, respectively).  Suppose there exist two different Pareto optimal power allocations $\mathbf{r}_a,\mathbf{r}_b\in\mathcal{R}$. It means that there exist $i,j\in\langle K\rangle$ such that $r_{a,i}<r_{b,i}$ and $r_{a,j}>r_{b,j}$ ($r_{a,i}$ and $r_{b,i}$ denote the $i$-th entry of $\mathbf{r}_a$ and $\mathbf{r}_b$, respectively), which is clearly a contradiction that there exists a globally optimal solution. 
%\end{remark}

%\begin{remark}
In a general $K$-user compound interference channel, we have established that for any feasible GDoF tuple, we can always find its unique globally optimal power allocation by applying the GGPC algorithm to its  counterpart $K$-user regular interference channel. Notably, using this approach, both the running time and complexity only depend on the number of users $K$ and are irrelevant to the state number of each receiver in the compound channel.  
%\end{remark}

\section{Discussion}

\subsection{Performance of GDoF-based Power Control at Finite SNRs}\label{sec_sim}
 In Section \ref{s_result}, we have solved GDoF-based power control problems, i.e., given a feasible GDoF tuple, how to obtain its optimal power allocation. It is of interest to see its performance in the finite SNR setting. In the sequel, we show how to deal with the joint rate assignment and power control problem based on the developed GDoF-based power control algorithms, and compare its performance with that of the (finite) SINR-based approaches.  If for each receiver, it has one dominant interferer and all other interference is substantially weaker (e.g., all the other interference is below the noise floor), it is expected that in the finite SNR setting, the simple GDoF-based power control schemes should obtain comparable performance with its sophisticated SINR-based counterparts. Through two simple examples, we show that in some other cases, the performance of GDoF-based schemes may not degrade too much either, and can still be close to that of the SINR-based approaches.

In the following simulations, to compare the performance of the GDoF-based power control and the SINR-based approaches at finite SNRs, we fix the channel strength levels and only scale the nominal parameter $P$. Also note we always assume that each transmitter is subject to the unit power constraint, and the noise variance at each receiver is normalized to one.

\begin{figure}[h]
\centering
\subfigure[]{
\includegraphics[width= 6 cm]{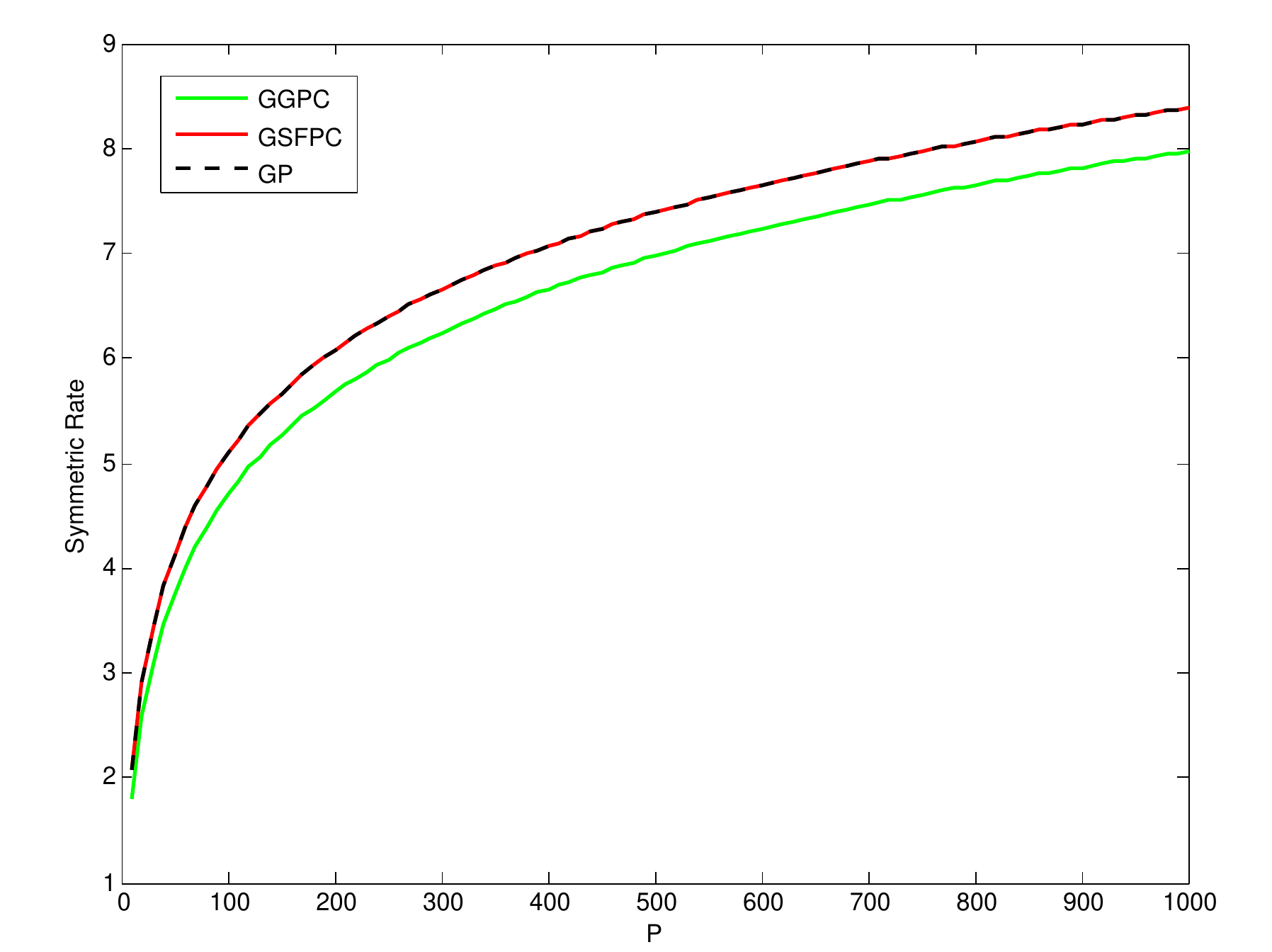}
\label{Sym_SymNet}}
\hspace{1in}
\subfigure[]{
\includegraphics[width= 6 cm]{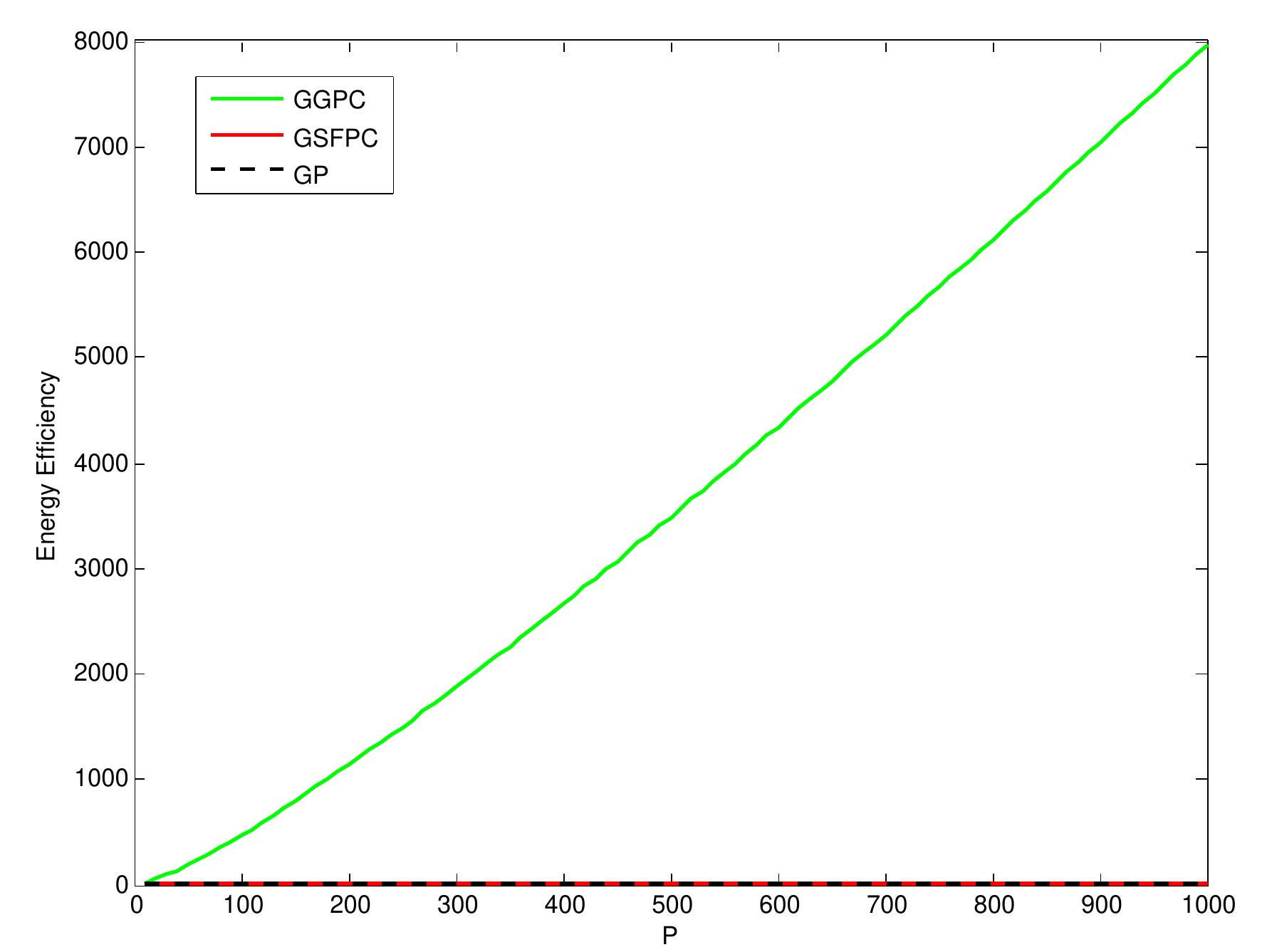}
\label{Eff_SymNet}}
\caption[]{
\subref{Sym_SymNet} The achievable symmetric rates under various power control schemes in the finite SNR setting for the $4$-user symmetric interference channel;  \subref{Eff_SymNet} The energy efficiency under various power control schemes in the finite SNR setting for the $4$-user symmetric interference channel.}
\end{figure}

\begin{example}
\emph{ First, consider the interference networks where each receiver has several interfering links with comparable strengths. Apparently, the extreme case is the symmetric setting. Take a fully symmetric $4$-user interference channel as an example, where all the direct links are with channel strength levels $2$ and all the cross links are with channel strength levels $1$. Obviously, in such a symmetric setting, the maximal symmetric GDoF is $1$ for each user, and it is easy to obtain the corresponding power allocations through GSFPC and GGPC.  From the SINR perspective, we use geometric programming (GP) \cite{Chiang_GP} to obtain the optimal rate.  The maximal symmetric achievable rates under various power control schemes are depicted in Fig.~\ref{Sym_SymNet}. It is shown that GSFPC and GP obtain the exactly same rates, which are achieved by full power transmission at all transmitters. Compared with GSFPC and GP, GGPC suffers a rate loss of $5\%$ when $P=1000$. Note that GP and GSFPC consume more transmitter powers than GGPC. Thus one may also be interested in the energy efficiency (bits per Joule) of these schemes, which can be defined as the ratio between the achievable sum rate and the total transmit power expense naturally.  From Fig.~\ref{Eff_SymNet}, we see that the GGPC is significantly superior to the other two schemes in terms of energy efficiency. }
%They are also the highest achievable symmetric rates through the TIN scheme for this fully symmetric network \cite{Hanly_Symmetric}. 
\end{example}

\begin{figure}[h]
\begin{center}
 \includegraphics[width= 4.6 cm]{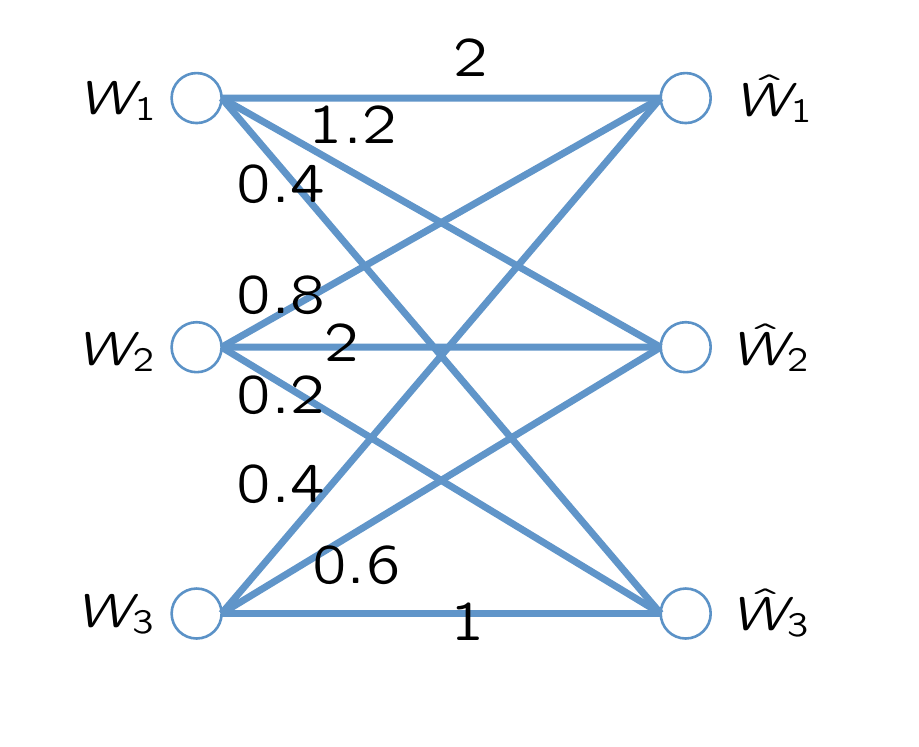}
 \caption{A $3$-user interference channel, where the value on each link denotes the channel strength level. }
\label{sim_ex}
\end{center}
\end{figure}

\begin{example} \emph{ Next, consider a more practical asymmetric $3$-user interference channel in Fig.~\ref{sim_ex}, where for each receiver, the channel strength level of the stronger interfering link is twice of that of the weaker one. Since the channel is TIN-optimal, it is easy to obtain that the GDoF region of this channel includes all the tuples $(d_1,d_2,d_3)$ satisfying
\begin{align*}
0\leq d_1 &\leq 2\\
0\leq d_2 &\leq 2\\
0 \leq d_3 &\leq 1\\
d_1+d_2&\leq 2\\
d_2+d_3&\leq 2.2\\
d_1+d_3&\leq 2.2\\
d_1+d_2+d_3&\leq 3.2 
\end{align*}
which is achievable by power control and TIN.}

\emph{First, consider the maximal achievable sum rate. For the SINR-based power control, we choose the SINR approximation power control (SAPC) scheme developed in \cite{Chiang_Sumrate}. For the GDoF-based power control, it is easy to verify that the sum-GDoF of this channel is $3$. Taking into account the fairness issue, we set the target GDoF tuple as $(1,1,1)$ and then use GSFPC and GGPC to get the power allocations, respectively. We also obtain the achievable sum rate without power control, i.e., the case where every user transmits at its full power. The achievable sum rates under the above schemes are given in Fig.~\ref{sum_rate}. It can be seen that GSFPC and GGPC perform very close to SAPC and provide significant benefits over the full power transmission. Compared with SAPC, when $P=10000$, GSFPC and GGPC only suffer a sum rate loss of $0.55\%$ and $1.61\%$, respectively. The energy efficiency of these power control schemes are also presented in Fig.~\ref{Eng_Eff}, demonstrating that the GGPC is most efficient. As $P$ further increases, the obtained sum rates are illustrated in Fig.~\ref{sum_rate_Plimit}, which indicates that SAPC, GSFPC and GGPC achieve the same sum-GDoF and full power transmission suffers a sum-GDoF loss. It is notable that for a $K$-user interference channel,  in general there are multiple (usually infinite) GDoF tuples achieving the sum-GDoF. In the finite SNR setting, these GDoF tuples enable different achievable sum rates.  How to choose an appropriate GDoF tuple to maximize the sum rate in finite SNR setting is open.}

\begin{figure}[h]
\centering
\subfigure[]{
\includegraphics[width= 6 cm]{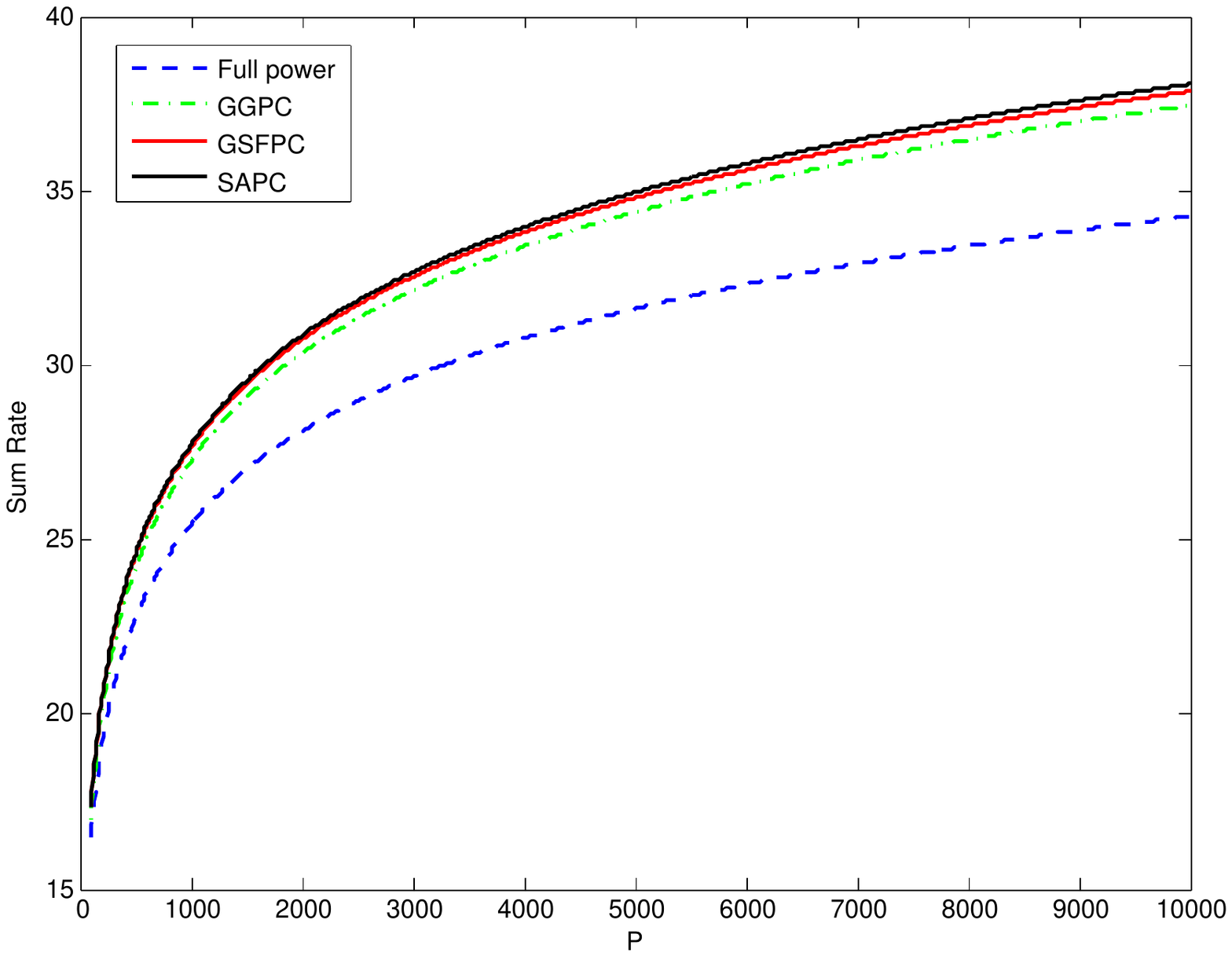}
\label{sum_rate}}
\hspace{1in}
\subfigure[]{
\includegraphics[width= 6 cm]{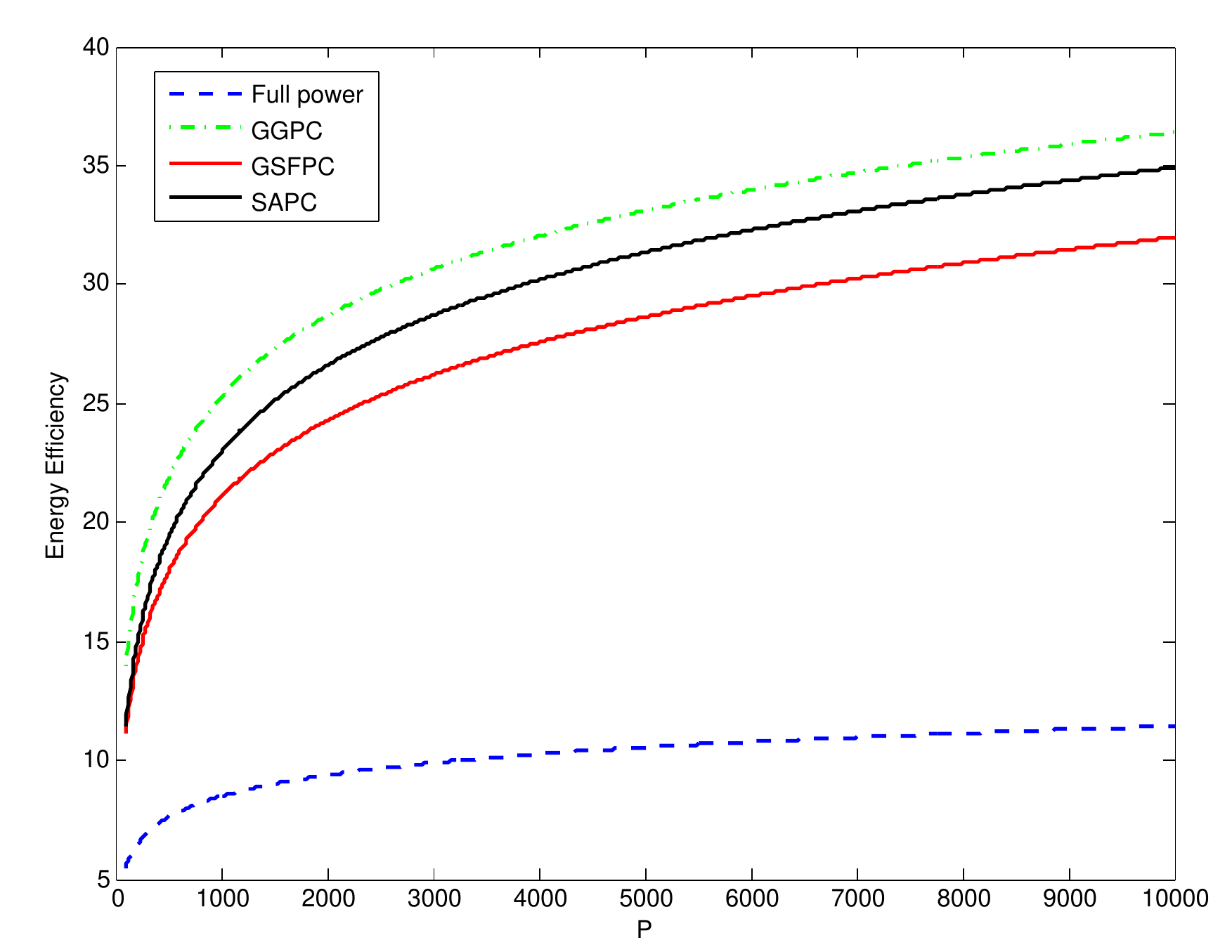}
\label{Eng_Eff}}
\hspace{1in}
\subfigure[]{
\includegraphics[width= 6 cm]{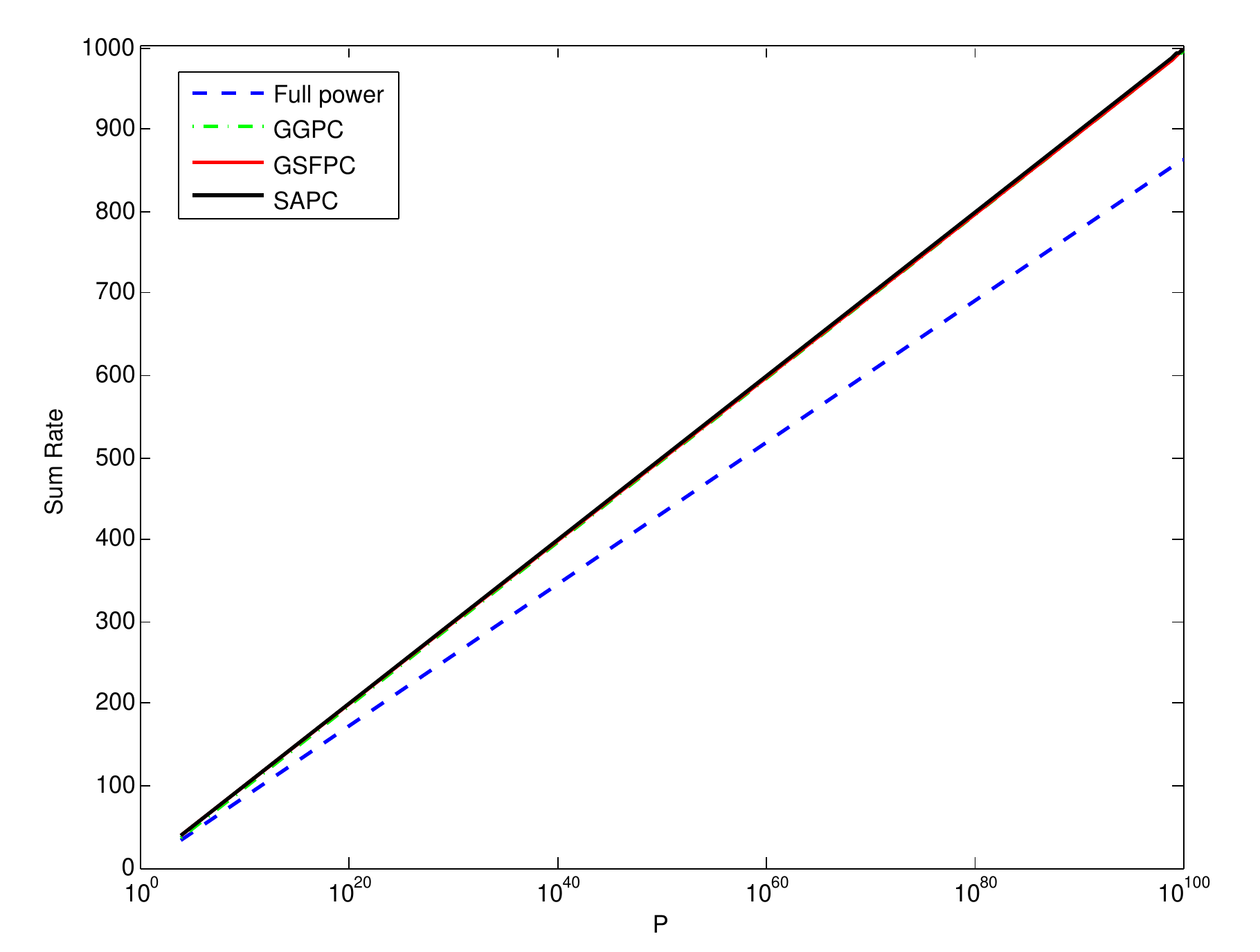}
\label{sum_rate_Plimit}}
\caption[]{
\subref{sum_rate} The achievable sum rates under various power control schemes in the finite SNR setting;  \subref{Eng_Eff} The energy efficiency under various power control schemes in the finite SNR setting; \subref{sum_rate_Plimit} The achievable sum rates under various power control schemes as P (in log scale) increases.}
\end{figure}

\emph{Next, consider the maximal achievable symmetric rate.  The achievable symmetric rates under various power control schemes are depicted in Fig.~\ref{sym_rate}. Compared with GP, GGPC suffers a loss less than $3\%$ when $P=10000$. Note in the GDoF-based power control, each user obtains the same GDoF in the high SNR limit. However, in the finite SNR setting, the same GDoF value do not mean the same achievable rate, i.e., the achievable rate of each user is unbalanced. For instance, when $P=10000$, using GGPC, the achievable rates of User $1$, $2$, and $3$ are $12.27$, $12.29$ and $12.89$, respectively. Hence the symmetric rate obtained by GGPC is the minimal one, which is $12.27$. While in GP, when $P=10000$, all users obtain the same rate $12.64$. With $P$ further increasing, from Fig.~\ref{sym_rate_Plimit}, we find that GSFPC, GGPC and GP obtain the same symmetric GDoF.}

\begin{figure}[h]
\centering
\subfigure[]{
\includegraphics[width= 6 cm]{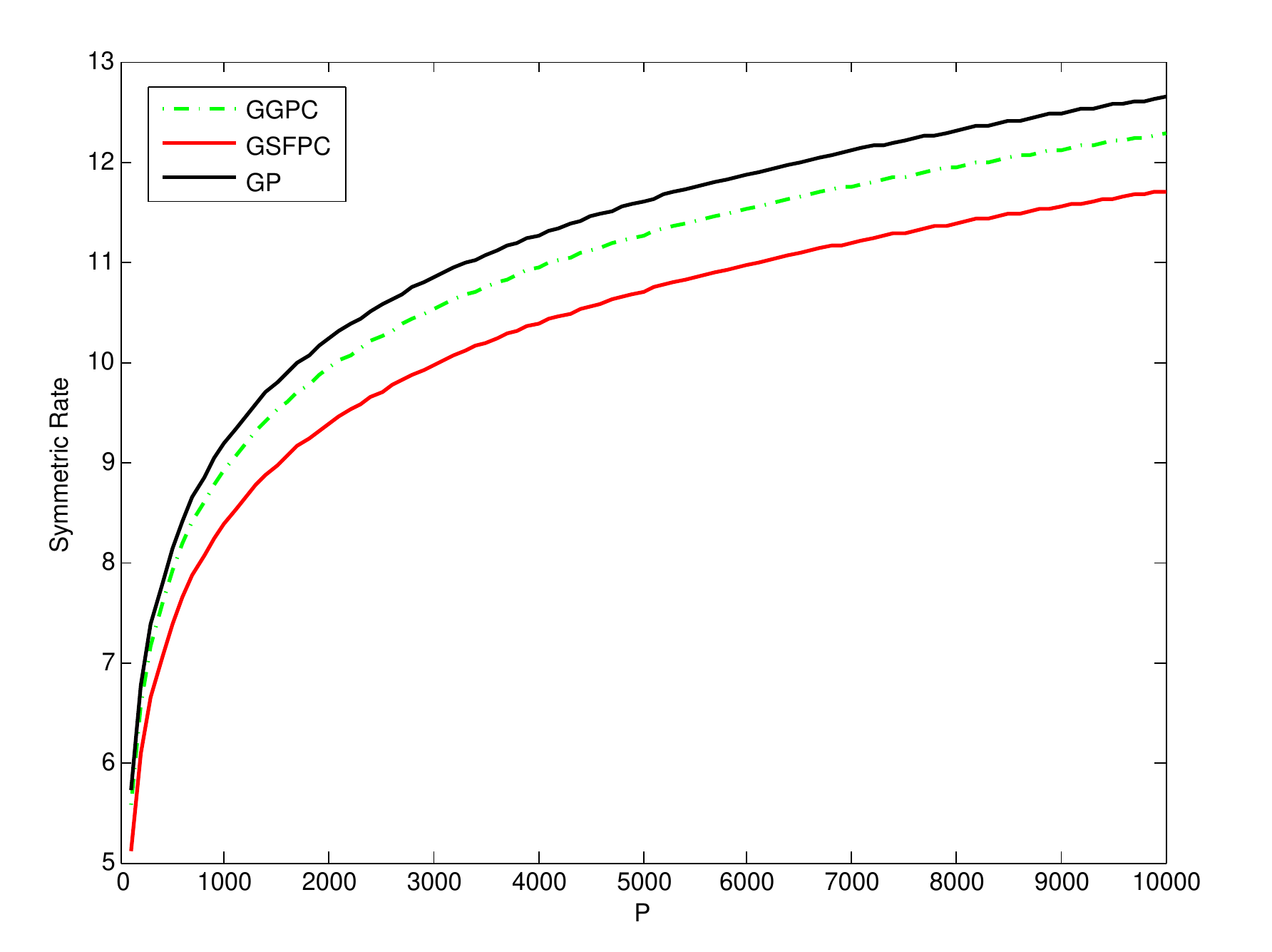}
\label{sym_rate}}
\hspace{1in}
\subfigure[]{
\includegraphics[width= 6 cm]{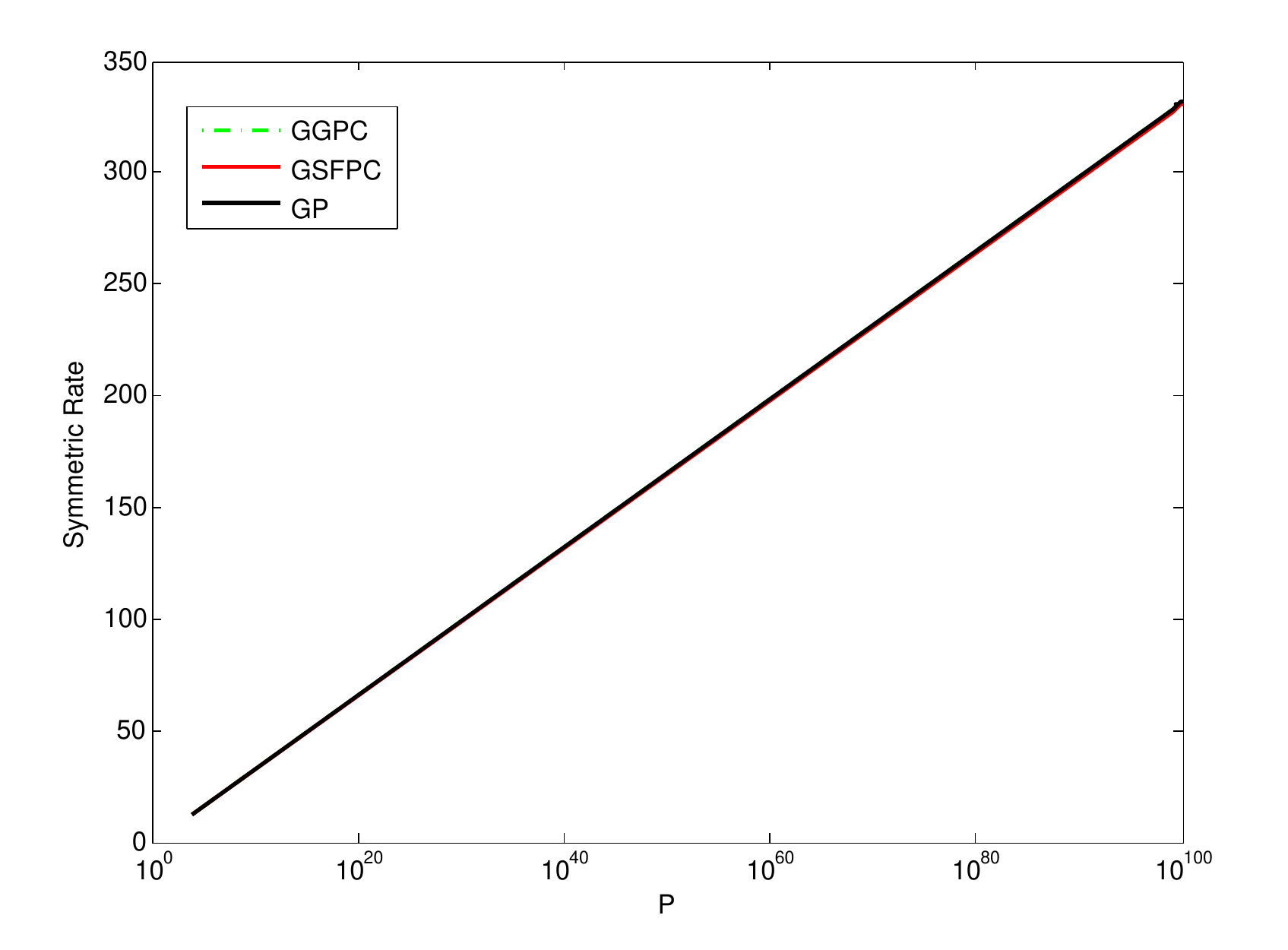}
\label{sym_rate_Plimit}}
\caption[]{
\subref{sym_rate} The achievable symmetric rates under various power control schemes in the finite SNR setting;  \subref{sym_rate_Plimit} The achievable symmetric rates under various power control schemes as P  (in log scale)  increases.}
\end{figure}

\end{example}

\subsection{More on Compound Interference Channel and its Regular Counterpart}\label{sec_counterexp}
In this part, we first show that if the compound interference channel $\mathcal{IC}_C$ satisfies the TIN-optimality condition (\ref{TIN_cond}), its regular counterpart $\mathcal{IC}_R$ is also TIN-optimal.  Define 
\begin{align}\label{tin_def}
\bar{l}_{k,j}=\arg \min_{l_k}\{\alpha_{kk}^{[l_k]}-\alpha_{kj}^{[l_k]}\} 
\end{align}
Since (\ref{TIN_cond}) is satisfied, for any $i,j,k\in\langle K\rangle$ and $i,j\neq k$, we have 
\begin{align}
&\alpha_{kk}^{[\bar{l}_{k,j}]} \geq \alpha_{kj}^{[\bar{l}_{k,j}]}+\alpha_{ik}^{[\bar{l}_{i,k}]}\\
\Rightarrow & (\alpha_{kk}^{[\bar{l}_{k,j}]}-\alpha_{kj}^{[\bar{l}_{k,j}]})+(\alpha_{ii}^{[\bar{l}_{i,k}]}-\alpha_{ik}^{[\bar{l}_{i,k}]})\geq \alpha_{ii}^{[\bar{l}_{i,k}]}\\
\Rightarrow & \min_{l_k}\{\alpha_{kk}^{[l_{k}]}-\alpha_{kj}^{[l_{k}]}\}+\min_{l_i}\{\alpha_{ii}^{[l_{i}]}-\alpha_{ik}^{[l_{i}]}\}\geq \min_{l_i}\{\alpha_{ii}^{[l_i]}\}\label{tin_11}\\
\Rightarrow & \min_{l_k}\{\alpha_{kk}^{[l_k]}\}\geq \bigg(\min_{l_k}\{\alpha_{kk}^{[l_k]}\}-\min_{l_k}\{\alpha_{kk}^{[l_{k}]}-\alpha_{kj}^{[l_{k}]}\}\bigg)+\bigg(\min_{l_i}\{\alpha_{ii}^{[l_i]}\}-\min_{l_i}\{\alpha_{ii}^{[l_{i}]}-\alpha_{ik}^{[l_{i}]}\}\bigg)\\
\Rightarrow & \bar{\alpha}_{kk}\geq \bar{\alpha}_{kj}+\bar{\alpha}_{ik}\label{tin_12}
\end{align}
where (\ref{tin_11}) follows due to the definition of $\bar{l}_{k,j}$ in $(\ref{tin_def})$ and $ \alpha_{ii}^{[\bar{l}_{i,k}]}\geq \min_{l_i}\{\alpha_{ii}^{[l_i]}\}$. Since (\ref{tin_12}) holds for any  $i,j,k\in\langle K\rangle$ and $i,j\neq k$, we finally obtain 
\begin{align}
\bar{\alpha}_{kk}\geq \max_{j:j\neq k}\{ \bar{\alpha}_{kj}\}+\max_{i:i\neq k} \{\bar{\alpha}_{ik}\}, ~~ \forall k\in\langle K \rangle 
\end{align}
which indicates  that $\mathcal{IC}_R$ is a TIN-optimal interference channel. 

Another natural question one may ask is if $\mathcal{IC}_R$ satisfies the TIN-optimality condition identified in \cite{Geng_TIN}, whether the original compound interference channel $\mathcal{IC}_C$ is also TIN-optimal. In other words, can we identify a sufficient condition on the optimality of TIN for the compound interference channel based on its regular counterpart? In the following, we give an example to show that the answer is negative. 

\begin{figure}[h]
\centering
\subfigure[]{
\includegraphics[width= 5 cm]{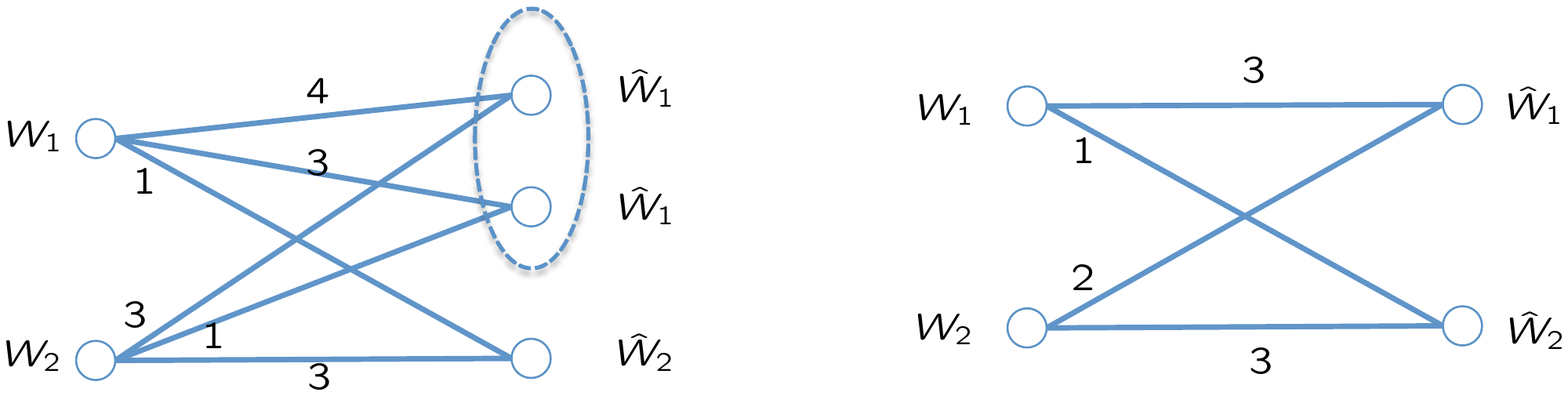}
\label{ADT_com}}
\hspace{1in}
\subfigure[]{
\includegraphics[width=5 cm]{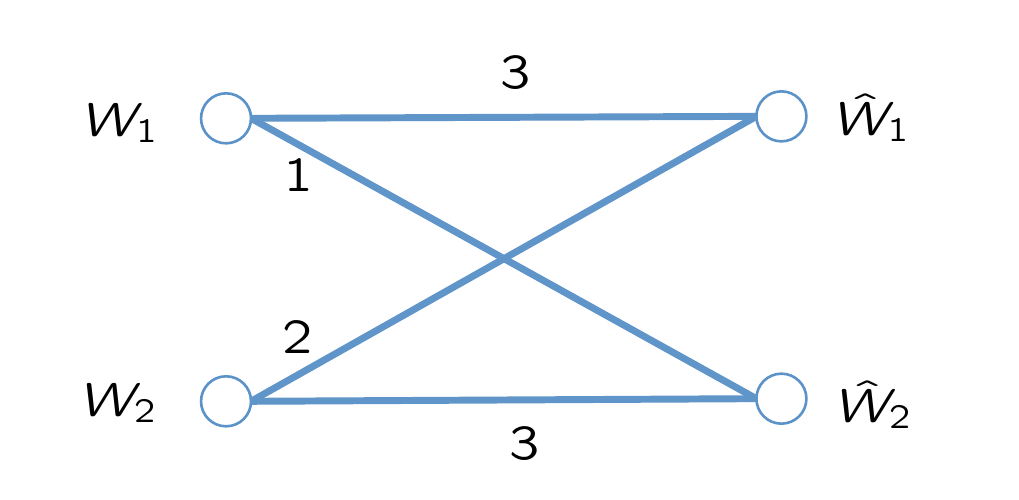}
\label{ADT_reg}}
\subfigure[]{
\includegraphics[width= 5 cm]{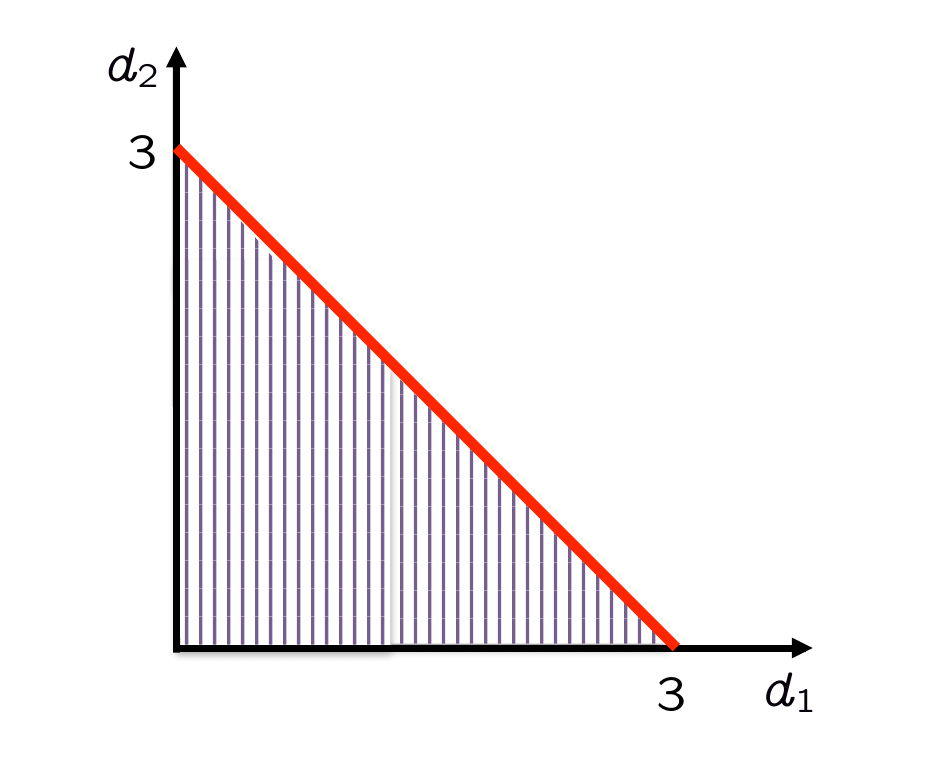}
\label{ADT_TIN_Region}}
\subfigure[]{
\includegraphics[width= 5.3 cm]{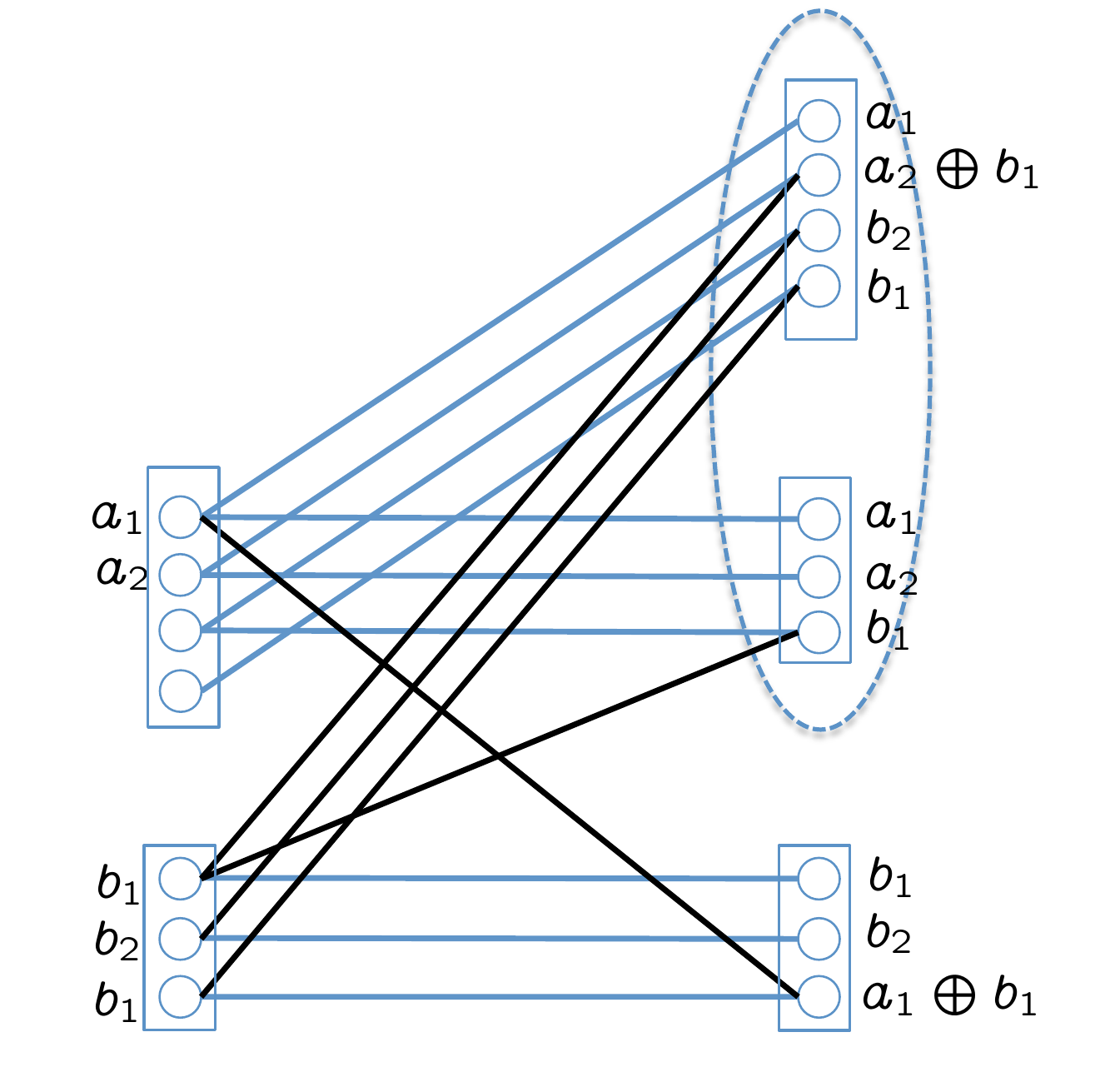}
\label{ADT}}
\caption[]{
\subref{ADT_com} A $2$-user compound interference channel $\mathcal{IC}_C$, where the value on each link is equal to its channel strength level;  \subref{ADT_reg} The counterpart regular interference channel $\mathcal{IC}_R$, where the value on each link is equal to its channel strength level; \subref{ADT_TIN_Region} The TIN region of $\mathcal{IC}_C$ and $\mathcal{IC}_R$; \subref{ADT} An achievable scheme for the ADT linear deterministic model of $\mathcal{IC}_C$ in which each user achieves two bits.}
\end{figure}

Consider the $2$-user compound interference channel depicted in Fig.~\ref{ADT_com}. Its counterpart regular interference channel is shown in Fig.~\ref{ADT_reg}, which satisfies the TIN-optimality condition identified in \cite{Geng_TIN}.  It is easy to verify that the shadowed region in Fig.~\ref{ADT_TIN_Region} is the TIN region of both channels. Next, we show that in the $2$-user compound interference channel in Fig.~\ref{ADT_com}, each user can achieve $2$ GDoF, thus the compound channel is not TIN-optimal. To convey the key idea of the achievability, we provide the achievable scheme for the corresponding ADT linear deterministic model \cite{ADT} in Fig.~\ref{ADT}, which can be translated to the Gaussian case as well.  As shown in Fig.~\ref{ADT}, $a_1$ and $a_2$ are sent on the top two levels of Transmitter $1$, and $b_1$, $b_2$ and $b_1$ are sent out from Transmitter $2$. In the first state of Receiver $1$, the receiver first decodes the interference $b_1$ at the least-significant signal level and then subtract it from the second top signal level to decode the desired signal $a_2$. Thus the first state of Receiver $1$ can decode both desired signals $a_1$ and $a_2$.  It is easy to check that the second state of Receiver $2$ can decode $a_1$ and $a_2$, and Receiver $2$ can decode $b_1$ and $b_2$. Hence each receiver can achieve $2$ bits in the ADT model, which indicates that in the corresponding Gaussian setting each receiver can achieve $2$ GDoF.

\subsection{Shortest Path Results for Non-Pareto Optimal GDoF Tuples}

\begin{figure}[h]
\begin{center}
 \includegraphics[width= 12 cm]{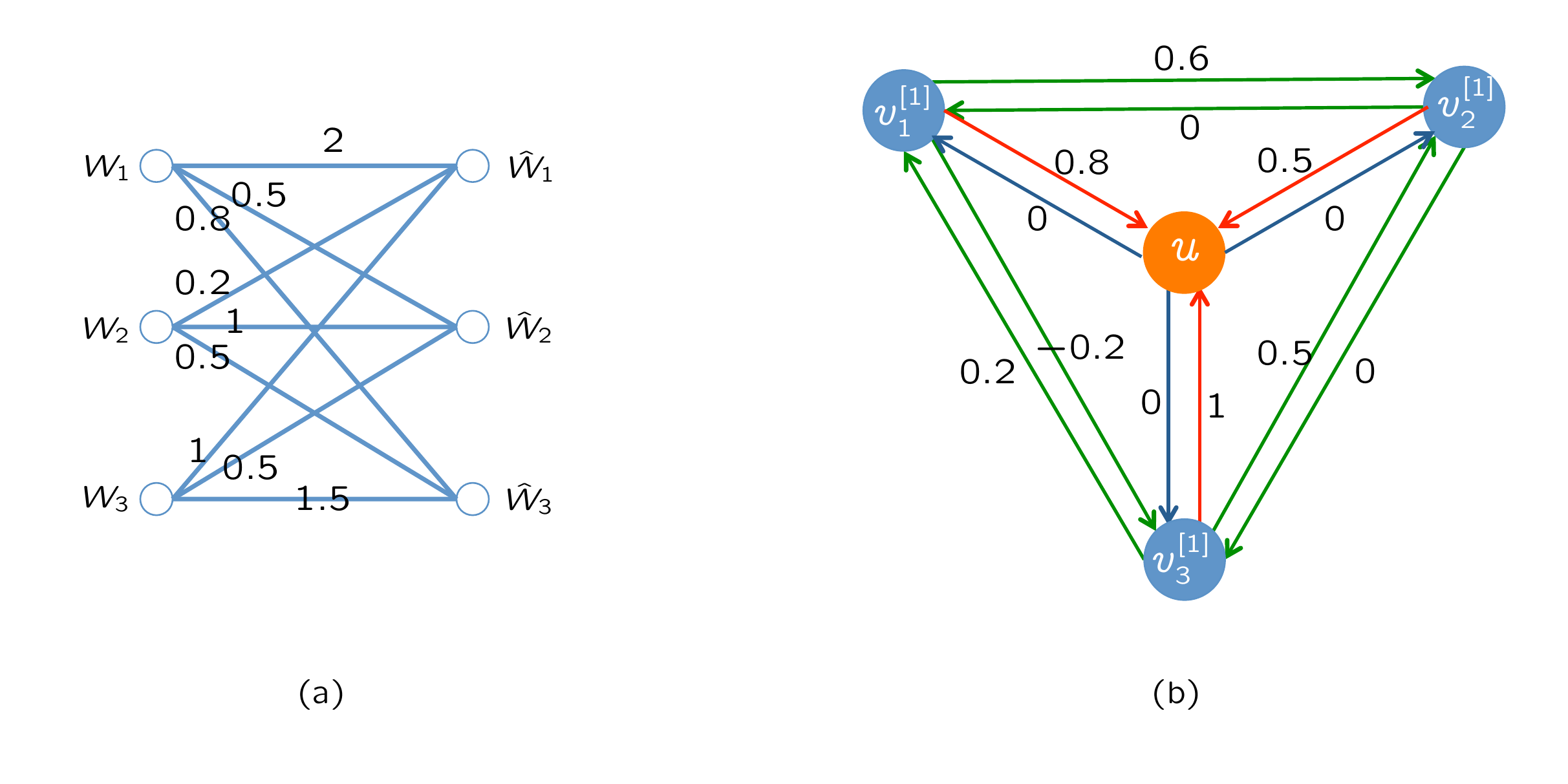}
 \caption{(a) A $3$-user interference channel, where the value on each link denotes its channel strength level; (b) The corresponding potential graph $D_p$ when $(d_1,d_2,d_3)=(1.2,0.5,0.5)$.}
\label{ex_dis2}
\end{center}
\end{figure}

In this section, we give an example to show that for a general $K$-user interference channel, even a target GDoF tuple is not Pareto optimal in the polyhedral TIN region, the shortest path results in the potential graph may still give a locally optimal power control solution.  Consider the $3$-user interference channel as shown in Fig.~\ref{ex_dis2}(a). The target GDoF tuple is $(1.2, 0.5, 0.5)$, which is not Pareto optimal for the polyhedral TIN region. From its corresponding potential graph $D_p$ depicted in Fig.~\ref{ex_dis2}(b), it is easy to obtain that the shortest path result is $(0, 0, -0.2)$, which is a locally optimal power control solution for the target GDoF tuple. 

%Specifically, in this example, the locally optimality of the obtained power allocation solution comes from three edges $(v_2^{[1]}\rightarrow v_1^{[1]})$, $(v_1^{[1]}\rightarrow v_3^{[1]})$ and $(v_3^{[1]}\rightarrow v_1^{[1]})$. In Fig.~\ref{ex_dis2}(b), from $(v_2^{[1]}\rightarrow v_1^{[1]})$, we have
%\begin{align*}
%l(v_2^{[1]},v_1^{[1]})=l_{1,dst}-l_{2,dst}&\Leftrightarrow \alpha_{22}-d_2-\alpha_{21}=l_{1,dst}-l_{2,dst}\\
%&\Leftrightarrow d_2=\alpha_{22}+l_{2,dst}-(\alpha_{21}+l_{1,dst})\\
%&\Leftrightarrow d_2=\alpha_{22}+l_{2,dst}-\max\{\alpha_{21}+l_{1,dst},\alpha_{23}+l_{3,dst},0\}
%\end{align*}
%Similarly, from the edges $(v_1^{[1]}\rightarrow v_3^{[1]})$ and $(v_3^{[1]}\rightarrow v_1^{[1]})$, we can obtain
%\begin{align*}
%l(v_1^{[1]},v_3^{[1]})=l_{3,dst}-l_{1,dst}&\Leftrightarrow d_1=\alpha_{11}+l_{1,dst}-\max\{\alpha_{12}+l_{2,dst},\alpha_{13}+l_{3,dst},0\}\\
%l(v_3^{[1]},v_1^{[1]})=l_{1,dst}-l_{3,dst}&\Leftrightarrow d_3=\alpha_{33}+l_{3,dst}-\max\{\alpha_{31}+l_{1,dst},\alpha_{32}+l_{2,dst},0\}\\
%\end{align*}

\subsection{Additional Insights on Globally Optimal Power Allocations for Compound Channels}

In this section, we give two alternative approaches to get the globally optimal power allocation for compound interference channels and offer some additional insights on the globally optimal solution. A main observation is that for any feasible GDoF tuple, the globally optimal power allocation for the compound channel is the globally optimal solution for one of its potential network states.

First, we extend the GGPC algorithm given in Section \ref{s_result} to the compound setting and obtain the GGPC-C algorithm, which is specified as follows. %at the top of page 26. 
% \\Compound algorithm
\begin{algorithm}
\caption{GDoF-based globally-optimal power control for compound interference channels (GGPC-C)}
\begin{algorithmic}
\STATE 1) Initialize: set $\mathcal{I}=\langle K\rangle$, $\mathcal{M}=\phi$, and $r_i(0)=l_{i,dst}$ for $i\in\langle K\rangle$;

\STATE 2) Update:
\begin{alignat}{2}
&\Delta(n)=\min_i\bigg\{\min_{l_i}\Big\{r_i(n)+\alpha_{ii}^{[l_i]}-d_i-\max_{m\neq i}\{0,\alpha_{im}^{[l_i]}+r_m(n)\}\Big\}\bigg\},\:\:&&i\in \mathcal{I}, m\in \mathcal{M}\\
%&\Delta(n)=\min_i\{\max_{j\neq i}\{0,\alpha_{ij}+r_j(n)\}-\max_{m\neq i}\{0,\alpha_{im}+r_m(n)\}\},\:\:&&i\in I, j\in [K], m\in M\\
&\mathcal{N}=\arg\min_i\bigg\{\min_{l_i}\Big\{r_i(n)+\alpha_{ii}^{[l_i]}-d_i-\max_{m\neq i}\{0,\alpha_{im}^{[l_i]}+r_m(n)\}\Big\}\bigg\},\:\:&&i\in \mathcal{I}, m\in \mathcal{M}\\
%&k=\arg\min_i\{\max_{j\neq i}\{0,\alpha_{ij}+r_j(n)\}-\max_{m\neq i}\{0,\alpha_{im}+r_m(n)\}\},\:\:&&i\in I, j\in [K], m\in M\\
&r_i(n+1)=r_i(n)-\Delta(n),&&i\in \mathcal{I}\\
&r_m(n+1)=r_m(n),&& m\in \mathcal{M}\\
&\mathcal{I}=\mathcal{I}\backslash \mathcal{N}, \mathcal{M}=\mathcal{M}\cup \mathcal{N} 
\end{alignat}

\STATE where $n$ indexes discrete time slots. The update phase terminates when $\mathcal{I}=\phi$.
\end{algorithmic}
\end{algorithm}

 For the GGPC-C algorithm, we have the following theorem. 
\begin{theorem}\label{GPUC_converge}
In a $K$-user compound interference channel, for any achievable GDoF tuple in its polyhedral TIN region $\mathcal{P}$, the GGPC-C algorithm obtains the globally optimal transmit power allocation.  
\end{theorem}

The proof of Theorem \ref{GPUC_converge} is essentially the same as that of Theorem \ref{GPU_converge} and relegated to Appendix \ref{app_GPUC}. We find that for any feasible GDoF tuple, at the end of the GGPC-C algorithm, for every user there exists at least one state achieving the exact target GDoF value,\footnote{Note that for different feasible target GDoF tuples, the state of each user that achieves the exact target GDoF value might be different.} and each of the other states achieves a GDoF value no less than the target one. Picking out that state achieving the exact target value for each user (which apparently results in a possible network realization of the original compound channel, and for different target GDoF tuples, the obtained network realization might be different), it is not hard to argue that the solution obtained from the GGPC-C algorithm is globally optimal in this network realization for the target GDoF tuple. As mentioned earlier, it implies that for any feasible GDoF tuple, the globally optimal power allocation for the compound channel is the globally optimal solution for one of its possible network realizations.  We can further show that for a given feasible  GDoF tuple, among the globally optimal solutions for all the possible network realizations, there is only one unique solution which can achieve a GDoF tuple dominating the target one in each possible network realization.\footnote{We  prove this argument through contradiction. For a target GDoF tuple, denote the set of the globally optimal solutions for all the possible network solutions as $\mathcal{S}_g$. Assume $\mathbf{s}_A, \mathbf{s}_B\in\mathcal{S}_g$ are the globally optimal power allocations for two network realizations $\mathcal{A}$ and $\mathcal{B}$, respectively, and $\mathbf{s}_A\neq \mathbf{s}_B$. Further assume that using $\mathbf{s}_A$ and $\mathbf{s}_B$ as power allocations, all network realizations achieve GDoF tuples dominating the target one. In the network realization $\mathcal{A}$, since for the target GDoF tuple $\mathbf{s}_A$ is the globally optimal power allocation and $\mathbf{s}_B$ is an acceptable power allocation, we have $\mathbf{s}_A\leq \mathbf{s}_B$. Similarly, in the network realization $\mathcal{B}$, we have $\mathbf{s}_B\leq \mathbf{s}_A$. Finally we obtain $\mathbf{s}_A= \mathbf{s}_B$, which contradicts the assumption that $\mathbf{s}_A\neq \mathbf{s}_B$.} Based on the above observations, we can also apply the GGPC algorithm to each of the possible network realizations individually, and then find the globally optimal power allocation for the original compound channel. However, both methods given in this section incur higher computational complexity compared with the approach presented in Section \ref{s_result}.

\section{Conclusion}
In this work, we first generalize the optimality of TIN to compound interference channels. We show that for a $K$-user compound Gaussian interference channel, if in each possible network state, the channel satisfies the TIN-optimality condition identified in \cite{Geng_TIN},  then its GDoF region is the intersection of the GDoF regions of all possible network states, which is achievable via power control and TIN. For general compound interference channels which may be not TIN-optimal, the TIN region is also fully characterized. Next, the power control problem is investigated from the GDoF perspective for compound networks. Notably, we demonstrate that to solve the GDoF-based power control problem for a $K$-user interference channel with arbitrary number of states for each receiver, we only need to construct its counterpart $K$-user regular interference channel that has the same TIN region as the target compound channel, and solve the power control problem in this new channel.  For general $K$-user compound interference channels,  we develop a centralized power control scheme to achieve all Pareto optimal GDoF tuples. Finally, an iterative power control algorithm with at most $K$ updates is proposed to obtain the globally optimal transmit power allocations for all feasible GDoF tuples. 

Combining the TIN region characterization with the GDoF-based power control algorithms, this work also solves the joint GDoF assignment and power control problem for general $K$-user compound interference channels. For the TIN-optimal channels, the obtained result guarantees a constant gap to the information-theoretically optimal performance (e.g., the sum rate) at any finite SNR.  As a byproduct, this work provides an alternative perspective to deal with the challenging joint rate assignment and power control problem. Simple simulations indicate that this simple GDoF-based approach may attain close performance to its sophisticated SINR-based counterpart in the finite SNR setting. Substantial efforts are still needed to illustrate the performance of GDoF-based power control schemes for general network settings in finite SNRs. Also, developing fully distributed GDoF-based power control algorithm is another interesting future direction from both theoretical and practical points of view.

\appendix
  \renewcommand{\appendixname}{Appendix~\Alph{section}}

\section{Proof of Theorem \ref{T1_IC}} \label{S1_IC}
 \subsection{Proof for a Simple Example}\label{S1_IC1}
Before proceeding to the proof for the general case, let us consider a relatively simple example, a $2$-user compound interference channel with $L_1=2$ and $L_2=1$ as shown in Fig. \ref{ex1}(a), to convey the key argument. Assume the condition (\ref{TIN_cond}) is satisfied, which means that in each network realization, the channel is TIN-optimal, i.e., 
\begin{align}
\alpha_{11}^{[1]}&\geq \alpha_{12}^{[1]}+\alpha_{21}^{[1]}\label{cond1}\\
\alpha_{22}^{[1]}&\geq \alpha_{12}^{[1]}+\alpha_{21}^{[1]} \label{cond2}
\end{align}
and 
\begin{align}
\alpha_{11}^{[2]}&\geq \alpha_{12}^{[2]}+\alpha_{21}^{[1]}\label{cond3}\\
\alpha_{22}^{[1]}&\geq \alpha_{12}^{[2]}+\alpha_{21}^{[1]} \label{cond4}
\end{align}

For the converse, we first consider each of the two possible network realizations individually. Since in each network realization, the channel satisfies the TIN-optimality condition, it is not hard to characterize its GDoF region \cite{Geng_TIN}. Taking the intersection of these two GDoF regions, we obtain the desired outer bounds. 

Next, we prove the achievability. Assume the power allocated to Transmitter $1$ and $2$ are $P^{r_1}$ and $P^{r_2}$, respectively.  Due to the unit power constraint, $r_1,r_2\leq 0$. By treating interference as noise, User 1 and 2 can achieve any rates $R_1$ and $R_2$ such that  
\begin{align}
R_1&\leq\min\bigg\{\log\bigg(1+\frac{P^{r_1}\times P^{\alpha_{11}^{[1]}}}{1+P^{r_2}\times P^{\alpha_{12}^{[1]}}}\bigg)
,\log\bigg(1+\frac{P^{r_1}\times P^{\alpha_{11}^{[2]}}}{1+P^{r_2}\times P^{\alpha_{12}^{[2]}}}\bigg)
\bigg\}\\
R_2&\leq\log\bigg(1+\frac{P^{r_2}\times P^{\alpha_{22}^{[1]}}}{1+P^{r_1}\times P^{\alpha_{21}^{[1]}}}\bigg)
\end{align}  
In the GDoF sense, we have 
\begin{align}
0\leq d_1&\leq\min\big\{\max\{0,\alpha_{11}^{[1]}+r_1-(\alpha_{12}^{[1]}+r_2)^+\},\max\{0,\alpha_{11}^{[2]}+r_1-(\alpha_{12}^{[2]}+r_2)^+\}\big\}\label{Rop1}\\
0\leq d_2&\leq \max\{0,\alpha_{22}^{[1]}+r_2-(\alpha_{21}^{[1]}+r_1)^+\}\label{Rop2}
\end{align}
 
Applying the polyhedral TIN scheme, we can ignore the terms of $\max\{0,.\}$ in (\ref{Rop1}) and (\ref{Rop2}) and obtain
\begin{align}
0\leq d_1&\leq\min\{\alpha_{11}^{[1]}+r_1-(\alpha_{12}^{[1]}+r_2)^+,\alpha_{11}^{[2]}+r_1-(\alpha_{12}^{[2]}+r_2)^+\}\label{d_11}\\
0\leq d_2&\leq \alpha_{22}^{[1]}+r_2-(\alpha_{21}^{[1]}+r_1)^+\label{d_21}
\end{align}
provided that the right hand sides of (\ref{d_11})-(\ref{d_21}) are non-negative. The polyhedral TIN region $\mathcal{P}$ is the set of all GDoF tuples $(d_1,d_2)$ for which there exist $r_1$ and $r_2$ such that
\begin{align}
d_1&\geq 0 \label{ie1}\\
d_2 &\geq 0\\
r_1&\leq 0\\
r_2&\leq 0\\
d_1&\leq \alpha_{11}^{[1]}+r_1\\
d_1&\leq(\alpha_{11}^{[1]}-\alpha_{12}^{[1]})+(r_1-r_2)\\
d_1&\leq \alpha_{11}^{[2]}+r_1\\
d_1&\leq (\alpha_{11}^{[2]}-\alpha_{12}^{[2]})+(r_1-r_2)\\
d_2&\leq \alpha_{22}^{[1]}+r_2\\
d_2&\leq (\alpha_{22}^{[1]}-\alpha_{21}^{[1]})+(r_2-r_1)\label{ie2}
\end{align}
Later our results will show that when (\ref{cond1})-(\ref{cond4}) hold, the above modification incurs no loss for the achievable GDoF region via power control and TIN.

%\begin{remark}
%Note that from (\ref{ie1})-(\ref{ie2}), to obtain the polyhedra TIN region $\mathcal{P}$ which is only associated with $d_i$'s and $\alpha_{ki}^{[l_k]}$'s, essentially we just need to apply FME to eliminate the power exponent variables $r_1$ and $r_2$. However, as $K$ and $L_k$, $k\in\langle K\rangle$ increase, FME incurs high complexity. Instead, the following argument based on the potential graph $D_p$ can be easily extended to the general case. 
%\end{remark}

Set $r_1^{[1]}=r_1^{[2]}=r_1$ and $r_2^{[1]}=r_2$. It is easy to check that the polyhedral TIN region $\mathcal{P}$ is also characterized by the following linear inequalities
\begin{align}
d_1&\geq 0\\
d_2 &\geq 0\\
r_1^{[1]}&\leq 0  \label{ie3}\\
r_1^{[2]}&\leq 0\\
r_2^{[1]}&\leq 0\\
r_1^{[1]}-r_1^{[2]}&\leq 0\\
r_1^{[2]}-r_1^{[1]}&\leq 0\\
-r_1^{[1]}&\leq \alpha_{11}^{[1]}-d_1\\
r_2^{[1]}-r_1^{[1]}&\leq(\alpha_{11}^{[1]}-\alpha_{12}^{[1]})-d_1\\
-r_1^{[2]}&\leq \alpha_{11}^{[2]}-d_1\\
r_2^{[1]}-r_1^{[2]}&\leq (\alpha_{11}^{[2]}-\alpha_{12}^{[2]})-d_1\\
-r_2^{[1]}&\leq \alpha_{22}^{[1]}-d_2\\
r_1^{[1]}-r_2^{[1]}&\leq (\alpha_{22}^{[1]}-\alpha_{21}^{[1]})-d_2\\
r_1^{[2]}-r_2^{[1]}&\leq (\alpha_{22}^{[1]}-\alpha_{21}^{[1]})-d_2\label{ie4}
\end{align}
In other words, for a GDoF tuple $(d_1,d_2)\in\mathbb{R}^2_+$ (the non-negative orthant of the $2$-dimensional Euclidean space), it is in the polyhedral TIN region $\mathcal{P}$ if and only if there exists $r_1^{[1]}, r_1^{[2]}, r_2^{[1]}$ such that (\ref{ie3})-(\ref{ie4}) hold.

Next, recall that in the potential graph $D_p$ for this $2$-user compound interference channel (see Fig.~\ref{potential_graph1}), the length of each edge is defined as follows 

\begin{align*}
l(v_1^{[1]},v_1^{[2]})&=l(v_1^{[2]},v_1^{[1]})=0\\ 
l(v_1^{[1]},v_2^{[1]})&=(\alpha_{11}^{[1]}-\alpha_{12}^{[1]})-d_1\\
l(v_1^{[2]},v_2^{[1]})&=(\alpha_{11}^{[2]}-\alpha_{12}^{[2]})-d_1\\
l(v_2^{[1]},v_1^{[1]})&=l(v_2^{[1]},v_1^{[2]})=(\alpha_{22}^{[1]}-\alpha_{21}^{[1]})-d_2\\
l(v_i^{[l_i]},u)&=\alpha_{ii}^{[l_i]}-d_i,~~~\forall i\in[2], \forall l_i\in\langle L_i\rangle\\
l(u,v_i^{[l_i]})&=0,~~~~~~~~~~~~~\forall i\in[2], \forall l_i\in\langle L_i\rangle
\end{align*}

By definition \cite{Potential}, for a graph a function $p:V\rightarrow \mathbb{R}$ is called a potential if for every two vertices $a,b\in V$ such that $(a,b)\in E$, $l(a,b)\geq p(b)-p(a)$. Note these inequalities depend upon the difference between potential function values only. Therefore, without loss of generality, if there exists a valid potential function for the potential graph $D_p$, we can make one vertex, e.g., vertex $u$, ground, i.e., $p(u)=0$. Then let $r_i^{[l_i]}:=p(v_i^{[l_i]})$.  It is easy to check that these potential function values should satisfy the inequalities (\ref{ie3}) - (\ref{ie4}). In other words, \emph{in the $2$-user compound interference channel in Fig.~\ref{ex1}(a), for a GDoF tuple $(d_1,d_2)\in\mathbb{R}^2_+$, it is in the region  $\mathcal{P}$ if and only if there exists a valid potential function for the potential graph $D_p$ in Fig.~\ref{potential_graph1}}.

Based on the potential theorem in \cite{Potential}, which says that \emph{there exists a potential function for a  digraph $D$ if and only if each directed circuit in $D$ has a non-negative length}, we conclude that for a GDoF tuple $(d_1,d_2)\in\mathbb{R}^2_+$, it is in the region $\mathcal{P}$ if and only if the length of each directed circuit in the potential graph $D_p$ is non-negative.  Considering all the circuits in Fig.~\ref{potential_graph1}, requiring each of them to have a non-negative length, and removing all the redundant inequalities, we end up with 
\begin{align}
d_1&\leq \alpha_{11}^{[1]}\\
d_1&\leq \alpha_{11}^{[2]}\\
d_2&\leq \alpha_{22}^{[1]}\\
d_1+d_2&\leq(\alpha_{11}^{[1]}+\alpha_{22}^{[1]})-(\alpha_{21}^{[1]}+\alpha_{12}^{[1]})\\
d_1+d_2&\leq(\alpha_{11}^{[2]}+\alpha_{22}^{[1]})-(\alpha_{21}^{[1]}+\alpha_{12}^{[2]})
\end{align}
Explicitly adding the non-negative constraint on $d_i$'s, finally we obtain the polyhedral TIN region $\mathcal{P}$, which is the set of all the GDoF tuples $(d_1,d_2)$ satisfying
\begin{align*}
0\leq d_1&\leq \alpha_{11}^{[1]}\\
0\leq d_1&\leq \alpha_{11}^{[2]}\\
0\leq d_2&\leq \alpha_{22}^{[1]}\\
d_1+d_2&\leq(\alpha_{11}^{[1]}+\alpha_{22}^{[1]})-(\alpha_{21}^{[1]}+\alpha_{12}^{[1]})\\
d_1+d_2&\leq(\alpha_{11}^{[2]}+\alpha_{22}^{[1]})-(\alpha_{21}^{[1]}+\alpha_{12}^{[2]})
\end{align*}
Note that the polyhedral TIN region $\mathcal{P}$ is exactly the intersection of the polyhedral TIN regions of the two possible network realizations.  
Under conditions (\ref{cond1})-(\ref{cond4}), $\mathcal{P}$ is the intersection of the GDoF regions of the two possible network realizations and matches the outer bounds. Thus we complete the proof. 

\subsection{Proof for General Cases}
Now let us consider the general $K$-user compound interference channel, where the condition (\ref{TIN_cond}) is satisfied. For the converse, again we first consider each possible network realization individually. Since in each network realization, the channel satisfies the TIN-optimality condition, it is not hard to characterize its GDoF region \cite{Geng_TIN}. Taking the intersection of the GDoF regions of all the possible network realizations, we have the desired outer bounds. 

Next, consider the achievability. Recall that by allocating the power $P^{r_k}$, $r_k\leq 0$, to Transmitter $k$ and treating interference as noise at each receiver,  User $k$ can achieve any rate $R_k$ such that 
\begin{align*}
R_k\leq\min_{l_k\in\langle L_k\rangle}\bigg\{\log\bigg(1+\frac{P^{r_k}\times P^{\alpha_{kk}^{[l_k]}}}{1+\sum_{j=1,j\neq k}^KP^{r_j}\times P^{\alpha_{kj}^{[l_k]}}}\bigg)\bigg\},~~\forall k\in\langle K\rangle
\end{align*}  
In the GDoF sense, we have 
\begin{align*}
0\leq d_k\leq\min_{l_k\in\langle L_k\rangle}\big\{ \max\{0,\alpha_{kk}^{[l_k]}+r_k-(\max_{j:j\neq k}\{\alpha_{kj}^{[l_k]}+r_j\})^+\}\big\},~~\forall k\in\langle K\rangle
%d_1&=\min\big\{\max\{0,\alpha_{11}^{[1]}+r_1-(\alpha_{12}^{[1]}+r_2)^+\},\max\{0,\alpha_{11}^{[2]}+r_1-(\alpha_{12}^{[2]}+r_2)^+\}\big\}\label{Rop1}\\
%d_2&=\max\{0,\alpha_{22}^{[1]}+r_2-(\alpha_{21}^{[1]}+r_1)^+\}\label{Rop2}
\end{align*}

Applying the polyhedral TIN scheme, we can ignore the terms of $\max\{0,.\}$ in the above equation and obtain
\begin{align}
0\leq d_k&\leq \min_{l_k\in\langle L_k\rangle}\big\{ \alpha_{kk}^{[l_k]}+r_k-(\max_{j:j\neq k}\{\alpha_{kj}^{[l_k]}+r_j\})^+\big\},~~\forall k\in\langle K\rangle\label{d_k1}
\end{align}
provided that the right hand side of (\ref{d_k1}) is non-negative for all users $k\in\langle K\rangle$. The obtained polyhedral TIN region $\mathcal{P}$ is the set of all GDoF tuples $(d_1,d_2,...,d_K)$ for which there exist $r_k$'s, $k\in\langle K\rangle$, such that

\begin{alignat}{2}
d_k&\geq 0,&&\forall k\in\langle K\rangle\label{eqq3}\\
r_k&\leq 0,&&\forall k\in\langle K\rangle\\
d_k\leq \alpha_{kk}^{[l_k]}+r_k \Leftrightarrow r_k&\geq d_k-\alpha_{kk}^{[l_k]},&&\forall k\in\langle K\rangle,\forall l_k\in\langle L_k\rangle\label{eqq4}\\
d_k\leq \alpha_{kk}^{[l_k]}+r_k-(\alpha_{kj}^{[l_k]}+r_j) \Leftrightarrow r_k-r_j&\geq (\alpha_{kj}^{[l_k]}-\alpha_{kk}^{[l_k]})+d_k,\:\:&&\forall k,j\in\langle K\rangle, k\neq j, \forall l_k\in\langle L_k\rangle\label{eqq5}
\end{alignat}
Similar to the previous section, later our results will show that when (\ref{TIN_cond}) holds, the above modification incurs no loss for the achievable GDoF region via power control and TIN. 

Setting
\begin{align}
r_k^{[l_k]}=r_k,~~\forall k\in\langle K\rangle, \forall l_k\in\langle L_k\rangle,
\end{align}
it is easy to check that the set of the above inequalities (\ref{eqq3})-(\ref{eqq5}) is equivalent to the following ones
\begin{alignat}{2}
d_k&\geq 0,&&\forall k\in\langle K\rangle\label{eqq6_1}\\
r_k^{[l_k]}&\leq 0,&&\forall k\in\langle K\rangle,\forall l_k\in\langle L_k\rangle\label{eqq6}\\
r_k^{[l_k]}-r_k^{[l_k']}&\leq 0,&&\forall k\in\langle K\rangle, \forall l_k,l_k'\in\langle L_k\rangle, l_k\neq l_k'\\
r_k^{[l_k]}&\geq d_k-\alpha_{kk}^{[l_k]},&&\forall k\in\langle K\rangle,\forall l_k\in\langle L_k\rangle\\
r_k^{[l_k]}-r_j^{[l_j]}&\geq (\alpha_{kj}^{[l_k]}-\alpha_{kk}^{[l_k]})+d_k,\:\:&&\forall k,j\in\langle K\rangle, k\neq j, \forall l_k\in\langle L_k\rangle,\forall l_j\in\langle L_j\rangle\label{eqq7}
\end{alignat}
Therefore, the polyhedral TIN region $\mathcal{P}$ is also characterized by (\ref{eqq6_1})-(\ref{eqq7}). In other words, for a GDoF tuple $(d_1,d_2,...,d_K)\in\mathbb{R}^K_+$, it is in the polyhedral TIN region $\mathcal{P}$ if and only if there exists $r_k^{[l_k]}$, $k\in\langle K\rangle$, $l_k\in\langle L_k\rangle$, such that (\ref{eqq6})-(\ref{eqq7}) hold. 

Next, recall that in the potential graph $D_p$ for the $K$-user compound interference channel, the length $l(e)$ assigned to each edge $e\in E$ is as follows
\begin{alignat*}{2}
l(v_k^{[l_k]},v_k^{[l_k']})&=0,&&\forall k\in\langle K\rangle, \forall l_k,l_k'\in\langle L_k\rangle, l_k\neq l_k'\\
l(v_k^{[l_k]},v_j^{[l_j]})&=(\alpha_{kk}^{[l_k]}-\alpha_{kj}^{[l_k]})-d_k,~~~~~&&\forall k,j\in\langle K\rangle, k\neq j, \forall l_k\in\langle L_k\rangle,\forall l_j\in\langle L_j\rangle\\
l(v_k^{[l_k]},u)&=\alpha_{kk}^{[l_k]}-d_k,&&\forall k\in\langle K\rangle, \forall l_k\in\langle L_k\rangle\\
l(u,v_k^{[l_k]})&=0,&&\forall k\in\langle K\rangle, \forall l_k\in\langle L_k\rangle
\end{alignat*}

Without loss of generality, if there exists a valid potential function for the potential graph $D_p$, we can make the vertex $u$ ground, i.e., $p(u)=0$. Then let $r_k^{[l_k]}:=p(v_k^{[l_k]})$, $\forall k\in\langle K\rangle$. It is not hard to check that the potential function values should satisfy the inequalities (\ref{eqq6}) - (\ref{eqq7}). In other words, \emph{in a $K$-user compound interference channel, for a GDoF tuple $(d_1,d_2,...,d_K)\in\mathbb{R}^K_+$, it is in the region $\mathcal{P}$ if and only if there exists a valid potential function for its potential graph $D_p$}.

Again, based on the potential theorem,  we conclude that  for a GDoF tuple $(d_1,d_2,...,d_K)\in\mathbb{R}^K_+$, it is in the region $\mathcal{P}$ if and only if the length of each directed circuit in $D_p$ is non-negative. Therefore, to characterize the polyhedral TIN region $\mathcal{P}$, the only job left is to make sure the lengths of all the directed circuits in $D_p$ are no less than $0$.  We categorize all the directed circuits of $D_p$ into the following three classes:

\begin{itemize}
\item Circuits in the form of $(u\rightarrow v_k^{[l_k]}\rightarrow u)$. For these circuits, the non-negative length condition gives us
\begin{align}\label{eqq16}
\alpha_{kk}^{[l_k]}-d_k\geq 0\Leftrightarrow d_k\leq \alpha_{kk}^{[l_k]}
\end{align}

\item Circuits in the form $(v_{i_0}^{[l_{i_0}]}\rightarrow v_{i_1}^{[l_{i_1}]}\rightarrow...\rightarrow v_{i_m}^{[l_{i_m}]})$, where $i_0=i_m$ and $(i_1,i_2,...,i_m)\in\Pi_K$. For these circuits, the non-negative length condition becomes
\begin{equation}\label{eqq17}
\begin{aligned}
\sum_{j=0}^{m-1} (\alpha_{i_j i_j}^{[l_{i_j}]}-\alpha_{i_j i_{j+1}}^{[l_{i_j}]}-d_{i_j})\geq 0 \Leftrightarrow&
\sum_{j=0}^{m-1} d_{i_j} \leq \sum_{j=0}^{m-1} (\alpha_{i_j i_j}^{[l_{i_j}]}-\alpha_{i_j i_{j+1}}^{[l_{i_j}]})
% \overset{(a)}{\Leftrightarrow}& \sum_{j=1}^{m}d_{i_j}\leq\sum_{j=1}^{m}(\alpha_{i_ji_j}^{[l_{i_j}]}-\alpha_{i_{j-1}i_j}^{[l_{i_{j-1}}]}).
\end{aligned}
\end{equation}
%where in step $(a)$ we just reorder the terms in the right hand side and use the fact that $i_m=i_0$.

\item All the other circuits. For these remaining circuits, it is not hard to check that given (\ref{eqq16}) and (\ref{eqq17}), the inequalities derived from the non-negative length condition are all redundant.
%\begin{align}\label{eqq18}
%\sum_{j=1}^{m-1} (\alpha_{i_j i_j}^{[l_{i_j}]}-d_{i_j}-\alpha_{i_j i_{j+1}}^{[l_{i_j}]})+(\alpha_{i_m i_m}^{[l_{i_m}]}-d_{i_m})\geq 0.
%\end{align}

%Note 
\end{itemize}

Consequently, we end up with the conditions (\ref{eqq16})-(\ref{eqq17}). Next, explicitly adding the non-negative constraint on $d_i$'s in  (\ref{eqq16})-(\ref{eqq17}), we obtain the polyhedral TIN region $\mathcal{P}$, which turns out to be the intersection of the polyhedral TIN regions for all the possible network realizations and fully characterized by (\ref{eqq111}) and (\ref{eqq112}). Clearly, under condition (\ref{TIN_cond}), the polyhedral TIN region $\mathcal{P}$ coincides with the derived outer bounds.  Thus we complete the proof.

\section{Proof of Theorem \ref{th-equ}}\label{app_equ} 
To prove $\mathcal{IC}_C$ and $\mathcal{IC}_R$ have the same TIN region, we only need to show that with the same power allocation $\mathbf{r}^*=(r_1^*,r_2^*,...,r_K^*)$, $\mathcal{IC}_C$ and $\mathcal{IC}_R$ obtain the same GDoF tuple.

In the compound channel $\mathcal{IC}_C$, when the power allocation is $(r_1^*,r_2^*,...,r_K^*)$, we have
\begin{align}
d_k^{\dag}=&\min_{l_k\in\langle L_k\rangle}\big\{ \alpha_{kk}^{[l_k]}+r_k^*-(\max_{j\neq k}\{\alpha_{kj}^{[l_k]}+r_j^*\})^+\big\}\\
=&\min_{l_k\in\langle L_k\rangle}\big\{ \alpha_{kk}^{[l_k]}+r_k^*-\max_{j\neq k}\{\alpha_{kj}^{[l_k]}+r_j^*,0\}\big\}\\
=&r_k^*+\min_{l_k\in\langle L_k\rangle}\big\{ \alpha_{kk}^{[l_k]}-\max_{j\neq k}\{\alpha_{kj}^{[l_k]}+r_j^*,0\}\big\}\\
=&r_k^*+\min_{l_k\in\langle L_k\rangle}\big\{ \alpha_{kk}^{[l_k]}+\min_{j\neq k}\{-\alpha_{kj}^{[l_k]}-r_j^*,0\}\big\}\\
=&r_k^*+\min_{l_k\in\langle L_k\rangle}\big\{ \min_{j\neq k}\{\alpha_{kk}^{[l_k]}-\alpha_{kj}^{[l_k]}-r_j^*,\alpha_{kk}^{[l_k]}\}\big\}\label{sec_dkA}
\end{align}
For User $k\in\langle K\rangle$ in $\mathcal{IC}_C$, the achievable GDoF value is 
\begin{align}
d_k&=\min_{l_k\in\langle L_k\rangle}\big\{\max\{ 0,\alpha_{kk}^{[l_k]}+r_k-(\max_{j:j\neq k}\{\alpha_{kj}^{[l_k]}+r_j\})^+\}\big\}\\
&=\max\{0,d_k^{\dag}\}\\
&=\max\bigg\{0,r_k^*+\min_{l_k\in\langle L_k\rangle}\big\{ \min_{j\neq k}\{\alpha_{kk}^{[l_k]}-\alpha_{kj}^{[l_k]}-r_j^*,\alpha_{kk}^{[l_k]}\}\big\}\bigg\}\label{sec_dkA1}
\end{align}

In its regular counterpart $\mathcal{IC}_R$, with the same power allocation $(r_1^*,r_2^*,...,r_K^*)$, we have
\begin{align}
d_k^{\dag\dag}&=\bar{\alpha}_{kk}+r_k^*-(\max_{j\neq k}\{\bar{\alpha}_{kj}+r_j^*\})^+\\
&=\bar{\alpha}_{kk}+r_k^*-\max_{j\neq k}\{\bar{\alpha}_{kj}+r_j^*, 0\}\\
&=\bar{\alpha}_{kk}+r_k^*+\min_{j\neq k}\{-\bar{\alpha}_{kj}-r_j^*, 0\}\\
&=r_k^*+\min_{j\neq k}\{\bar{\alpha}_{kk}-\bar{\alpha}_{kj}-r_j^*, \bar{\alpha}_{kk}\}\\
&=r_k^*+\min_{j\neq k}\big\{\min_{l_k\in\langle L_k\rangle}\{\alpha_{kk}^{[l_k]}-\alpha_{kj}^{[l_k]}-r_j^*\}, \min_{l_k\in\langle L_k\rangle}\{\alpha_{kk}^{[l_k]}\}\big\}\\
&=r_k^*+\min_{j\neq k}\big\{\min_{l_k\in\langle L_k\rangle}\{\alpha_{kk}^{[l_k]}-\alpha_{kj}^{[l_k]}-r_j^*, \alpha_{kk}^{[l_k]}\}\big\}\\
&=r_k^*+\min_{l_k\in\langle L_k\rangle}\big\{ \min_{j\neq k}\{\alpha_{kk}^{[l_k]}-\alpha_{kj}^{[l_k]}-r_j^*,\alpha_{kk}^{[l_k]}\}\big\}\label{sec_dkB}
\end{align}
For User $k\in\langle K\rangle$ in $\mathcal{IC}_R$, the achievable GDoF value is 
\begin{align}
d_k&=\max\{0,d_k^{\dag\dag}\}\\
&=\max\bigg\{0,r_k^*+\min_{l_k\in\langle L_k\rangle}\big\{ \min_{j\neq k}\{\alpha_{kk}^{[l_k]}-\alpha_{kj}^{[l_k]}-r_j^*,\alpha_{kk}^{[l_k]}\}\big\}\bigg\}\label{sec_dkB1}
\end{align}
Comparing (\ref{sec_dkA1}) and (\ref{sec_dkB1}), we establish that $\mathcal{IC}_C$ and $\mathcal{IC}_R$ have the same TIN region.

From the above proof, we also obtain that $\mathcal{IC}_C$ and $\mathcal{IC}_R$ have the same polyhedral TIN region. Denote by $\mathcal{P}_C$ and $\mathcal{P}_R$ the polyhedral TIN regions of $\mathcal{IC}_C$ and $\mathcal{IC}_R$ respectively, $\mathcal{S}_C$ and $\mathcal{S}_R$ the sets of all valid power exponent vectors for the polyhedral TIN scheme of  $\mathcal{IC}_C$ and $\mathcal{IC}_R$ respectively. In $\mathcal{IC}_C$, with any power allocation $\mathbf{r}^*=(r_1^*,r_2^*,...,r_K^*)\in\mathcal{S}_C$, the obtained GDoF value of User $k\in\langle K\rangle$ is $d_k^{\dag}$ in (\ref{sec_dkA}) and $d_k^{\dag}\geq 0$. And the obtained GDoF tuple is $\mathbf{d}=\{d_1^{\dag},d_2^{\dag},...,d_K^{\dag}\}\in\mathcal{P}_C$. According to (\ref{sec_dkA}) and (\ref{sec_dkB}), in $\mathcal{IC}_R$ with the same power allocation $\mathbf{r}^*$, for User $k\in\langle K\rangle$ the obtained GDoF value is $d_k^{\dag\dag}=d_k^{\dag}\geq 0$. Thus in $\mathcal{IC}_R$, $\mathbf{r}^*\in\mathcal{S}_R$ and $\mathbf{d}\in\mathcal{P}_R$. So we obtain that $\mathcal{P}_C\subseteq\mathcal{P}_R$. Similarly, we can also argue that $\mathcal{P}_R\subseteq\mathcal{P}_C$. Therefore, we have $\mathcal{P}_C=\mathcal{P}_R$. 

Finally, through contradiction we prove that for any feasible GDoF tuple in the polyhedral TIN region $\mathcal{P}$, $\mathcal{IC}_C$ and $\mathcal{IC}_R$ have the same set of locally optimal power allocations. For any feasible GDoF tuple $\mathbf{d}$ in the polyhedral TIN region, denote the sets of locally optimal solutions of $\mathcal{IC}_C$ and $\mathcal{IC}_R$ as $\mathcal{L}_C$ and $\mathcal{L}_R$, respectively. We assume that  $\mathcal{L}_C\neq \mathcal{L}_R$. Then we have a power allocation $\mathbf{r}$ such that $\mathbf{r}\in\mathcal{L}_C$ and $\mathbf{r}\not\in\mathcal{L}_R$ (or $\mathbf{r}\not\in\mathcal{L}_C$ and $\mathbf{r}\in\mathcal{L}_R$), which clearly contradicts the fact that with the same power allocation, $\mathcal{IC}_C$ and $\mathcal{IC}_R$ obtain the same GDoF tuple. Thus we complete the proof.

\section{Proof of Theorem \ref{PGPC_NP}}
In a $K$-user compound interference channel, we start with any achievable GDoF tuple $(d_1,d_2,...,d_K)$ in the polyhedral TIN region $\mathcal{P}$. According to the proof of Theorem \ref{T1_IC},  for all the achievable GDoF tuples in $\mathcal{P}$, there exist valid potential functions in the potential graph $D_p$, and all the directed circuits in $D_p$ have non-negative lengths. It is easy to argue that if in $D_p$  each directed circuit has a non-negative length, letting $p(v)$ ($\forall v\in V$) be the shortest length of any path ending at $v$ (starting from \emph{any} vertex) satisfies the required condition for the potential functions. A natural choice for the potential function value is the length of the shortest path starting from the grounded vertex $u$ to each vertex in $D_p$ , i.e., 
\begin{align}\label{p_constraint}
p(v_k^{[l_k]})=l_{k,\mathrm{dst}}\leq 0,~~ \forall k\in\langle K\rangle, \forall l_k\in\langle L_k\rangle,  
\end{align} 
and
\begin{align}
p(u)=0,
\end{align}
According to the definition of potential function, i.e., for any edge $(a,b)\in E$, $l(a,b)\geq p(b)-p(a)$, from (\ref{eq:samelength}) and (\ref{eq_lst1}) we obtain 
\begin{alignat}{2}
d_k&\leq \alpha_{kk}^{[l_k]}+l_{k,\mathrm{dst}}-(\alpha_{kj}^{[l_k]}+l_{j,\mathrm{dst}}),~~&&\forall k,j\in\langle K\rangle, k\neq j, \forall l_k\in\langle L_k\rangle, \label{ind1_1}\\
d_k&\leq \alpha_{kk}^{[l_k]}+l_{k,\mathrm{dst}},&&\forall k\in\langle K\rangle, \forall l_k\in\langle L_k\rangle \label{ind1_2}
\end{alignat}
From (\ref{ind1_1}) and (\ref{ind1_2}), we have
\begin{align}\label{ind_1}
d_k\leq\min_{l_k\in\langle L_k\rangle}\big\{ \alpha_{kk}^{[l_k]}+l_{k,\mathrm{dst}}-(\max_{j:j\neq k}\{\alpha_{kj}^{[l_k]}+l_{j,\mathrm{dst}}\})^+\big\},~~\forall k\in\langle K\rangle
\end{align}

Using $(l_{1,\mathrm{dst}},l_{2,\mathrm{dst}},...,l_{K,\mathrm{dst}})$ as the transmit power exponent vector, which satisfies the unit power constraint for each user according to (\ref{p_constraint}), and treating interference as noise at each receiver, the achievable GDoF of User $k$ is 
\begin{align}
\hat{d}_k&=\min_{l_k\in\langle L_k\rangle}\big\{ \max\{0,\alpha_{kk}^{[l_k]}+l_{k,\mathrm{dst}}-(\max_{j:j\neq k}\{\alpha_{kj}^{[l_k]}+l_{j,\mathrm{dst}}\})^+\}\big\}\\
&\geq \min_{l_k\in\langle L_k\rangle}\big\{ \alpha_{kk}^{[l_k]}+l_{k,\mathrm{dst}}-(\max_{j:j\neq k}\{\alpha_{kj}^{[l_k]}+l_{j,\mathrm{dst}}\})^+\big\}\\
&\geq d_k
\end{align}
which holds for all $k\in\langle K\rangle$ and thus completes the proof. 

\section{GDoF-based Synchronous Fixed-point Power Control}\label{app_FP}
The GSFPC algorithm for general $K$-user compound interference channels is specified as follows.

%\footnote{In the multiple multicast setting, the GSFPC algorithm can be implemented in a distributed fashion except for the initialize phase, which is equivalent to the centralized feasibility test in the conventional SINR-based power control algorithms as stated before. In the update phase, each receiver only needs the local measurement of its desired signal and interference strengths and sends this information back to its corresponding transmitter to complete the update.  We can also apply the GSFPC to its counterpart regular channel and get the same locally optimal power allocation with less complexity. However, in that case the update phase cannot be implemented in a decentralized way.}. 

\begin{algorithm}
\caption{GDoF-based synchronous fixed-point power control (GSFPC)}
\begin{algorithmic}
\STATE 1) Initialize: set $r_k(0)=l_{k,dst}$ for $k\in\langle K\rangle$;

\STATE 2) Update:
\begin{align}\label{fx}
r_k(n+1)=d_k-\min_{l_k\in\langle L_k\rangle}\big\{ \alpha_{kk}^{[l_k]}-(\max_{j:j\neq k}\{\alpha_{kj}^{[l_k]}+r_j(n)\})^+\big\},~~k\in\langle K\rangle
\end{align}
\STATE  where $n$ indexes discrete time slots.
\end{algorithmic}
\end{algorithm}

Note that for the conventional SINR-based fixed-point power control algorithms, the convergence is usually proved through the framework of standard interference function developed in \cite{Yates_SIF}, which is not applicable to the GDoF setting. The following theorem demonstrates the convergence of the GSFPC algorithm.

 \begin{theorem}\label{FP_PC}
In a $K$-user compound interference channel, for any achievable GDoF tuple in its polyhedral TIN region $\mathcal{P}$, the GSFPC algorithm converges to a locally optimal transmit power allocation.  Further, if there are multiple locally optimal power allocations $\bf{r}^*$ satisfying $\bf{r}^*$$\leq (l_{1,dst},l_{2,dst},...,l_{K,dst})$, denote by $\mathcal{R}_l$ the set of all such locally optimal solutions.  Then, $\mathbf{r}^{**}$, the locally optimal power allocation obtained from the GSFPC algorithm, satisfies $\mathbf{r}^{**}\geq\mathbf{r}^*$, $\forall \mathbf{r}^*\in\mathcal{R}_l$. 
\end{theorem}

\emph{Proof}:  Define $\mathbf{r}(n)=(r_1(n),r_2(n),...,r_K(n))$, and rewrite (\ref{fx}) in the vector form
\begin{align}
\mathbf{r}(n+1)=f(\mathbf{r}(n))
\end{align}
where $f: \mathbb{R}^K\rightarrow \mathbb{R}^K$ is the update function. 

We first prove the convergence of this fixed-point algorithm. To this end, we go through the following steps:
\begin{itemize}
\item First, prove that the sequence $\{\mathbf{r}(n)\}_{n=0}^{\infty}$ is decreasing. It is proved by induction. According to (\ref{ind_1}), we have
\begin{align}
&d_k\leq\min_{l_k\in\langle L_k\rangle}\big\{ \alpha_{kk}^{[l_k]}+l_{k,\mathrm{dst}}-(\max_{j:j\neq k}\{\alpha_{kj}^{[l_k]}+l_{j,\mathrm{dst}}\})^+\big\}\\
\Leftrightarrow & l_{k,\mathrm{dst}}\geq d_k-\min_{l_k\in\langle L_k\rangle}\big\{ \alpha_{kk}^{[l_k]}-(\max_{j:j\neq k}\{\alpha_{kj}^{[l_k]}+l_{j,\mathrm{dst}}\})^+\big\}\\
\Leftrightarrow & r_k(0)\geq d_k-\min_{l_k\in\langle L_k\rangle}\big\{ \alpha_{kk}^{[l_k]}-(\max_{j:j\neq k}\{\alpha_{kj}^{[l_k]}+r_j(0)\})^+\big\}
\end{align}
Writing the above inequality in the vector form, we have 
\begin{align}\label{r_nonempty}
\mathbf{r}(0)\geq f(\mathbf{r}(0))=\mathbf{r}(1)
\end{align}
Then, assume $\mathbf{r}(k)\leq \mathbf{r}(k-1)$. As $f$ is an increasing function, we have
\begin{align}
\mathbf{r}(k+1)=f(\mathbf{r}(k))\leq f(\mathbf{r}(k-1))=\mathbf{r}(k)
\end{align} 
Therefore,  $\{\mathbf{r}(n)\}_{n=0}^{\infty}$ is a decreasing sequence.

\item Second, prove there exists a fixed point $\mathbf{r}^*=(r_1^*,r_2^*,...,r_K^*)\leq \mathbf{r}(0)$ such that $\mathbf{r}^*=f(\mathbf{r}^*)$. It is easy to verify that the following conditions are satisfied: (i) $f$ is a  continuous increasing function, as the maximum/minimum of continuous functions is still continuous; (ii) From (\ref{r_nonempty}), the set $\mathcal{S}_{\mathbf{r}}=\{\mathbf{r}:\mathbf{r}\geq f(\mathbf{r})\}$ is nonempty; (iii) The set $\mathcal{S}_{\mathbf{r}}$ is bounded from below.  Therefore, according to the fixed point theorem in \cite{Fixed_Point} (Proposition 6 in \cite{Fixed_Point}), we know  that for $\mathbf{r}(0)\in\mathcal{S}_{\mathbf{r}}$, there exists a $\mathbf{r}^*\leq \mathbf{r}(0)$ such that $\mathbf{r}^*=f(\mathbf{r}^*)$.

\item Finally, prove that the sequence $\{\mathbf{r}(n)\}_{n=0}^{\infty}$ is bounded by induction. According to the second step, we have known that $\mathbf{r}(0)\geq \mathbf{r}^*$. Next, assume $\mathbf{r}(k)\geq \mathbf{r}^*$. We have 
\begin{align}
\mathbf{r}(k+1)=f(\mathbf{r}(k))\geq f(\mathbf{r}^*)=\mathbf{r}^*
\end{align}
Therefore, $\{\mathbf{r}(n)\}_{n=0}^{\infty}$ is bounded from below by $\mathbf{r}^*$.
\end{itemize}
Based on the above steps, we establish that $\{\mathbf{r}(n)\}_{n=0}^{\infty}$ is decreasing and bounded, thus convergent.  As a fixed-point algorithm is able to estimate a fixed point if and only if the sequence $\{\mathbf{r}(n)\}_{n=0}^{\infty}$ converges, and clearly the fixed point solution we obtained is a locally optimal power allocation for the target GDoF tuple, we complete the proof for the convergence of the GSFPC algorithm.

 It is also easy to prove that $\bf{r}^{**}$, the locally optimal power allocation obtained from the GSFPC algorithm, satisfies $\mathbf{r}^{**}\geq\mathbf{r}^*$, $\forall \mathbf{r}^*\in\mathcal{R}_l$.  Let $\mathbf{r}^*$ be any locally optimal power allocation satisfying $\mathbf{r}^*\leq\mathbf{r}(0)$. As in the proof above, we have $\{\mathbf{r}(n)\}_{n=0}^{\infty}$ is bounded from below by $\mathbf{r}^*$. Therefore, we establish that $\mathbf{r}^{**}=\lim_{n\rightarrow\infty}\mathbf{r}(n)\geq \mathbf{r}^*$, $\forall \mathbf{r}^*\in\mathcal{R}_l$. \hfill $\Box$

\bigskip

We also make another interesting observation for the GSFPC algorithm: for any Pareto optimal GDoF tuple in the polyhedral TIN region, in its locally optimal power allocation obtained from the GSFPC algorithm, i.e., $(l_{1,dst},l_{2,dst},...,l_{K,dst})$, there is at least one user transmitting at its full power. To prove it, we only need to show that for any feasible GDoF tuple, in its potential graph, there exists at least one $l_{k,\mathrm{dst}}=0$, $k\in\langle K\rangle$. From Remark \ref{remark_equ}, we have known that for a compound interference channel, the length of the shortest path from $u$ to $v_i^{[l_{m_i}]}$ ($\forall i\in\langle K\rangle$) in its potential graph $D_p$ is equal to that of the shortest path from $u$ to $\bar{v}_i$ in the potential graph $\bar{D}_p$ of its regular counterpart.  Thus we only need to consider $\bar{D}_p$. We prove the above result by contradiction. We assume that for a GDoF tuple $(d_1,d_2,...,d_K)$ in the polyhedral TIN region $\mathcal{P}$, all $l_{i,\mathrm{dst}}$'s in the potential graph $\bar{D}_p$ are strictly less than $0$. In $\bar{D}_p$, removing the vertex $u$ and all its connected edges, we obtain a complete digraph $D^*$. It is easy to verify that if in $\bar{D}_p$, $l_{i,\mathrm{dst}}<0$, then for the vertex $\bar{v}_i$, in the new graph $D^*$, there exists a vertex $\bar{v}_j$ such that the length of the shortest path from $\bar{v}_j$ to $\bar{v}_i$ is negative, where $ i,j\in\langle K\rangle$ and $i\neq j$. Otherwise, in the original graph $\bar{D}_p$ there must exist a negative-length circuit consisting of ($\bar{v}_k$, $u$) with a negative length and $(u,\bar{v}_k)$ with length $0$, $k\in\langle K \rangle$, which contradicts that the target GDoF tuple is achievable through the polyhedral TIN scheme. Next, in $D^*$, we start with an arbitrary vertex $\bar{v}_{m_1}$, $m_1\in\langle K\rangle$.  We assume the length of the shortest path from $\bar{v}_{m_2}$ to $\bar{v}_{m_1}$ is negative, where $m_2\in\langle K\rangle\backslash\{m_1\}$. Then for the vertex $\bar{v}_{m_2}$, to avoid negative-length circuits, there must exist a vertex $\bar{v}_{m_3}$ such that the length of the shortest path from this vertex to $\bar{v}_{m_2}$ is negative, where $m_3\in\langle K\rangle\backslash\{m_1,m_2\}$. We repeat the above procedure. Specifically,  for the vertex $\bar{v}_{m_i}$, to avoid negative-length circuits , there must exist a vertex $\bar{v}_{m_{i+1}}$ such that the length of the shortest path from this vertex to $\bar{v}_{m_i}$ is negative, where $m_{i+1}\in\langle K\rangle\backslash\{m_1,m_2,...,m_i\}$. Finally, for the vertex $\bar{v}_{m_K}$, there exists a vertex $\bar{v}_{m_j}$ such that the length of the shortest path from this vertex to $\bar{v}_{m_K}$ is negative, where $m_j\in\{m_1,m_2,...,m_{K-1}\}$, which results in a negative-length circuit in $D^*$ and thus in $\bar{D}_p$. It is clearly a contradiction to the fact that the target GDoF tuple is achievable through the polyhedral TIN scheme. Hence we complete the proof. 

%{\color{red}The above observation has some interesting implications. For example, }

\section{Proof of Theorem \ref{GPU_converge}}
For clarity, we first review how the GGPC algorithm works. In the initialize step, we obtain the initial power allocation $r_i(0)=l_{i,dst}$, $\forall i\in\langle K \rangle$, with which we achieve a GDoF tuple $(d_1(0),d_2(0),...d_K(0))$ dominating the target GDoF tuple $(d_1,d_2,...,d_K)$, according to Theorem \ref{PGPC_NP}. Note that at this point, the effective noise floor at each receiver is $0$.  In the first update step, each transmitter reduces its power by 
\begin{align*}
\Delta(0)=\min_{i}\{r_i(0)+\alpha_{ii}-d_i\},
\end{align*}
which is no less than $0$, as $r_i(0)+\alpha_{ii}\geq d_i(0)\geq d_i$, $\forall i\in\langle K \rangle$. Without loss of generality, we assume 
\begin{align*}
\arg\min_i\{r_i(0)+\alpha_{ii}-d_i\} = 1 
\end{align*}
After the first update, the transmit power of User 1 becomes
\begin{align}\label{T4_0}
r_1(1)=r_1(0)-\Delta(0) = d_1 - \alpha_{11} 
\end{align}
It is easy to check that with the updated power allocation, the achievable GDoF value of User $1$ is  
\begin{align}
d_1(1)=&\max\big\{0,r_1(0)-\Delta(0)+\alpha_{11}-\max_{j\neq 1} \{0,r_j(0)-\Delta(0)+\alpha_{1j}\}\big\}\\
=&\max\big\{0,r_1(0)-[r_1(0)+\alpha_{11}-d_1]+\alpha_{11}\label{T4_1}\big\}\\
=&\max\{0,d_1\}\\
=&d_1
\end{align}
where (\ref{T4_1}) follows 
\begin{align}
&d_1(0)\geq d_1\\
\Rightarrow& r_1(0)+\alpha_{11}-\max_{j\neq 1}\{0,r_j(0)+\alpha_{1j}\}\geq d_1\\
\Rightarrow & r_1(0)+\alpha_{11}-d_1\geq \max_{j\neq 1}\{0,r_j(0)+\alpha_{1j}\}\\
\Rightarrow & \Delta(0)  \geq \max_{j\neq 1}\{0,r_j(0)+\alpha_{1j}\}\\
\Rightarrow &  \Delta(0)  \geq \max_{j\neq 1}\{r_j(0)+\alpha_{1j}\}\\
\Rightarrow & \max_{j\neq 1} \{0,r_j(0)-\Delta(0)+\alpha_{1j}\} = 0 \label{T4_2}
\end{align}
From (\ref{T4_2}), we observe that after the first update,  for Receiver $1$ the interference from others is all below the effective noise floor, and Transmitter $1$ cannot further reduce its power  in (\ref{T4_0}) due to the effective noise floor. In other words, User $1$ cannot further lower its transmit power, since under the target GDoF constraint, it  ``hits'' the effective noise floor. While for other users $j\neq 1$, we have 
\begin{align}
d_j(1)=\max\big\{0,r_j(0)-\Delta(0)+\alpha_{jj}-\max_{i\neq j} \{0,r_i(0)-\Delta(0)+\alpha_{ji}\}\big\} \label{T4_3}
\end{align}
We consider the following two cases,
\begin{itemize}
\item $\max_{i\neq j}\{r_i(0)-\Delta(0)+\alpha_{ji}\} \leq 0$: In this case, from (\ref{T4_3}), we have 
\begin{align*}
d_j(1)=&\max\big\{0,r_j(0)-\Delta(0)+\alpha_{jj}\big\}\\
=&\max\big\{0,r_j(0)+\alpha_{jj}-\min_{i}\{r_i(0)+\alpha_{ii}-d_i\}\big\}\\
\geq & \max\big\{0,r_j(0)+\alpha_{jj}-[r_j(0)+\alpha_{jj}-d_j]\big\}\\
=&\max\{0,d_j\}\\
=&d_j
\end{align*} 
\item $\max_{i\neq j}\{r_i(0)-\Delta(0)+\alpha_{ji}\} > 0$: In this case, from (\ref{T4_3}), we obtain 
\begin{align*}
d_j(1)&=\max\big\{0,r_j(0)-\Delta(0)+\alpha_{jj}-\max_{i\neq j} \{r_i(0)-\Delta(0)+\alpha_{ji}\}\big\}\\
&=\max\big\{0,r_j(0)+\alpha_{jj}-\max_{i\neq j} \{r_i(0)+\alpha_{ji}\}\big\}\\
&=\max\big\{0,r_j(0)+\alpha_{jj}-\max_{i\neq j} \{0,r_i(0)+\alpha_{ji}\}\big\}\\
&=\max\{0,d_j(0)\}\\
&=d_j(0)\\
&\geq d_j
\end{align*}
\end{itemize}
Combining the above two cases, we establish that after the first update, User $j\neq 1$ still achieves an acceptable GDoF value. To sum up, after the first update, User 1 obtains the target GDoF value $d_1$ and achieves its transmit power limit due to the effective noise floor, and the achievable GDoF tuple $(d_1(1),d_2(1),...,d_K(1))$ dominates the target one. 

Next, in the second update, we fix the transmit power of User 1 and only update the transmit powers of others. Notably, in the following updates the fixed power of User 1 exerts a constant interference level to other users. Therefore, at this point the effective noise floor of Receiver $j\neq 1$ becomes $\max\{0,\alpha_{j1}+r_1(1)\}$. Also note in the sequel updates, all the other users can only reduce their transmit powers, so the achieved GDoF value of User 1 always remains as $d_1$. In the following, we just need to repeat the above argument for each update till the powers of all users are fixed by the GGPC algorithm. Specifically, in each update, transmit powers of certain users are reduced to the limits to achieve the target GDoF values, since they ``hit'' their effective noise floors, respectively, which are decided by the users whose transmit powers are fixed in the previous updates. And after each update, we always end up with an acceptable GDoF tuple which dominates the target one. 

Now we can prove Theorem \ref{GPU_converge} by contradiction. We assume that for a target GDoF tuple $(d_1,d_2,...,d_k)$,  the solution $\mathbf{r}^*=(r_1^*,r_2^*,...,r_K^*)$ obtained from the GGPC algorithm is not globally optimal. Then there is another power allocation $\mathbf{r}^{\dag}=(r_1^{\dag},r_2^{\dag},...,r_K^{\dag})$ which can also achieve the target GDoF tuple, and there exists at least one $i_0\in\langle K\rangle$ such that $r_{i_0}^{\dag}<r_{i_0}^*$.

In the GGPC algorithm, before the update step deciding the final power $r_i^*$ for User $i\in\langle K\rangle$, the set of users whose transmit powers have been fixed is denoted as $\mathcal{U}_i$. For example, if $r_{i}^*$ is determined in the first update step, $\mathcal{U}_i=\phi$. 

We assume that in the GGPC algorithm, User $i_0$'s final power allocation $r_{i_0}^*$ is limited by a user $i_1\in\mathcal{U}_{i_0}$. In other words, after the previous updates, among all the users belonging to the set $\mathcal{U}_{i_0}$, the User $i_1\in\mathcal{U}_{i_0}$, whose transmit power is $r_{i_1}^*$, gives the strongest interference to User $i_0$, and this strongest interference level is larger than $0$.\footnote{When there are multiple users giving the same strongest interference to User $i_0$, we can pick up User $i_1$ as any one of them to proceed the proof. } Clearly, if after the previous power updates, the strongest interference level to user $i_0$ from the users in $\mathcal{U}_{i_0}$ is no larger than $0$, then according to the GGPC algorithm, $r_{i_0}^*=d_{i_0}-\alpha_{i_0i_0}$. In this case, $r_{i_0}^*$ cannot be reduced any more to maintain the desired GDoF value $d_{i_0}$, which contradicts that $r_{i_0}^{\dag}<r_{i_0}^*$ can still achieve $d_{i_0}$.  Since $r_{i_0}^*$ can be reduced without affecting the achieved GDoF value $d_{i_0}$, the transmit power of User $i_1$ should also be reduced in order to lower the interference to User $i_0$, thus we have $r_{i_1}^{\dag}<r_{i_1}^*$. Apply the same argument to User $i_1$. Similarly we will have a user $i_2\in\mathcal{U}_{i_1}$ such that $r_{i_2}^{\dag}<r_{i_2}^*$. Repeat the same argument. Specifically, for User $i_{n-1}$, we have known that $r_{i_{n-1}}^{\dag}<r_{i_{n-1}}^*$. Then among all the users in $\mathcal{U}_{i_{n-1}}$, there exists a user $i_n\in\mathcal{U}_{i_{n-1}}$ giving the strongest interference to user $i_{n-1}$, and this strongest interference level is larger than $0$. As $r_{i_{n-1}}^*$ can be reduced without affecting the achieved GDoF value $d_{i_{n-1}}$, the transmit power of User $i_n$ should also be reduced in order to lower the interference to User $i_{n-1}$, hence we have $r_{i_n}^{\dag}<r_{i_n}^*$.  Assume the user whose final power allocation is determined in the first update and holds fixed afterwards is $i_m$. Finally we have $r_{i_m}^{\dag}<r_{i_m}^*$. Recall that in the GGPC algorithm, after the first update, we have $r_{i_m}^*=d_{i_m}-\alpha_{i_mi_m}$, which cannot be reduced further to maintain the target GDoF value. We obtain a contradiction that  with $r_{i_m}^{\dag}<r_{i_m}^*$ we can still achieve the target GDoF value for User $i_m$. Thus we complete the proof.

 \section{Proof of Theorem \ref{GPUC_converge}} \label{app_GPUC}
 The GGPC-C algorithm works similarly to the GGPC algorithm. Therefore, Theorem \ref{GPUC_converge} can also be proved by contradiction. We assume that for a target GDoF tuple $(d_1,d_2,...,d_K)$, the solution $\mathbf{r}^*=(r_1^*,r_2^*,...,r_K^*)$ obtained from the GGPC-C algorithm is not globally optimal. Then there is another power allocation $\mathbf{r}^{\dag}=(r_1^{\dag},r_2^{\dag},...,r_K^{\dag})$ which can also achieve the target GDoF tuple, and there exists at least one $i_0\in\langle K\rangle$ such that $r_{i_0}^{\dag}<r_{i_0}^*$. 

In the GGPC-C algorithm, before the update step deciding the final power $r_i^*$ for User $i\in\langle K\rangle$, the set of users whose transmit powers have been fixed is denoted as $\mathcal{U}_i$. We assume that in the GGPC-C algorithm, User $i_0$'s final power allocation $r_{i_0}^*$ is limited by a user $i_1\in\mathcal{U}_{i_0}$. In other words, after the previous updates, among all the users belonging to the set $\mathcal{U}_{i_0}$, the User $i_1\in\mathcal{U}_{i_0}$, whose transmit power is $r_{i_1}^*$, gives the strongest interference to the state $\bar{l}_{i_0}$ of User $i_0$, where 
\begin{align*}
\bar{l}_{i_0}=\arg\min_{l_{i_0}}\big\{\alpha_{i_0i_0}^{[l_{i_0}]}-\max_{i_j}\{0,\alpha_{i_0i_j}^{[l_{i_0}]}+r_{i_j}^*\}\big\}, ~~i_j\in\mathcal{U}_{i_0},
\end{align*}
and this strongest interference level is larger than $0$. (Also recall that after the GGPC-C algorithm, the state $\bar{l}_{i_0}$ of User $i_0$ achieves the target GDoF value $d_{i_0}$ exactly.) Clearly, if after the previous power updates, the strength level of the strongest interference to state $\bar{l}_{i_0}$ of User $i_0$ from the users in $\mathcal{U}_{i_0}$ is no larger than $0$, then $\bar{l}_{i_0}=\arg\min_{l_{i_0}}\{\alpha_{l_{i_0}l_{i_0}}^{[l_{i_0}]}\}$ and we have $r_{i_0}^*=d_{i_0}-\min_{l_{i_0}}\{\alpha_{i_0i_0}^{[l_{i_0}]}\}$ according to the GGPC-C algorithm. In this case, $r_{i_0}^*$ cannot be reduced any more to maintain the desired GDoF value $d_{i_0}$, which contradicts that $r_{i_0}^{\dag}<r_{i_0}^*$ can still achieve $d_{i_0}$.  Since $r_{i_0}^*$ can be reduced without affecting the achieved GDoF value $d_{i_0}$, the transmit power of User $i_1$ should also be reduced in order to lower the interference to state $\bar{l}_{i_0}$ of User $i_0$, thus we have $r_{i_1}^{\dag}<r_{i_1}^*$. Apply the same argument to User $i_1$. Similarly we will have a user $i_2\in\mathcal{U}_{i_1}$ such that $r_{i_2}^{\dag}<r_{i_2}^*$. Repeat the same argument. Specifically, for User $i_{n-1}$, we have known that $r_{i_{n-1}}^{\dag}<r_{i_{n-1}}^*$. Then among all the users in $\mathcal{U}_{i_{n-1}}$, there exists a user $i_n\in\mathcal{U}_{i_{n-1}}$ giving the strongest interference to state $\bar{l}_{i_{n-1}}$ of User $i_{n-1}$, where 
\begin{align*}
\bar{l}_{i_{n-1}}=\arg\min_{l_{i_{n-1}}}\big\{\alpha_{i_{n-1}i_{n-1}}^{[l_{i_{n-1}}]}-\max_{i_j}\{0,\alpha_{i_{n-1}i_j}^{[l_{i_{n-1}}]}+r_{i_j}^*\}\big\}, ~~i_j\in\mathcal{U}_{i_{n-1}},
\end{align*}
and this strongest interference level is larger than $0$. As $r_{i_{n-1}}^*$ can be reduced without affecting the achieved GDoF value $d_{i_{n-1}}$, the transmit power of User $i_n$ should also be reduced in order to lower the interference to state $\bar{l}_{i_{n-1}}$ of User $i_{n-1}$, hence we have $r_{i_n}^{\dag}<r_{i_n}^*$.  Assume the user whose final power allocation is determined in the first update and holds fixed afterwards is $i_m$. Finally we have $r_{i_m}^{\dag}<r_{i_m}^*$. Recall that in the GGPC-C algorithm, after the first update, we have $r_{i_m}^*=d_{i_m}-\min_{l_{i_m}}\{\alpha_{i_mi_m}^{[l_{i_m}]}\}$, which cannot be reduced further to maintain the target GDoF value. We obtain a contradiction that  with $r_{i_m}^{\dag}<r_{i_m}^*$ we can still achieve the target GDoF value for User $i_m$. Thus we complete the proof.

\end{document}